\documentclass[a4paper,11pt]{article}
\pdfoutput=1

\usepackage{jheppub}
\usepackage{graphicx}
\usepackage{amsmath}
\usepackage{xspace}
\usepackage{color}
\usepackage[utf8]{inputenc} 

\newcommand{\eq}[1]{eq.~\eqref{eq:#1}}
\newcommand{\eqs}[2]{eqs.~\eqref{eq:#1} and \eqref{eq:#2}}

\renewcommand{\sec}[1]{section~\ref{#1}}

\newcommand{\app}[1]{appendix~\ref{#1}}
\newcommand{\fig}[1]{figure~\ref{fig:#1}}

\newcommand{\Tab}[1]{table~\ref{tab:#1}}
\newcommand{\mycites}[1]{refs.~\cite{#1}}
\newcommand{\mycite}[1]{ref.~\cite{#1}}

\newcommand{\CP}{$\mathcal{CP}$}
\newcommand{\msbar}{$\overline{\text{MS}}$}
\newcommand{\Zsym}{$\mathbb{Z}_2$ }
\newcommand{\real}[1]{\mathrm{Re}\left(#1\right)}
\newcommand{\imag}[1]{\mathrm{Im}\left(#1\right)}
\newcommand{\df}{\mathrm{d}}

\newcommand{\nn}{\nonumber\\}
\newcommand{\abs}[1]{\lvert#1\rvert}

\newcommand{\ordo}[1]{\mathcal{O}\left(#1\right)}

\newcommand{\braket}[1]{\left\langle #1\right\rangle}
\newcommand{\code}[1]{\texttt{#1}}

\usepackage{feynmp-auto}

\allowdisplaybreaks[3]

\title{\Zsym breaking effects in 2-loop RG evolution of 2HDM}

\author[a]{Joel Oredsson,}
\author[a]{Johan Rathsman}

\emailAdd{joel.oredsson@thep.lu.se}
\emailAdd{johan.rathsman@thep.lu.se}

\affiliation[a]{Department of Astronomy and Theoretical Physics, Lund University, S\"{o}lvegatan 14A 223 62 Lund, Sweden}

\abstract{
We investigate the effects of a \Zsym symmetry in the \CP-conserving Two-Higgs-Doublet-Model (2HDM); which is often imposed to prevent Flavor-Changing-Neutral-Currents (FCNCs) at tree-level.
Specifically, we analyze how a breaking of the \Zsym symmetry spreads during renormalization group evolution; employing general 2-loop renormalization group equations that we have derived.
Evolving the model from the electroweak to the Planck scale, we find that while the case of an exact \Zsym symmetric 2HDM is very constrained, a soft breaking of the \Zsym symmetry extends the valid parameter space regions.
The effects of a hard \Zsym breaking in the scalar sector as well as the stability of the flavor alignment ansatz are also investigated.
We find that while a hard breaking of the \Zsym symmetry in the potential is problematic, since it speeds up the growth of quartic couplings, the generated FCNCs are heavily suppressed.
Conversely, we also find that hard \Zsym breaking in the Yukawa sector at most gives moderate \Zsym breaking in the potential; whereas the FCNCs can become quite sizable far away from the \Zsym symmetric regions.
}


\begin{document}

\preprint{
\begin{flushright}
LU TP 18-31\\
\today
\end{flushright}
}

\maketitle

\section{Introduction}\label{sec:intro}

Since the discovery of a 125 GeV scalar particle at the Large Hadron Collider (LHC), by the ATLAS \cite{Aad:2012tfa} and CMS \cite{Chatrchyan:2012xdj} collaborations, the quest to decipher its true nature has begun.
So far, it closely resembles the Standard Model (SM) Higgs boson \cite{Khachatryan:2016vau}; making the minimal SM electroweak sector a viable description.
However, a thorough experimental investigation is required to determine the exact dynamics of electroweak symmetry breaking.

On the theoretical side, there are still arguments for a non-minimal scalar sector.
One of the simplest extensions of the SM is to add another Higgs doublet.
There are numerous motivations for doing so, e.g.\ SUSY models need another Higgs doublet to cancel anomalies; realistic models of baryogenesis cannot be constructed out of the SM; and the vacuum metastability of the SM can be rendered stable.
It is also interesting in its own right to investigate the effects of having an extended scalar sector, much like there are three generations of fermions in the SM.

Two-Higgs-doublet models (2HDMs) exhibit a rich phenomenology with three new scalar particles and potential sources of \CP~violation, to name only a few of its features.
It has been studied for a long time and we refer to \mycite{Branco:2011iw} for a recent review.

An immediate problem when introducing an additional Higgs doublet to the SM is that the Yukawa sector, in its general form, gives rise to flavor changing neutral currents (FCNCs) at tree-level, that are severely constrained by experiments.
These arise since it is in general not possible to diagonalize the Yukawa matrices for both Higgs doublets simultaneously, when going to the fermion mass eigenbasis.
One mechanism, often employed in the literature, that solves this problem is to impose a discrete \Zsym symmetry on the 2HDM as proposed in \mycite{Glashow:1976nt,Paschos:1976ay}.

In this paper, we are interested in the effects of breaking such a \Zsym symmetry.
More specifically, we investigate how stable an approximate \Zsym symmetry of the 2HDM is in cases where it is explicitly broken by a small amount.
To investigate these effects, we employ a renormalization group (RG) analysis of the parameters of the model.
A small breaking of the \Zsym symmetry at one energy scale will, in general, spread during the RG evolution and generate additional \Zsym breaking parameters.
Our main aim of this work is to give a quantitative estimate of how large these \Zsym breaking parameters can be when requiring a model that is stable, unitary and perturbative up to some energy scale, where potentially some new physics will be present.
Experimental constraints are also taken into consideration by making sure that the model is within the bounds that are implemented in \code{HiggsBounds} \cite{Bechtle:2008jh,Bechtle:2011sb,Bechtle:2013wla} and \code{HiggsSignals} \cite{Bechtle:2013xfa}.
We look both at the cases with a softly broken as well as a hard broken \Zsym scalar potential.
The softly broken \Zsym scenario is very common in the literature and is also stable; in the sense that no dimensionless \Zsym symmetry breaking parameters are generated during the RG running. 
A hard breaking in the scalar potential, on the other hand, can have severe effects on the RG flow of the quartic couplings and potentially induce problematically large FCNCs in the Yukawa sector.

As a scenario of \Zsym breaking in the Yukawa sector, we look at the ansatz of flavor alignment in the 2HDM \cite{Pich:2009sp,Penuelas:2017ikk}.
Although the flavor alignment solves the problem of FCNCs at a specific energy scale, the ansatz is not stable during RG evolution \cite{Jung:2010ik,Ferreira:2010xe,Botella:2015yfa};
which also induces \Zsym breaking parameters in the scalar potential.

There has been plenty of work on the 2HDM's scalar and Yukawa sectors using the leading order 1-loop RGEs \cite{Bijnens:2011gd,Dev:2014yca, Ferreira:2015rha, Gori:2017qwg,Basler:2017nzu, Botella:2018gzy}.
More recently, even 2-loop RGEs have been employed \cite{Chowdhury:2015yja, Braathen:2017jvs, Krauss:2018thf}.
Also, 3-loop RGEs involving only the scalar quartic couplings have been derived in \mycite{Bednyakov:2018cmx}.
 
At 2-loop order, the RG analysis becomes more complicated since there is a stronger coupling between the scalar and Yukawa sectors; the quartic couplings enter the Yukawa beta functions first at 2-loop order.
This also makes it more interesting when looking at how a small breaking of the \Zsym symmetry can spread during RG evolution.
In \mycite{Chowdhury:2015yja, Braathen:2017jvs, Krauss:2018thf}, the authors employ 2-loop RGEs in their analysis, but on a \CP-conserving 2HDM, with a softly broken \Zsym symmetry.
We have derived the full set of 2-loop RGEs of the general, potentially complex, 2HDM and implemented them in an open-source \code{C++} program called \code{2HDME} \cite{Oredsson:2018vio}.
\\ \\
The structure of this paper is as follows:
In \sec{2HDMmodel}, we review the needed theory of the 2HDM. 
We discuss the technical details of the RG evolution in \sec{RGevolution}; how the RGEs are derived and the algorithm to perform the evolution.
The RGEs are used in parameter scans that we describe and present the results of in \sec{paramScan}.
Finally, we present our conclusions in \sec{conclusion}.

\section{The 2HDM}\label{2HDMmodel}

Here, we will  briefly review the content of the most general renormalizable 2HDM, i.e.\ the standard model, $SU(3)_c\times SU(2)_W\times U(1)_Y$, with an additional Higgs doublet. 
The 2HDM has been studied extensively and for a thorough review see \mycite{Branco:2011iw}.

The 2HDM contains two hypercharge $+1$ complex scalar $SU(2)$ doublets, $\Phi_1$ and $\Phi_2$.
First of all, since the scalar fields have identical quantum numbers, one can always perform a field redefinition of the scalar fields, i.e.\ a non-singular complex transformation $\Phi_a\to B_{a\bar{b}}\Phi_b$\footnote{The bar notation keep tracks of complex conjugation. That is, replacing a barred index to an unbarred corresponds to complex conjugation \cite{Davidson:2005cw,Haber:2006ue, Haber:2010bw}.},
where the matrix $B$ depends on 8 real parameters.
One uses four of these parameters to transform to canonical diagonal kinetic terms.
The Lagrangian of the 2HDM exhibits a $U(2)$ Higgs flavor symmetry, $\Phi_a \rightarrow U_{a\bar{b}} \Phi_b$; since the Lagrangian keeps the same form after such a transformation.
We will denote 2HDMs related by such Higgs flavor transformations as different bases of the 2HDM.
It is therefore very important, when investigating the general 2HDM, to work with basis-independent quantities.
A thorough basis-independent treatment of 2HDM has been developed in \mycites{Davidson:2005cw,Haber:2006ue, Haber:2010bw} and we will, mostly, follow their notational conventions.

\subsection{Generic basis}

The most general 2HDM gauge invariant renormalizable scalar potential can be written
\begin{align}\label{eq:GenericPotential}
		-\mathcal{L}_V=&m_{11}^2\Phi_1^\dagger\Phi_{1}+m_{22}^2\Phi_2^\dagger\Phi_{2}-(m_{12}^2\Phi_1^\dagger\Phi_{2}+\text{h.c.})+\frac{1}{2}\lambda_1\left(\Phi_1^\dagger\Phi_{1}\right)^2+\frac{1}{2}\lambda_2\left(\Phi_2^\dagger\Phi_{2}\right)^2\nonumber\\
		&+\lambda_3\left(\Phi_1^\dagger\Phi_{1}\right)\left(\Phi_2^\dagger\Phi_2\right)+\lambda_4\left(\Phi_1^\dagger\Phi_{2}\right)\left(\Phi_2^\dagger\Phi_1\right)
		\nonumber\\
		&+\left[\frac{1}{2}\lambda_5\left(\Phi_1^\dagger\Phi_2\right)^2+\lambda_6\left(\Phi_1^\dagger\Phi_1\right)\left(\Phi_1^\dagger\Phi_2\right)+\lambda_7\left(\Phi_2^\dagger\Phi_2\right)\left(\Phi_1^\dagger\Phi_2\right)+\text{h.c.}\right],
\end{align}
where $m_{12}^2,~\lambda_5,~\lambda_6$ and $\lambda_7$ are potentially complex while all the other parameters are real; resulting in a total of 14 degrees of freedom.
Three of these will be removed by the tadpole equations and one can be removed by a re-phasing of the second Higgs doublet.
Thus, the most general 2HDM potential exhibits 10 physical degrees of freedom; or 11 if one includes the SM Higgs VEV\footnote{We denote the $SU(3)_c\times SU(2)_W\times U(1)_Y$ gauge couplings as $g_3, g_2, g_1$ respectively.}, $v=2m_W/g_2\approx 246$ GeV.

After electroweak symmetry breaking, $SU(2)\times U(1)_Y\rightarrow U(1)_{\text{em}}$, both the scalar fields acquire a VEV, which can be expressed in terms of a unit vector in the Higgs flavor space
\begin{align}\label{eq:vev}
	\braket{\Phi_a} = \frac{v}{\sqrt{2}} \left( \begin{array}{c} 0 \\ \hat{v}_a \end{array} \right), && \hat{v}_a \equiv \left( \begin{array}{c} c_\beta \\ s_\beta e^{i\xi} \end{array}\right),
\end{align}
where the unit vector is normalized to $\hat{v}_{\bar{a}}^* \hat{v}_a=1$.
By convention, we take $0\leq \beta \leq \pi/2$ and $0\leq \xi\leq 2\pi$.
Here, we have used up all our gauge freedom, when setting the VEV in the lower component of the doublets with a $SU(2)$ transformation and removing any phase in the $\Phi_1$ VEV with a $U(1)_Y$ transformation. 
We also define $\hat{w}_b \equiv \hat{v}_{\bar{a}}^*\epsilon_{ab}$, where $\epsilon_{12}=-\epsilon_{21}=1$ and $\epsilon_{11}=\epsilon_{22}=0$.

The angle $\beta$ can be defined in terms of the ratio of the Higgs fields,
\begin{align}
  \tan\beta \equiv \abs{\braket{\Phi_2}}/\abs{\braket{\Phi_1}}.
\end{align}
It should be noted that $\tan\beta$ is an unphysical parameter if the two Higgs fields have identical quantum numbers \cite{Haber:2006ue}; since there is no preferred basis in that case.
In subsequent sections it will however be promoted to a physical quantity when a \Zsym symmetry is imposed on the 2HDM in a particular basis.

The physical scalar degrees of freedom, after electroweak symmetry breaking, corresponds to three neutral and one $U(1)_{em}$ charged pair of Higgs bosons, $\{h_1,h_2,h_3,H^{\pm}\}$.
In the \CP-conserving case, the neutral mass eigenstates have definite \CP~properties.
Instead of $h_{1,2,3}$, we can then work with the two \CP~even states, denoted $h$ and $H$\footnote{Defined by which one is the lighter one; $m_h \leq m_H$.}, as well as a \CP~odd one, $A$.

\subsection{Higgs basis}

One particularly convenient basis is the Higgs basis \cite{Branco:1999fs,Davidson:2005cw}, where only one Higgs field gets a VEV.
The Higgs basis fields in terms of the previously defined generic basis fields are
\begin{align}
	H_1 \equiv \hat{v}_{\bar{a}}^* \Phi_a, && H_2 \equiv \hat{w}_{\bar{a}}^* \Phi_a,
\end{align}
which acquire the VEVs
\begin{align}
	\braket{H_1^0} = v/\sqrt{2} , && \braket{H_2^0} = 0.
\end{align}
The transformation between the bases is $H_a=\hat{U}_{a\bar{b}}\Phi_b$, with the inverse $\Phi_a = \hat{U}_{a\bar{b}}^\dagger H_b$, where 
\begin{align}\label{eq:UHiggsdef}
	\hat{U} = \left( \begin{array}{cc} \hat{v}_1^* & \hat{v}_2^* \\
				\hat{w}_1^* & \hat{w}_2^* \end{array}\right)
				= \left( \begin{array}{cc} c_\beta & e^{-i\xi}s_\beta \\
				-e^{i\xi}s_\beta & c_\beta \end{array}\right).
\end{align}
The scalar potential in the Higgs basis takes a similar form as in the generic basis,
\begin{align}\label{eq:HiggsPotential}
	-\mathcal{L}_V =~& Y_1 H_1^\dagger H_1 + Y_2 H_2^\dagger H_2 + \left(Y_3H_1^\dagger H_2+h.c.\right) + \frac{1}{2}Z_1(H_1^\dagger H_1)^2 + \frac{1}{2}Z_2(H_2^\dagger H_2)^2\nn
	&+ \frac{1}{2}Z_3(H_1^\dagger H_1)(H_2^\dagger H_2)+ \frac{1}{2}Z_4(H_1^\dagger H_2)(H_2^\dagger H_1)\nn
	&+\left\{\frac{1}{2}Z_5(H_1^\dagger H_2)^2 + \left[Z_6(H_1^\dagger H_1) + Z_7(H_2^\dagger H_2)\right]H_1^\dagger H_2 + \text{h.c.}\right\}.
\end{align}
Minimizing this potential leads to the tree-level tadpole equations
\begin{align}
	Y_1 = -\frac{1}{2}Z_1 v^2, && Y_3 = -\frac{1}{2}Z_6 v^2.
\end{align}

During a Higgs flavor transformation of the generic basis, $\Phi_a \rightarrow U_{a\bar{b}}\Phi_b$, the Higgs fields transform as \cite{Haber:2006ue}
\begin{align}
	H_1 \rightarrow H_1, && H_2 \rightarrow (\det U) H_2.
\end{align} 
Thus from inspection of the Higgs potential in \eq{HiggsPotential}, it follows that $Y_1, Y_2, Z_1, Z_2, Z_3, Z_4$ are invariant, while
\begin{align}
	\{Y_3, Z_6, Z_7\} \rightarrow (\det U)^{-1} \{Y_3, Z_6, Z_7\}, && Z_5\rightarrow (\det U)^{-2} Z_5
\end{align} 
are pseudo-invariants under the Higgs flavor transformation.
The Higgs basis is therefore unique up to a rephasing of $H_2$, arising from the freedom to perform a Higgs flavor transformation.

\subsection{Yukawa sector}\label{yukawaSector}

The Yukawa interactions in the generic basis are\footnote{For simplicity, we do not include any mechanism to give masses to the neutrinos.}
\begin{align}
	-\mathcal{L}_Y=&\bar{Q}_L^0\cdot\tilde{\Phi}_{\bar{a}}\eta_a^{U,0}U_R^0+\bar{Q}_L^0\cdot\Phi_a\eta_{\bar{a}}^{D,0\dagger}D_R^0 + \bar{L}_L^0\cdot\Phi_a\eta_{\bar{a}}^{L,0\dagger}E_R^0 + \text{h.c.}~,
\end{align}
where the left-handed fermion fields in the weak eigenbasis are
\begin{align}
	Q_L^0 \equiv \left( \begin{array}{c} U_L^0 \\ D_L^0 \end{array} \right), && L_L^0 \equiv \left( \begin{array}{c} \nu_L^0 \\ E_L^0 \end{array} \right)
\end{align}
and $\tilde{\Phi}\equiv i\sigma_2 \Phi^*$.
To go to the fermion mass eigenbasis, we perform the biunitary transformation
\begin{align}
	F_L \equiv V_L^F F_L^0, && F_R \equiv V_R^F F_R^0,
\end{align}
where $F\in \{U,D,E\}$ is each fermion species.
The CKM matrix is composed out of the left-handed transformation matrices, $V_{CKM} \equiv V_L^UV_L^{D\dagger}$,
and the Yukawa matrices in the mass eigenbasis are obtained with biunitary transformations
\begin{align}
	\eta_a^F \equiv& V_L^F\eta_a^{F,0}V_R^{F\dagger}.
\end{align}
Going to the Higgs basis, we get
\begin{align}
	-\mathcal{L}_Y =~& \bar{Q}_L \tilde{H}_1 \kappa^U U_R + \bar{Q}_L H_1 \kappa^{D\dagger} D_R + \bar{L}_L H_1 \kappa^{L\dagger} E_R\nn
	&+ \bar{Q}_L \tilde{H}_2 \rho^U U_R + \bar{Q}_L H_2 \rho^{D\dagger} D_R + \bar{L}_L H_2 \rho^{L\dagger} E_R + \text{h.c.},
\end{align}
where $\kappa^F$ are the diagonal mass matrices,
\begin{align}
	\kappa^U =~& \hat{v}_{\bar{a}}^*\eta_a^U=\frac{\sqrt{2}}{v}\text{diag}(m_u,m_c,m_t),\nn
	\kappa^D =~& \hat{v}_{\bar{a}}^*\eta_a^D=\frac{\sqrt{2}}{v}\text{diag}(m_d,m_s,m_b),\nn
	\kappa^L =~& \hat{v}_{\bar{a}}^*\eta_a^L=\frac{\sqrt{2}}{v}\text{diag}(m_e,m_\mu,m_\tau),
\end{align}
and $\rho^F=\hat{w}_{\bar{a}}^*\eta_a^F$ are arbitrary complex $3\times 3$ matrices.

If the $\rho^F$ matrices are non-diagonal, there are FCNCs present at tree-level.
There exist multiple solutions to the problem of FCNCs in the 2HDM.
One of them is the idea of alignment \cite{Pich:2009sp}, which makes the ansatz that $\eta_1^{F,0}$ and $\eta_2^{F,0}$ are proportional to each other; with the consequence that $\kappa^F$ and $\rho^F$ can be diagonalized simultaneously.
We will parameterize the alignment ansatz by setting
\begin{align}\label{eq:alignmentAnsatz}
	\rho^F = a^F \kappa^F,
\end{align}
where the alignment parameter $a^F$ could be a c-number, but we will restrict our analysis to real coefficients.

As is well known, the alignment ansatz is not stable under RG evolution \cite{Jung:2010ik,Ferreira:2010xe,Botella:2015yfa}, since there is no symmetry protecting it.
Thus, \eq{alignmentAnsatz} is only valid at a specific energy scale and during RG evolution FCNCs will be generated.

One ansatz, that allows for small FCNCs is the Cheng-Sher ansatz \cite{PhysRevD.35.3484} which parameterizes the non-diagonal Yukawa matrices as
\begin{align}\label{eq:chengSher}
	\rho^F_{ij} \equiv \lambda_{ij}^F\frac{\sqrt{2m_im_j}}{v}, && \text{(no implicit sum over $i,j$)}
\end{align}
where the $\lambda_{ij}^F$ should be order one.
We will use this parameterization when comparing generated FCNCs later on.

One can get limits on the non-diagonal elements from neutral meson mixing at low energy scales, see \cite{Bijnens:2011gd} and references therein.
Even though we will discuss FCNCs that are generated at high energy scales, we will use the limit $\lambda_{i\neq j}^F\leq 0.1$ as a general measure of sizable non-diagonal elements.

\subsubsection*{\Zsym symmetry}

Another solution to the FCNC problem is the idea of imposing a \Zsym symmetry \cite{Glashow:1976nt, Paschos:1976ay}, where the Higgs doublets have opposite charge.
By only coupling each right-handed fermion to one of the Higgs doublets, one can diagonalize all Yukawa matrices and end up with an alignment like in \eq{alignmentAnsatz}, but with the alignment parameters set to be equal to either $-\tan\beta$ or $\cot\beta$.
The \Zsym symmetric Yukawa sector is thus a special case of an aligned Yukawa sector; however, because of the symmetry, a \Zsym symmetric 2HDM is stable under RG evolution, in that no FCNCs are generated.

There are four different variations of this symmetry that are summarized in \Tab{Z2symmetries}.
It is also worth noting that the previously unphysical parameter $\tan\beta = v_2/v_1$ gets promoted to a physical degree of freedom, when imposing a \Zsym symmetry, see \mycite{Haber:2006ue} for details.
 
\begin{table}[h!]
		\centering
    		\begin{tabular}{|c|cccccc|}\hline
		Type	 & $U_R$	& $D_R$	& $L_R$ & $a^U$ & $a^D$ & $a^L$\\
		\hline 
		I & + & + & + &$\cot\beta$ & $\cot\beta$ & $\cot\beta$\\
		II & + & $-$ & $-$ &$\cot\beta$ & $-\tan\beta$ & $-\tan\beta$\\
		Y & + & $-$ & + &$\cot\beta$ & $-\tan\beta$ & $\cot\beta$\\
		X & + & + & $-$ &$\cot\beta$ & $\cot\beta$ & $-\tan\beta$\\ \hline
		\end{tabular}
		\caption{Different \Zsym symmetries that can be imposed on the 2HDM. $\Phi_1$ is odd($-1$) and $\Phi_2$ is even($+1$). For every type of \Zsym symmetry, the $\rho^F$ matrices become proportional to the diagonal mass matrices, $\rho^F = a^F \kappa^F$. }
		\label{tab:Z2symmetries}
\end{table}

Requiring the generic potential in \eq{GenericPotential} to be symmetric under a \Zsym symmetry fixes $m_{12}^2=\lambda_6=\lambda_7=0$.
Often in the literature, one relaxes the \Zsym symmetry by letting $m_{12}^2$ be non-zero; thus breaking the symmetry softly.
It should be noted that the mass parameters of the scalar potential do not enter the beta functions of the quartic or Yukawa couplings, thus a soft symmetry breaking does not alter the RG evolution of the dimensionless parameters.

One can also consider hard breaking of the \Zsym symmetry in the scalar potential, by having small $\lambda_{6,7}$.
This introduces additional terms in the RGEs and also induces FCNCs at different energy scales after 2-loop RGE running.
It is one of this article's main goal to investigate how severe these effects can be; see \sec{hardScan} for more details.

\subsection{\CP-invariant limit}

The scalar potential and vacuum are \CP-conserving if and only if \cite{Davidson:2005cw,Gunion:2005ja,Lavoura:1994fv,Botella:1994cs} 
\begin{align}
	\imag{Z_5^* Z_6^2} = \imag{Z_5^* Z_7^2} = \imag{Z_6^*Z_7} = 0.
\end{align}
If these conditions are satisfied, it is possible to perform a phase shift of $H_2$ to end up with a Higgs basis composed of purely real parameters making it a \textit{real basis}.

An exact \Zsym symmetric potential only has one potentially complex parameter, $\lambda_5$, that can be made real with a phase shift.
Also the phase of $\braket{\Phi_2}$ can be removed with a field redefinition; thus one ends up with a 2HDM with all real VEVs and couplings. 
Thus an exact \Zsym symmetry enforces \CP-invariance in the scalar sector.
As previously mentioned, one can relax the imposed \Zsym symmetry constraint by allowing for a softly broken symmetry from a non-zero $m_{12}^2$ parameter in the generic basis.
However, doing so does no longer guarantee \CP-invariance. 
To make sure that the model is still \CP-invariant, one can restrict the parameters such that \cite{Haber:2015pua}
\begin{align}
	\imag{\lambda_5^* (m_{12}^2)^2} = 0, &&\text{and}&& \lambda_5 \leq \frac{\abs{m_{12}^2}}{v_1v_2};
\end{align}
which will guarantee the absence of both explicit and spontaneous \CP-violation in the scalar sector.
It would be interesting to look at how \CP-violation affects the RGE running of the parameters, but this is beyond the scope of this work.
We will therefore assume \CP-invariance in the scalar sector; making it possible to always work in a real basis
\footnote{In principle, there is \CP-violation in the Yukawa sector arising from the CKM matrix, but this amount of \CP-violation is so small that it is irrelevant for our analysis.}.

\subsection{Hybrid basis}\label{hybridBasis}

When imposing a softly broken \Zsym symmetry on the \CP-conserving 2HDM, the number of quartic couplings is reduced to 5 real ones.
It is often very convenient to work with a basis where some quartic couplings are substituted for tree-level mass parameters, for physical clarity.
However, when doing numerical parameter scans, it can be computationally expensive to work with bases with many masses.
A bad choice of scalar masses easily corresponds to large quartic couplings that break perturbativity and unitarity. 
It is therefore a good idea to set up a basis with a mixture of scalar masses, angles and quartic couplings.
Such a hybrid basis is worked out in \mycite{Haber:2015pua} and is summarized in \Tab{hybridBasis}, which we will make use of in upcoming sections.

\begin{table}[h!]
		\centering
    	\begin{tabular}{|c|c|}\hline
			Parameter 	&  Description\\
			\hline
			$\{m_{h}, m_H\}$ 	& \CP~even neutral Higgs masses\\
			$\tan\beta$ 	& Ratio of VEVs in \Zsym symmetric basis\\
			$\cos(\beta-\alpha)$ & Mixing angle of neutral \CP~even mass matrix\\
			$\{Z_{4}, Z_5, Z_7\}$ & Real quartic Higgs couplings\\
			\hline
		\end{tabular}
		\caption{Hybrid basis for a softly broken \Zsym symmetric \CP-invariant 2HDM from \mycite{Haber:2015pua}. Imposing an exact \Zsym symmetry fixes $Z_7$ according to \eq{Z7sym}, with $\bar{m}^2=0$.} 
		\label{tab:hybridBasis}
\end{table}

The hybrid basis contains the two tree-level masses of the neutral \CP~even scalars,
\begin{align}
	m_{H,h}^2 = \frac{1}{2}\left[ m_A^2 + (Z_1 + Z_5)v^2 \pm \sqrt{\left[m_A^2 + (Z_5 - Z_1)v^2\right]^2 + 4Z_6^2v^4}\right],
\end{align}
where $m_A^2 = Y_2 + \frac{1}{2}(Z_3 + Z_4 - Z_5)v^2$ is the mass of the \CP-odd scalar and $m_h\leq m_H$.

In addition to $\beta$, the hybrid basis contains an angle $\alpha$ and we follow the convention that $0\leq \beta-\alpha \leq \pi$.
Here, we simply give the relations \cite{Haber:2015pua}
\begin{align}
  c_{\beta-\alpha} =~& \frac{-Z_6v^2}{\sqrt{(m_H^2 - m_h^2)(m_H^2 - Z_1v^2)}},\\
  s_{\beta-\alpha} =~& \frac{\abs{Z_6}v^2}{\sqrt{(m_H^2 - m_h^2)(Z_1v^2 - m_h^2)}},
\end{align}
which are valid for the case of $Z_6\neq 0$.

The three quartic couplings $Z_{4,5,7}$ are free parameters in the hybrid basis while
\begin{align}
    Z_1 =~& \frac{m_h^2}{v^2} s_{\beta-\alpha}^2 + \frac{m_H^2}{v^2} c_{\beta-\alpha}^2,\\ 
	Z_2 =~& Z_1 + 2(Z_6+Z_7)\cot 2\beta,\label{eq:Z2sym}\\
	Z_3 =~& Z_1 - Z_4 -Z_5 +2 Z_6\cot 2\beta-(Z_6-Z_7)\tan 2\beta,\label{eq:Z3sym}\\
	Z_6 =~& \frac{m_h^2 - m_H^2}{v^2} c_{\beta-\alpha} s_{\beta-\alpha}. 
\end{align}
There is also the relation
\begin{align}
	Z_7 v^2 =~&  Z_6v^2 + 2\cot 2 \beta\left(m_{H}^2s_{\beta-\alpha}^2 + m_{h}^2 c_{\beta-\alpha}^2 -\bar{m}^2\right),\label{eq:Z7sym}
\end{align}
where
\begin{align}
	\bar{m}^2 \equiv \frac{m_{12}^2}{c_\beta s_\beta}.
\end{align}
For an exact \Zsym symmetry, $\bar{m}^2=0$ gives that $Z_7$ is not a free parameter in this basis.

Finally, the mass of the charged Higgs boson is given by
\begin{align}
	m_{H^\pm}^2 = Y_2 +Z_3 v^2 /2.
\end{align}

\section{Renormalization group evolution}\label{RGevolution}

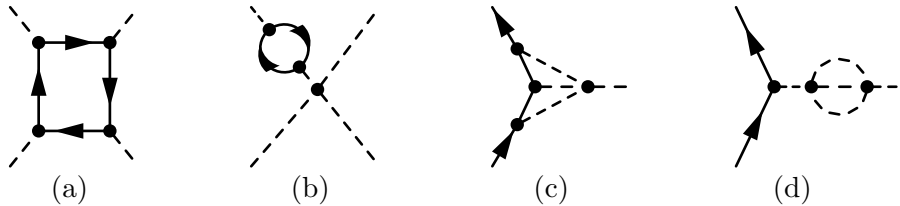
\begin{figure}[h!]
  \begin{center}
  \begin{tabular}{cccc}
  $
    \vcenter{\hbox{
      \begin{fmffile}{quartic1}
        \begin{fmfgraph*}(60,60)
          \fmfleft{i1,i2}
          \fmfright{o1,o2}
          \fmf{dashes}{i1,v1}
          \fmf{dashes}{i2,v2}
          \fmf{dashes}{v3,o1}
          \fmf{dashes}{v4,o2}
          \fmf{fermion,tension=0.4}{v1,v2,v4,v3,v1}
          \fmfdotn{v}{4}
        \end{fmfgraph*}
      \end{fmffile}
    }}
  $
  &
  $
    \vcenter{\hbox{
      \begin{fmffile}{quartic2}
        \begin{fmfgraph*}(60,60)
          \fmfleft{i1,i2}
          \fmfright{o1,o2}
          \fmf{dashes, tension=10}{i1,v1}
          \fmf{dashes}{i2,v2}
          \fmf{fermion, left, tension=0.3}{v2,v3,v2}
          \fmf{dashes}{v3,v1}
          \fmf{dashes}{v1,o1}
          \fmf{dashes, tension=10}{v1,o2}
          \fmfdotn{v}{3}
        \end{fmfgraph*}
      \end{fmffile}
    }}
    $
  &
  $
    \vcenter{\hbox{
      \begin{fmffile}{yuk1}
        \begin{fmfgraph*}(60,60)
          \fmfleft{i1,i2}
          \fmfright{o1}
          \fmf{fermion}{i1,v1}
          \fmf{plain}{v1,v2,v3}
          \fmf{fermion}{v3,i2}
          \fmf{dashes,tension=0.1}{v1,v4}
          \fmf{dashes,tension=0.1}{v2,v4}
          \fmf{dashes,tension=0.1}{v3,v4}
          \fmf{dashes,tension=0.4}{v4,o1}
          \fmf{phantom, tension=0.3}{v2,o1}
          \fmfdotn{v}{4}
        \end{fmfgraph*}
      \end{fmffile}
    }}
  $
  &
  $
    \vcenter{\hbox{
      \begin{fmffile}{yuk2}
        \begin{fmfgraph*}(60,60)
          \fmfleft{i1,i2}
          \fmfright{o1}
          \fmf{fermion}{i1,v1,i2}
          \fmf{dashes,tension=0.3}{v1,v2}
          \fmf{dashes,left,tension=0.1}{v2,v3,v2}
          \fmf{dashes,tension=0}{v2,v3}
          \fmf{dashes,tension=0.3}{v3,o1}
          \fmf{phantom, tension=0.5}{v1,o1}
          \fmfdotn{v}{3}
        \end{fmfgraph*}
      \end{fmffile}
    }}
  $
  \\
  (a) & (b) & (c) & (d) 
\end{tabular}
\end{center}
\caption{Diagrams that couple the scalar and Yukawa RGE sectors. The quartic couplings get Yukawa contributions in their beta functions from diagrams of the type in (a) and (b) already at 1-loop order. The quartics enter the Yukawa beta functions first at 2-loop order through diagrams like the ones in (c) and (d).}
\label{fig:rgeDiagrams}
  \end{figure}

The behavior of the parameters of the 2HDM during RG evolution is the focus of attention in this paper.
Even though there are many parameters in the general 2HDM that are essentially free, requiring ``good'' behavior when evolving the model in energy restricts the parameter space severely.
An RGE analysis is thus useful when looking for potential fine-tuning, stability of the model and valid energy ranges before new physics would need to come in.

A careful matching to physical observables should be done to get the most precise determination of the 2HDM's \msbar-parameters.
Including higher order corrections to observables can then be important, since it may shift the parameters by a non-trivial amount.
It is especially important when considering RG evolution of the model.
A common procedure in the literature is to do the matching at one loop order lower than the loop order of the RGEs;
there are however indications of that N-loop running should be combined with N-loop matching \cite{Braathen:2017jvs}.
In this work, we are more interested in the general behavior of the \msbar-parameters during RG evolution and not so much in the exact determination of the physical observables.
We will thus restrict ourselves to calculate 1-loop pole masses of the scalar particles as the only quantum corrections.
These corrections can indeed be very large in certain parts of the 2HDMs parameter space,  as will be shown in subsequent sections.

We will impose theoretical constraints, namely perturbativity, unitarity and stability, that are needed to arrive at a consistent model.
To check against experimental data from collider searches, we use \code{HiggsBounds} \cite{Bechtle:2008jh, Bechtle:2011sb, Bechtle:2013wla} and \code{HiggsSignals} \cite{Bechtle:2013xfa}.

\subsection*{Perturbativity}

Since our analysis of the RG running of the 2HDM uses the RGEs, calculated with perturbation theory, perturbativity must always be satisfied during the RG evolution; breaking it would render the whole analysis meaningless.
The couplings can therefore not be too large and we will impose the upper limit
\begin{align}
	\abs{\lambda_i} \leq 4\pi,
\end{align}
to specify perturbativity.
The RGEs make up a strongly coupled set of ordinary differential equations, which implies that once any coupling gets large others soon follow; unless there is some kind of cancellation occurring among the parameters.
Breaking of perturbativity at some energy scale can therefore be seen as an indication of a Landau pole at a scale not far from it.

\subsection*{Unitarity}

Unitarity of the model is very important for consistency, since it is needed for a well defined S-matrix, and enforcing it puts constraints on the eigenvalues of the S-matrix for scalar scattering.
At high energies, only the quartic couplings contribute to the amplitudes; they are therefore the ones affected by the unitarity limit.
In \mycite{Ginzburg:2005dt}, the conditions for tree-level unitarity in a general 2HDM has been worked out.
They take the form of upper limits on eigenvalues of scattering matrices, composed of the quartic couplings.
For completeness, we give these matrices in \app{unitarity}.

\subsection*{Bounded Potential}

To ensure that the vacuum is a stable minimum of the potential, we require that the potential is bounded from below.
This means that the potential should be positive when the field values go to infinity for any direction in field space.
The sufficient conditions for a general renormalizable 2HDM have been worked out in \mycite{Ivanov:2006yq, Ivanov:2007de}, which we will make use of.
Enforcing stability constrains the quartic couplings in the scalar potential and, for completeness, we state the conditions in \app{stability}.

In this work, we are only enforcing tree-level stability conditions on the potential.
Sometimes, this is a too strong constraint since loop corrections can render the potential stable.
A more formal test of stability would be to check the stability of the effective potential.
Thus, if one instead were to check the stability of the 1-loop effective potential, it would relax the stability constraint as in \mycite{Krauss:2018thf}; however, this is beyond the scope of this work.
In some cases, we will omit the stability constraint when investigating the maximal effects of symmetry breaking; in order not to miss any region of parameter space that might exhibit interesting behavior.

\subsection*{Experimental bounds}

Constraints coming from LEP, the Tevatron and the LHC are taken into consideration using \code{HiggsBounds} \cite{Bechtle:2008jh, Bechtle:2011sb, Bechtle:2013wla}; which uses experimental cross section limits to determine if a certain parameter point is excluded at a 95\% C.L.~.
To ensure that the lightest Higgs boson resembles the 125 GeV Higgs signal observed at the LHC, we use \code{HiggsSignals} \cite{Bechtle:2013xfa}.
It performs a statistical goodness-of-fit test by calculating the $\chi^2$ for each parameter point.
The $\chi^2$ is used to calculate a p-value with the number of degrees of freedom set to the number of free parameters in the scalar potential; we exclude points with a p-value less than 0.05.
We also use 2HDMC \cite{Eriksson:2009ws} to calculate the decay widths and cross sections needed for \code{HiggsBounds} and \code{HiggsSignals}.

\subsection{Derivation of 2-loop RGEs}

Although there are some works using 2-loop RGEs for analyzing 2HDMs \cite{Braathen:2017jvs,Chowdhury:2015yja, Krauss:2018thf}, they all work with different \Zsym symmetries and the RGEs used are therefore not applicable to our more general scenario
\footnote{Refs.\ \cite{Braathen:2017jvs, Krauss:2018thf} are using 2-loop RGEs derived by \code{SARAH} \cite{Staub:2008uz, Staub:2013tta} and \mycite{Chowdhury:2015yja} is using \code{PyR@TE} \cite{Lyonnet:2013dna}. As have been pointed out in \mycite{Bednyakov:2018cmx, Schienbein:2018fsw}, older versions of these software miss non-diagonal anomalous dimension terms in the case of models containing scalar fields with equal quantum numbers. The issue is resolved in \mycite{Schienbein:2018fsw} and newer versions of \code{SARAH} and \code{PyR@TE} should give the correct RGEs.}.

The general expressions for the 2-loop RGEs of massless parameters in any renormalizable gauge theory were first written down in the seminal papers of Machacek and Vaughn \cite{Machacek:1983tz,Machacek:1983fi,Machacek:1984zw}.
These were then supplemented with the 2-loop RGEs of massive parameters in \mycite{Luo:2002ti}, which is the source that we have used to derive the 1- and 2-loop RGEs for a general 2HDM. 
Care should be taken though, when working with quantum field theories with multiple indistinguishable scalar fields, since the formulas in  \mycite{Machacek:1983tz,Machacek:1983fi,Machacek:1984zw,Luo:2002ti} are written for the case of an irreducible representation of the scalar fields.
The anomalous dimension of scalar fields with equal quantum numbers is non-diagonal; the fields mix during RG evolution.
In 2HDMs, the fields obey the RGEs
\begin{align}
	\mu \frac{\df \Phi_i}{\df \mu} = \gamma_{ij}\Phi_j,
\end{align}
where $\gamma_{ij}$ is the 2-dimensional anomalous dimension.
Refs.\ \cite{Bednyakov:2018cmx, Schienbein:2018fsw} discuss this subtlety  at great length and we have independently reached the same conclusions.

As cross checks, we compared the SM limit with known SM RGEs as well as to \Zsym symmetric RGEs found in the literature.

In \mycite{Ginzburg:2008kr} it is argued that one needs to consider scalar kinetic mixing in RG analyses of quantum field theories with multiple scalar fields with equal quantum numbers.
We claim that this is not needed, since there is no physical parameter related to scalar kinetic mixing.
The issue can be resolved differently depending on the renormalization schemes; which we explain in more detail in \mycite{Bijnens:2018}.
Here, the scalar mixing phenomenon only manifests itself in the previously mentioned 2-dimensional non-diagonal anomalous dimension of the Higgs fields.
The anomalous dimension sets the evolution of the Higgs VEVs in our scheme; with the implication that $\tan\beta$ runs in energy.
Thus the Higgs flavor transformation matrix in \eq{UHiggsdef} is $\mu$-dependent.

In the general case of no \Zsym symmetry, the equations are very long and we will not write them down in this article.
They are instead available in the \code{C++} code \code{2HDME} \cite{Oredsson:2018vio}.

\subsection{Technical details of numerical code}

Evolving the 2HDM in energy corresponds to solving a system of 129\footnote{3 gauge couplings; 2 complex VEVs; 6 complex Yukawa matrices; 2 real and 1 complex mass parameters; 4 real and 3 complex quartic couplings.} coupled ordinary differential equations.
We have developed the \code{C++} code \code{2HDME} \cite{Oredsson:2018vio} to perform this task.
It uses \code{GSL} \cite{GSL} to solve the RGEs and the library \code{Eigen} \cite{eigenweb} for linear algebra operations.

To match the 2HDM to the $\mu$ dependent parameters of the SM at the top quark pole mass scale, $M_t=173.4$ GeV \cite{ATLAS:2014wva}, we used the following input as boundary conditions\footnote{For formal consistency, one should perform a, at least, 1-loop matching of the 2HDM to the experimental data. In practice, we only do this for the scalar potential, by requiring one Higgs boson with mass $\approx 125$ GeV. We assume that the effects of matching the other parameters are negligible for our analysis.}:
\begin{itemize}
	\item	The \msbar~fermion masses are used to fix the Yukawa matrix elements in the fermion mass eigenbasis. We use the ones from \mycite{Xing:2007fb}:
	 \begin{align}\label{eq:fermionMasses}
	 m_u =~& 1.22\text{ MeV}, &  m_c =~& 0.590\text{ GeV}, &  m_t =~& 162.2 \text{ GeV},\nn
     m_d =~& 2.76\text{ MeV}, &  m_s =~& 52\text{ MeV}, & m_b =~&  2.75\text{ GeV},\nn
     m_e =~& 0.485289396\text{ MeV}, & m_\mu =~& 0.1024673155\text{ GeV}, & m_\tau =~&  1.74215 \text{ GeV}.
     \end{align}
	
	\item Gauge couplings from \mycite{Buttazzo:2013uya}:
	\begin{align}
		g_1 =~& 0.3583,\nn
		g_2 =~& 0.64779,\nn
		g_3 =~& 1.1666,
	\end{align}
	for $U(1)_Y$, $SU(2)_W$ and $SU(3)_c$ respectively.
	
	\item For the CKM matrix, we use the standard parametrization
	\begin{align}\label{eq:CKM}
		V_{CKM} = \left( \begin{array}{ccc} c_{12}c_{13} & s_{12}c_{13} & s_{13}e^{-i\delta} \\ 
		-s_{12}c_{23} - c_{12}s_{23}s_{13}e^{i\delta} & c_{12}c_{23}-s_{12}s_{23}s_{13}e^{i\delta} & s_{23}c_{13} \\ 
		s_{12}s_{23} - c_{12}c_{23}s_{13}e^{i\delta} & -c_{12}s_{23}-s_{12}c_{23}s_{13}e^{i\delta}& c_{23}c_{13}\end{array}\right),
	\end{align}
	where the angles in terms of the Wolfenstein parameters are
	\begin{align}
		s_{12} =~& \lambda,\nn
		s_{23} =~& A\lambda^2,\nn
		s_{13}e^{i\delta} =~& \frac{A\lambda^3(\bar{\rho} + i\bar{\eta})\sqrt{1-A^2\lambda^4}}{\sqrt{1-\lambda^2}\left[1-A^2\lambda^4(\bar{\rho}+i\bar{\eta})\right]}.
	\end{align}
	The numerical values 
	\begin{align}
		\lambda =~& 0.22453,\nn
        A =~& 0.836,\nn
        \bar{\rho} =~& 0.122,\nn
        \bar{\eta} =~& 0.355,
	\end{align}
	are extracted from the PDG \cite{PDG}. 
	
	\item The SM Higgs VEV is taken to be $v= (\sqrt{2} G_F)^{-1/2} = 246.21971$ GeV \cite{Buttazzo:2013uya}.

\end{itemize} 
To calculate the loop corrected pole masses of the Higgs bosons, we make use of the Fortran code \code{SPheno} \cite{Porod:2003um,Porod:2011nf}, together with model files generated by \code{SARAH} \cite{Staub:2008uz,Staub:2013tta}.
We choose to work only at 1-loop order when calculating the loop corrected masses for computational efficiency.
This choice incorporates the largest loop corrections, but also treats all scalars equally.

\subsection{Evolution algorithm}\label{algorithm}

Here, we briefly describe the algorithm used when scanning the parameter space of the 2HDM:
\begin{itemize}
	\item Initialize the 2HDM with the SM values at $\mu = M_t$.
	The scalar potential is randomly generated in the hybrid basis; however, only potentials with a 1-loop corrected mass of 125$\pm$5 GeV for one of the scalar particles are accepted.
	Note that although we use the hybrid basis as a first input, we transform to and evolve the parameters in the generic basis.
	
	The Yukawa sector is set in the fermion mass eigenbasis.
	First, $\kappa^F$ is set with the fermion masses in \eq{fermionMasses} as input.
	The $\rho^F$ are then fixed to be proportional to the $\kappa^F$ matrices, either by a \Zsym symmetry or flavor ansatz.
	We then go to the fermion weak eigenbasis in the scalar generic basis with
	 \begin{align}
	   \eta_a^{U,0} =~& \hat{v}_a \kappa^U + \hat{w}_a \rho^U,\nn
	   \eta_a^{D,0} =~& \hat{v}_a \kappa^D V_{CKM}^\dagger + \hat{w}_a \rho^D V_{CKM}^\dagger,\nn
	   \eta_a^{L,0} =~& \hat{v}_a \kappa^L + \hat{w}_a \rho^L.
	 \end{align}
	 Here, we made the arbitrary choice to put the CKM matrix in the down sector by setting $V_L^D = V_{CKM}^\dagger$ and $V_L^U=\mathbb{I}$. 
	 This choice does not affect the results in any way.
	 
	 \item The parameters in the generic basis, i.e.\ the set $\{g_i, v_a, \eta_a^{F,0}, m_{ij}^2, \lambda_i \}$, are evolved according to the RGEs.
	 This includes the VEVs, $v_a \equiv  \hat{v}_a v$, which in our scheme obey the same RGE as the scalar fields, i.e.
	 \begin{align}
	 	\mu \frac{\df}{\df \mu} v_a = \gamma_{ab}v_b,
	 \end{align}
   where $\gamma_{ab}$ is defined in the Landau gauge. 
   A consequence of this is that $\tan\beta$ is $\mu$-dependent.
	 
	 \item As previously mentioned, the fact that $\tan\beta$ is $\mu$-dependent implies a $\mu$-dependent transformation of the generic basis to the Higgs basis with the matrix in \eq{UHiggsdef}. 
	 The parameters in the Higgs basis, $\{Y_i,Z_i, \kappa^{F,0}, \rho^{F,0}\}$, are at each step calculated with this $\hat{U}_{a\bar{b}}(\mu)$. 
	 Also, $\kappa^F$, $\rho^F$ and $V_{CKM}$ are calculated with biunitary transformations.
	 
   \item The RG evolution is stopped when perturbativity is broken or when the (reduced) Planck scale is reached, which we take to be $10^{18}$ GeV. 
   We define the breakdown scale, $\Lambda$, as the lowest energy scale where either perturbativity, unitarity or tree-level stability is violated.
   Note that this means that we continue the evolution even if the unitarity or stability constraints are broken.
   In this way, we can study each constraint individually and also loosen the stability constraint since this is applied at tree-level.
   
\end{itemize}

\subsection{RG evolution example}

\begin{figure}[h!]
\begin{center}
\begin{tabular}{cc}
\includegraphics[trim=0.5cm 1.5cm 1cm 0.5cm,clip,height=0.35\textwidth]{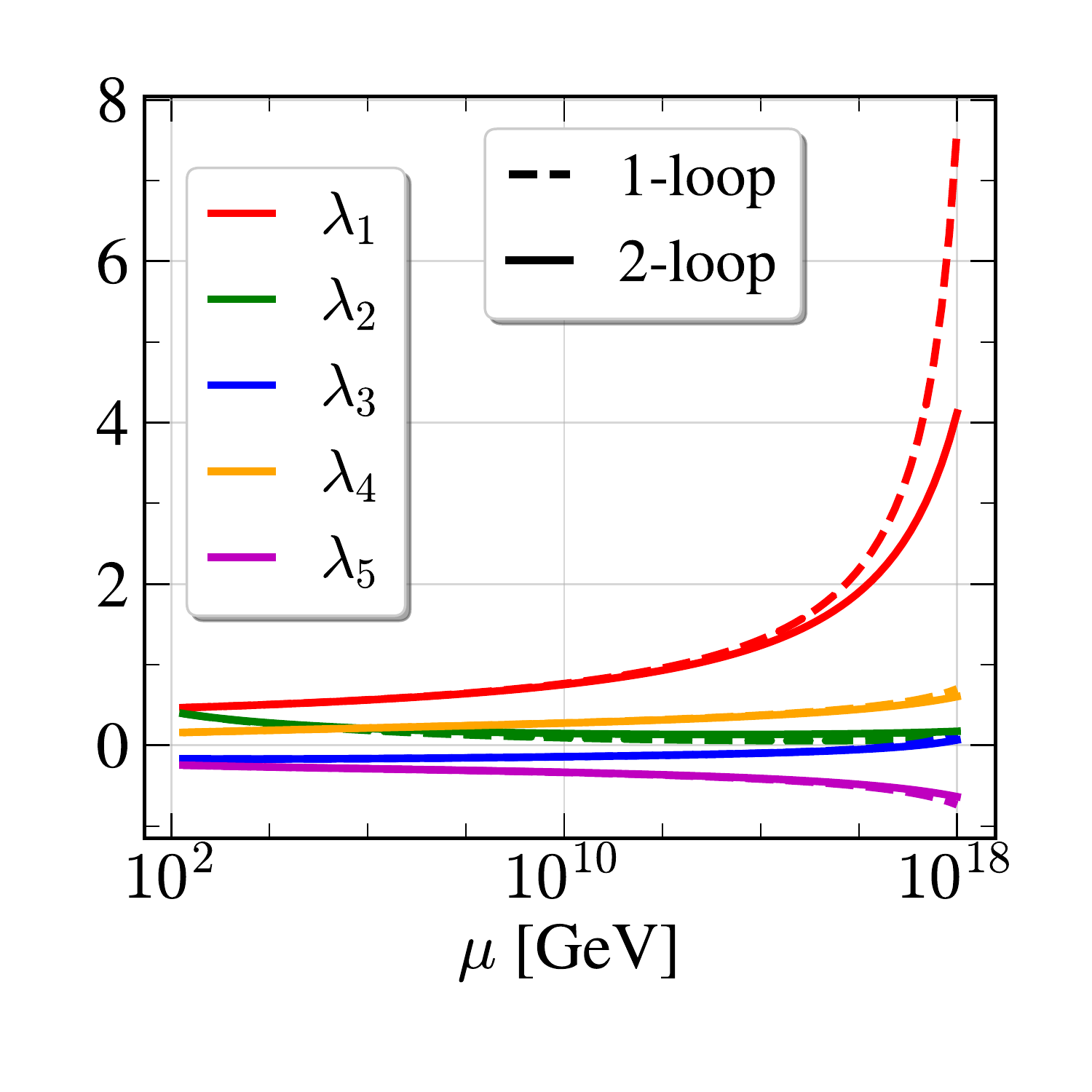}
\includegraphics[trim=0.5cm 1.5cm 1cm 0.5cm,clip,height=0.35\textwidth]{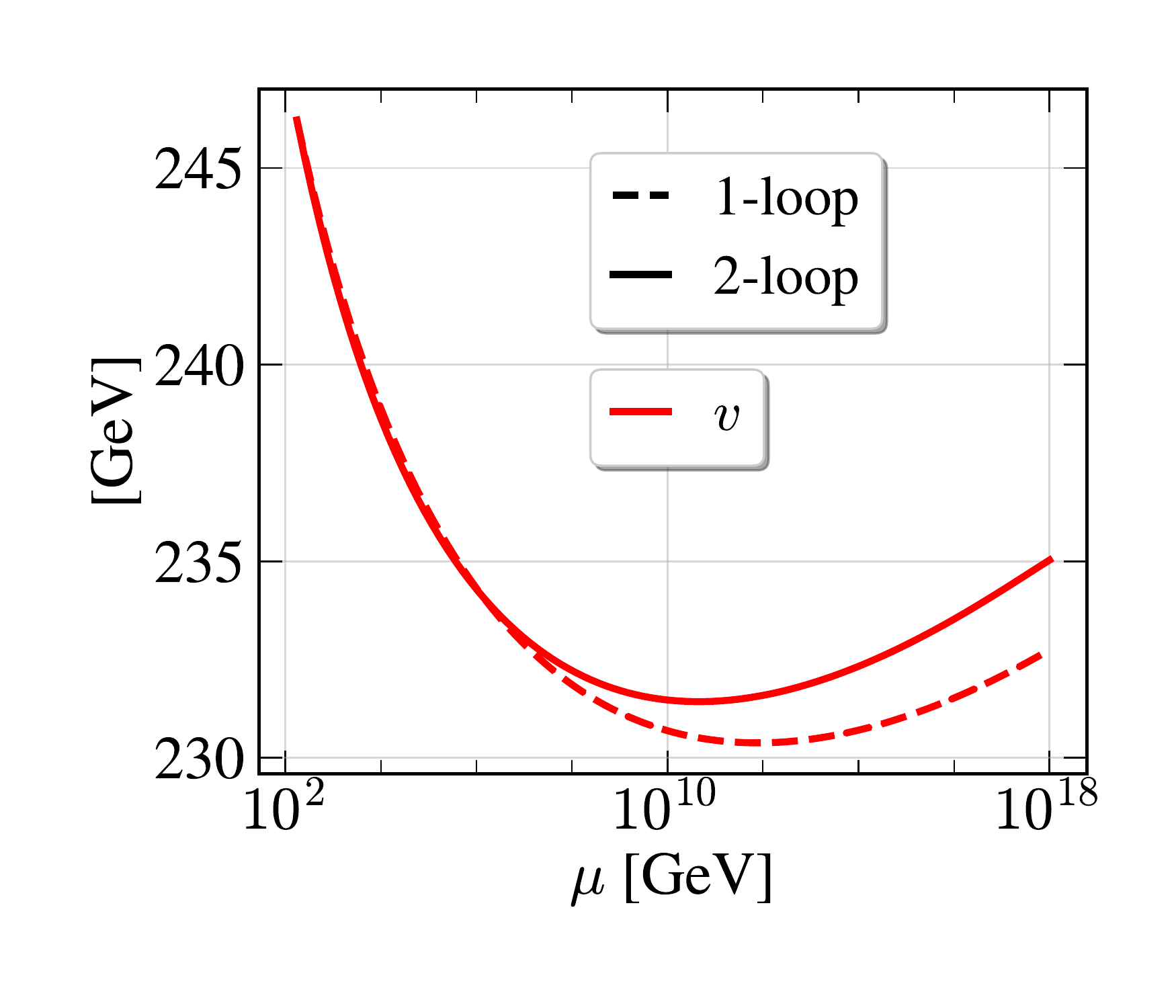}\\
\includegraphics[trim=0.5cm 1.5cm 1cm 0.5cm,clip,height=0.35\textwidth]{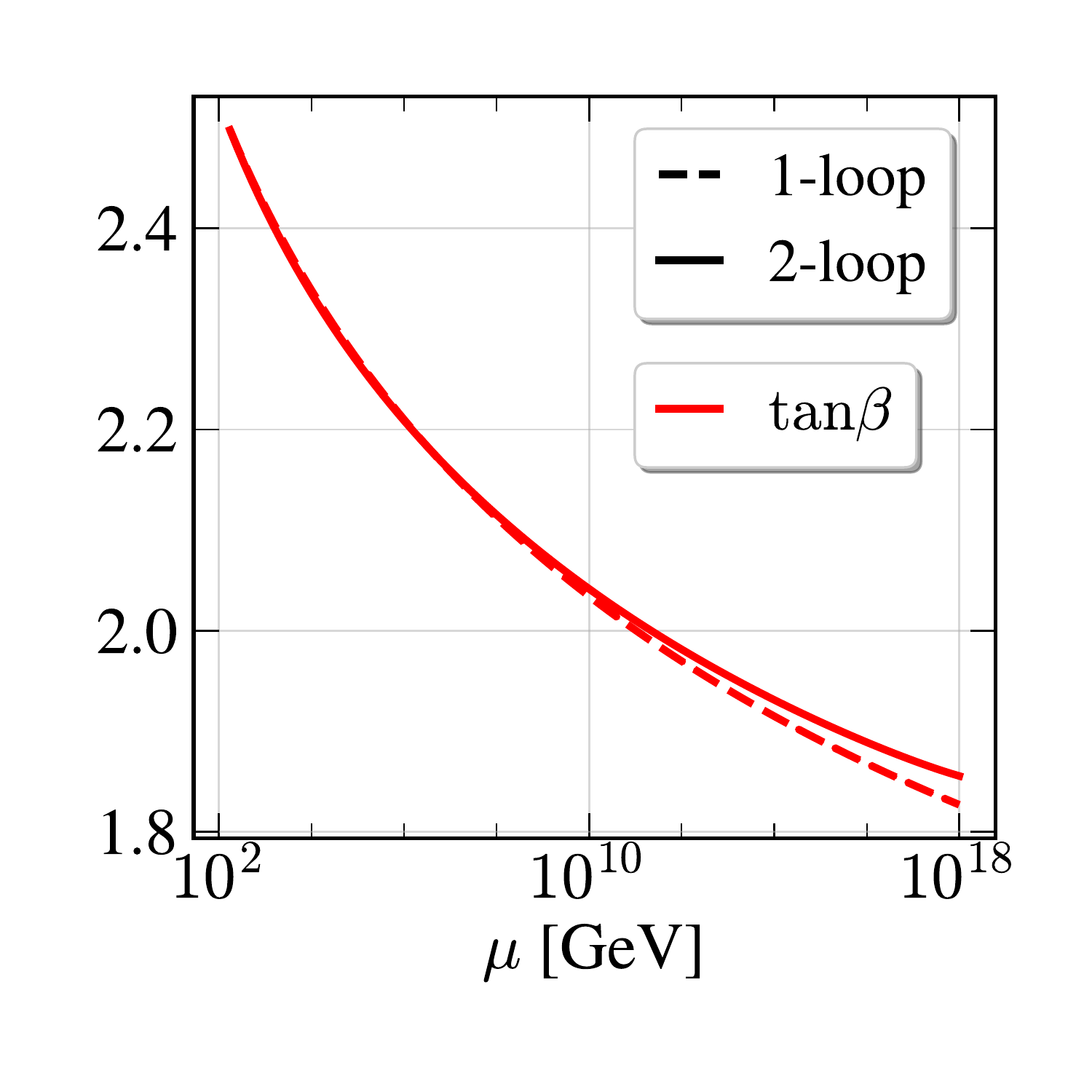}
\includegraphics[trim=0.5cm 1.5cm 1cm 0.5cm,clip,height=0.35\textwidth]{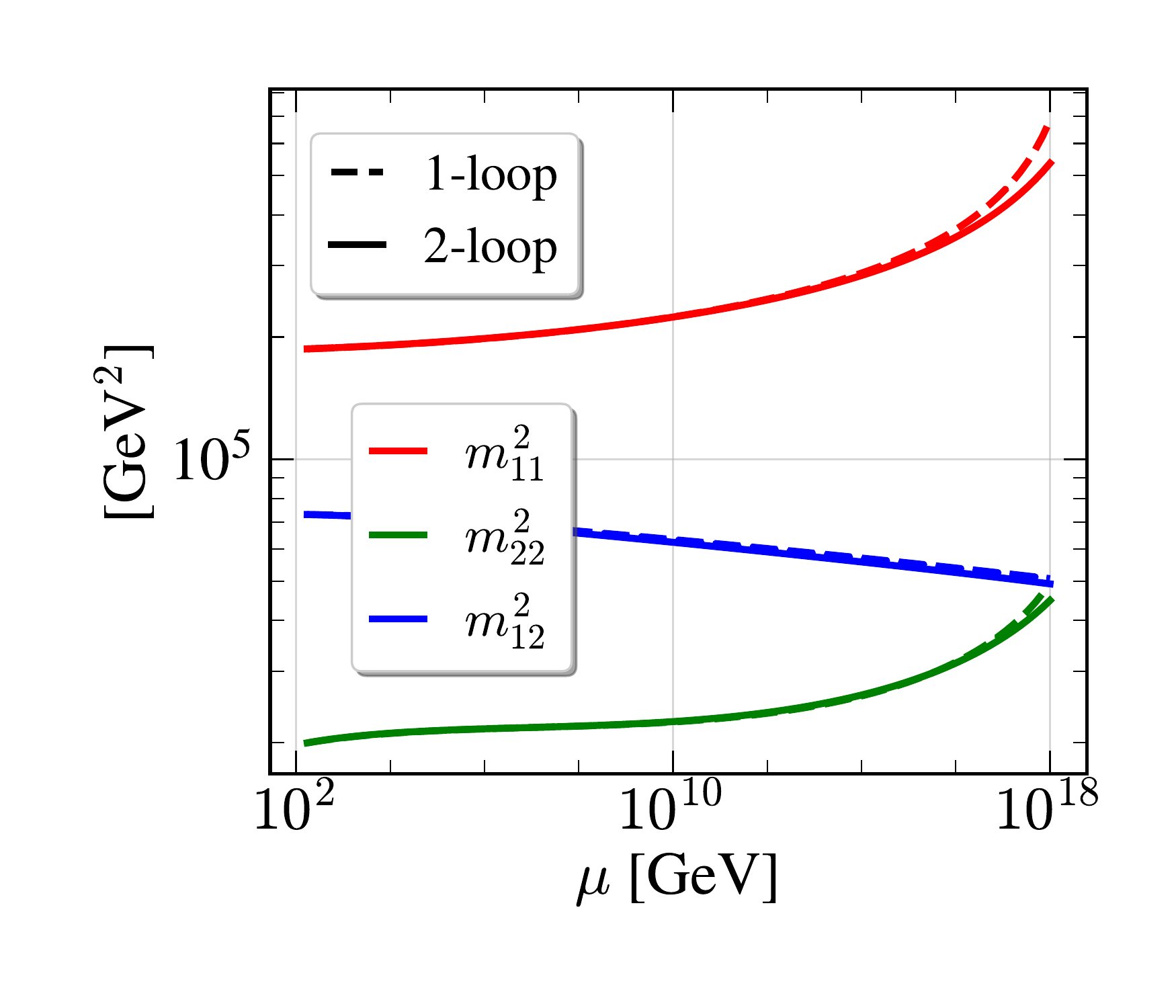}
\end{tabular}
\caption{Evolution of the type-I softly broken \Zsym parameter point in \eq{gen2}.}
\label{fig:gen2Demo}
\end{center}
\end{figure}

As a first example of the RG evolution of a 2HDM and to see the qualitative behavior of the parameters, we here evolve the type-I softly broken \Zsym parameter point\footnote{The mass parameters $M_{11}^2$ and $M_{22}^2$ are set by the tadpole equations.}
\begin{align}\label{eq:gen2}
  \tan\beta =~&  2.49772,\nn
  M_{12}^2 =~&   72993.4\text{ GeV}^2,\nn
  \lambda_1 =~&  0.467183,\nn
  \lambda_2 =~&  0.394626,\nn
  \lambda_3 =~&  -0.165783,\nn
  \lambda_4 =~& 0.159849,\nn
  \lambda_5 =~&  -0.245042,\nn
  \lambda_6 =~& \lambda_7 = 0.
\end{align}
which is perturbative, stable and unitary all the way to the Planck scale.
The scalar tree-level and 1-loop corrected masses for this parameter point are
\begin{align}
	m_h^{\text{tree}} =~& 119.356 \text{ GeV}, &&& m_h =~& 122.191 \text{ GeV}, \nn
	m_H^{\text{tree}} =~& 470.982 \text{ GeV}, &&& m_H =~& 469.629 \text{ GeV}, \nn
	m_A^{\text{tree}} =~& 475.809 \text{ GeV}, &&& m_A =~& 474.406 \text{ GeV}, \nn
	m_{H^\pm}^{\text{tree}} =~& 462.732 \text{ GeV}, &&& m_{H^\pm} =~& 461.325 \text{ GeV}.
\end{align}
A comparison of the 2-loop and 1-loop evolution of the parameters in the generic basis is shown in \fig{gen2Demo}.
This parameter point will also be the initial boundary condition of one of the alignment scans in \sec{yukScan}.

\section{Parameter space scan}\label{paramScan}

To probe the RG behavior of the 2HDM's parameter space, we construct random scans of the free parameters in scenarios with different levels of \Zsym symmetry.
Although \CP~violation would be very interesting to investigate, we restrict ourselves to \CP-invariant scalar potentials for simplicity, where a \textit{real basis} always exists.
The only source of \CP-violation is then the $\delta$ phase in the CKM matrix, defined in \eq{CKM}; however, its effects turn out to be negligible.

As explained in \sec{algorithm}, we employ the hybrid basis reviewed in \sec{hybridBasis} as a starting point in the generation of random parameter points.
The number of free parameters, however, depends on the nature of the \Zsym symmetry.
We will consider three different levels of \Zsym symmetry in the scalar potential:
\begin{itemize}
	\item \textbf{Scenario I}: 2HDM with an exact \Zsym symmetry\footnote{An exact \Zsym symmetric 2HDM also implies \CP-invariance.}. 
	This fixes $m_{12}^2=\lambda_6=\lambda_7=0$; which makes $Z_7$ no longer a free parameter since it is determined by \eq{Z7sym}.
	\item \textbf{Scenario II}: Softly broken \Zsym symmetry, with non-zero $m_{12}^2$. 
	In the hybrid basis, this corresponds to having $Z_7$ as a free parameter.
	\item \textbf{Scenario III}: Hard broken \Zsym symmetry in scalar sector that allows for small real values of $\lambda_6$ and $\lambda_7$.
	In the hybrid basis this means that $Z_2$ and $Z_3$ will deviate from \eqs{Z2sym}{Z3sym}.
\end{itemize}
There are 6, 7 and 9 degrees of freedom in each scenario respectively.

The tree-level mass can deviate significantly from the 1-loop corrected mass and we therefore start from a tree-level mass $m_h^{\text{tree}}\in [10,130]$ GeV.
For the heavier \CP~even Higgs boson, we scan a tree-level mass $m_H\in[150,1000]$ GeV.
To have $m_h$ to be SM-like, $c_{\beta-\alpha}$ should be close to zero; however, we use the generous range $c_{\beta-\alpha}\in[-0.5,0.5]$.
We also take $\tan\beta \in [1.1, 50]$\footnote{Although we actually generate random $\beta$ angles in the corresponding range from a flat distribution. The parameter region $\tan\beta =1$ is excluded for convenience, since it is singular in our parametrization of the potential; however, the singular behavior can be removed by re-parametrization.}.
The quartic couplings $Z_{4,5,7}$ are in the range -$\pi$ to $\pi$.

In the hard \Zsym breaking scenario III, we add deviations from the softly broken scenario in the generic basis and scan over the range $\lambda_{6,7}\in \left[10^{-5}, 2\right]$, although logarithmically distributed.

We have performed numerical scans of these three \Zsym scenarios; producing $10^5$ random parameter points for each one, that are perturbative, unitary and stable at the electroweak scale with a 1-loop corrected $m_h=125\pm 5$ GeV.
To take into account constraints from collider searches, the events are also filtered by only keeping points that are allowed by \code{HiggsBounds} and \code{HiggsSignals}.
As mentioned in \sec{algorithm}, these models are then evolved until perturbativity is lost or the Planck scale at $10^{18}$ GeV has been reached
and we define the breakdown energy of each parameter point, $\Lambda$, as the energy scale where either perturbativity, unitarity or stability is violated.

The specific type-choice of Yukawa symmetry has been checked not to matter for our conclusions about \Zsym symmetry breaking effects; the scans of scenario I-III are performed with a type I symmetry.

Lastly, we also investigate the effects of breaking the \Zsym in the Yukawa sector:
\begin{itemize}
  \item \textbf{Scenario IV.a}: 
  Starting from the softly broken \Zsym 2HDM defined in \eq{gen2}, we make an alignment ansatz $\rho^F = a^F \kappa^F$ and vary each $a^F\in\mathbb{R}$ separately, \textit{i.e.}\ one-by-one in the ranges $a^U\in[0.01, 10]$ and $a^D\in[-100,100]$; for larger values the corresponding Yukawa couplings become too large for perturbation theory to apply.
	
  \item \textbf{Scenario IV.b}: 
  We make a similar softly broken \Zsym parameter scan as in scenario II; however, we fix $\tan\beta=2$ and draw random $a^U\in[-1, 1]$ and $a^D\in[-50,50]$; which are limited to the most relevant ranges.
  For this scenario, we only consider models where neither perturbativity, unitarity nor stability is violated and we want to compare the generated \Zsym breaking parameters at a common scale.
  The scale choice is motivated by being able to find parameter points that are valid throughout the selected $a^F$ ranges.
  A higher scale would make the parameter scan more costly in terms of computer time, but we have checked that the dependence on the scale used is very small.
  In the end, we choose the scale to be $\mu=10^{10}$ GeV.
\end{itemize}
We impose the same theoretical and experimental constraints at the electroweak scale as in scenario I-III.

All of these scenarios correspond to a bottom-up running of the parameters.
Hence, we are only probing how large \Zsym breaking parameters can be at the EW scale and how fast an alignment ansatz breaks down.
Formally, there are of course no limits on the \Zsym breaking parameter at high energy scales; which have not been probed experimentally.
However, following \mycite{Bijnens:2011gd} we assume that their values should not differ by a large amount compared to the ones at the EW scale.
A large sensitivity, such that the \Zsym breaking parameters change with many orders of magnitude in the RGE evolution, would indicate that the underlying assumptions of the model are either fine-tuned or not stable.
It would also be interesting to look at top-down scenarios, where one starts of with small deviations from a \Zsym symmetric parameter point at a high scale and then see how large the \Zsym breaking effects are at the EW scale.
Such a scenario with a more symmetrical theory at higher scales would perhaps be deemed a more natural one, but we will leave such an investigation to future works.
However, our analysis provides an indication of how large such effects could be.
\newline 

\begin{table}[h!]
  \centering
      \begin{tabular}{|c|ccc|}\hline
  Scenario	 & Allowed by \code{HiggsBounds}	& Allowed by \code{HiggsSignals}	& Allowed by both \\
  \hline 
  I &  33 \% & 67 \% & 25 \% \\
  II &  49 \% & 69 \% & 41 \% \\
  III & 51 \% & 68 \% & 41 \% \\
  \hline
  \end{tabular}
  \caption{How many parameter points that pass the constraints from \code{HiggsBounds} and \code{HiggsSignals} in scenario I to III.}
  \label{tab:HBHSstats}
\end{table}

\begin{figure}[h!]
\begin{center}
\begin{tabular}{ccc}
\hspace{-1.2cm}
\includegraphics[trim=0cm 1cm 0.5cm 0cm,clip,height=0.35\textwidth]{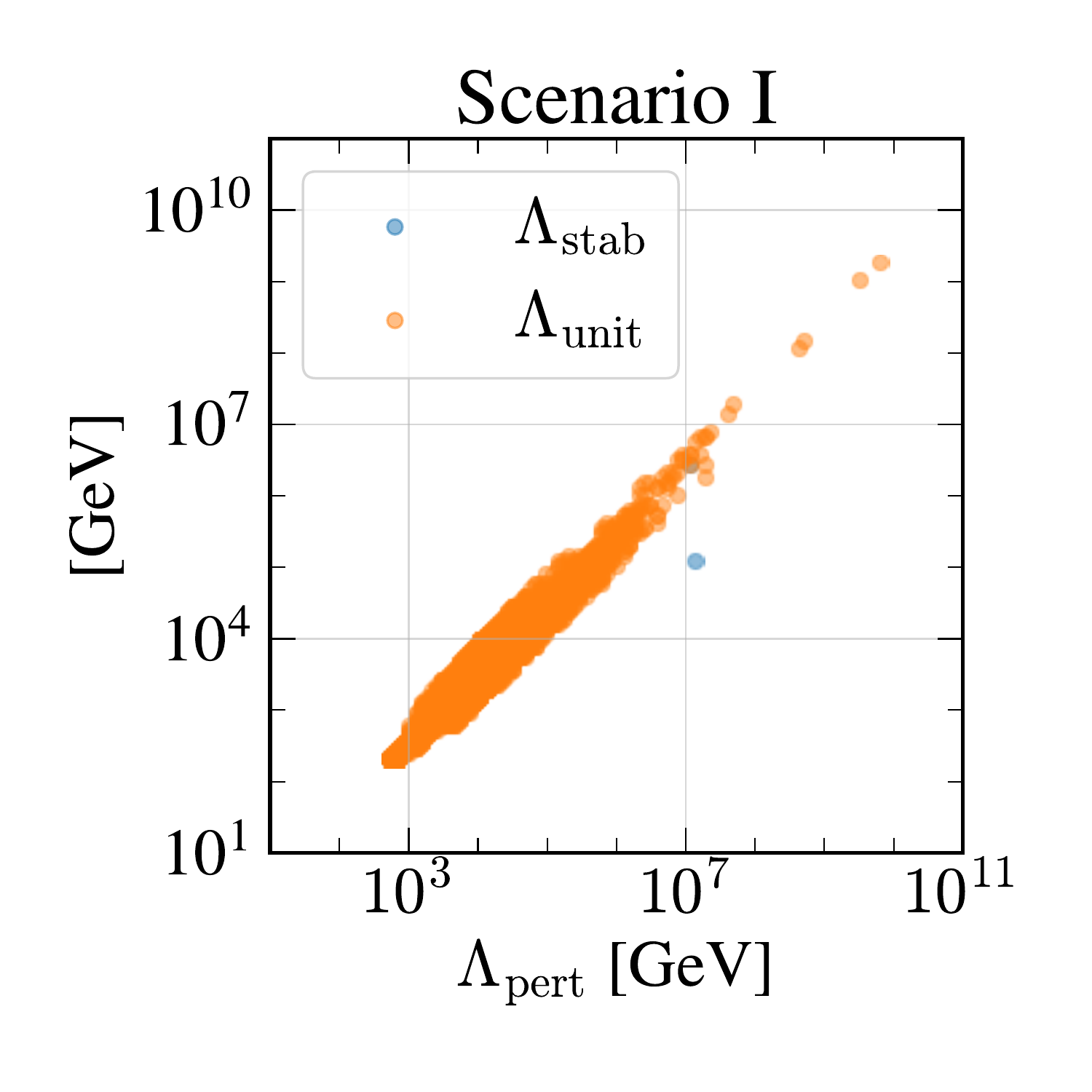} &
\includegraphics[trim=0.5cm 1cm 0.5cm 0cm,clip,height=0.35\textwidth]{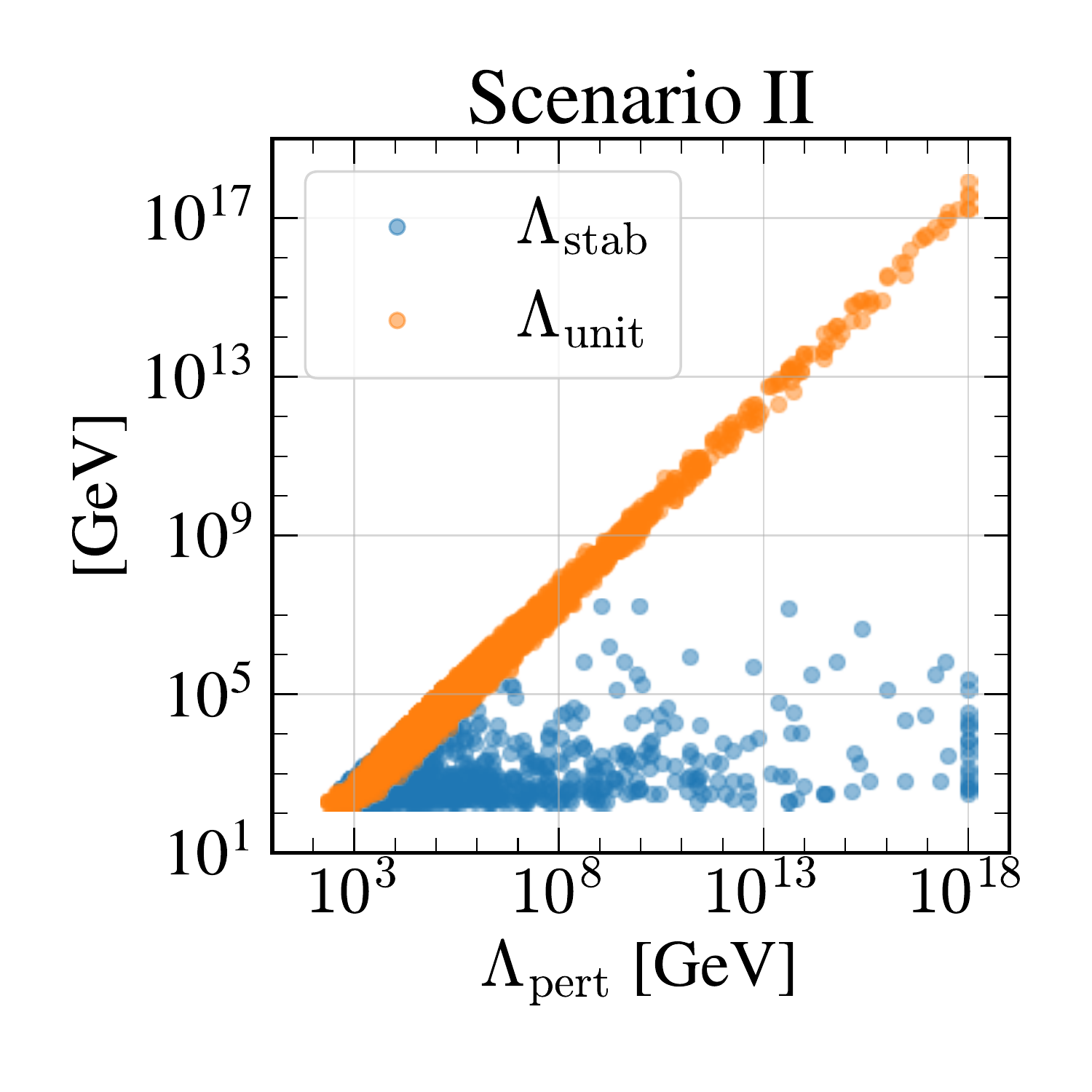}&
\includegraphics[trim=0.5cm 1cm 0.5cm 0cm,clip,height=0.35\textwidth]{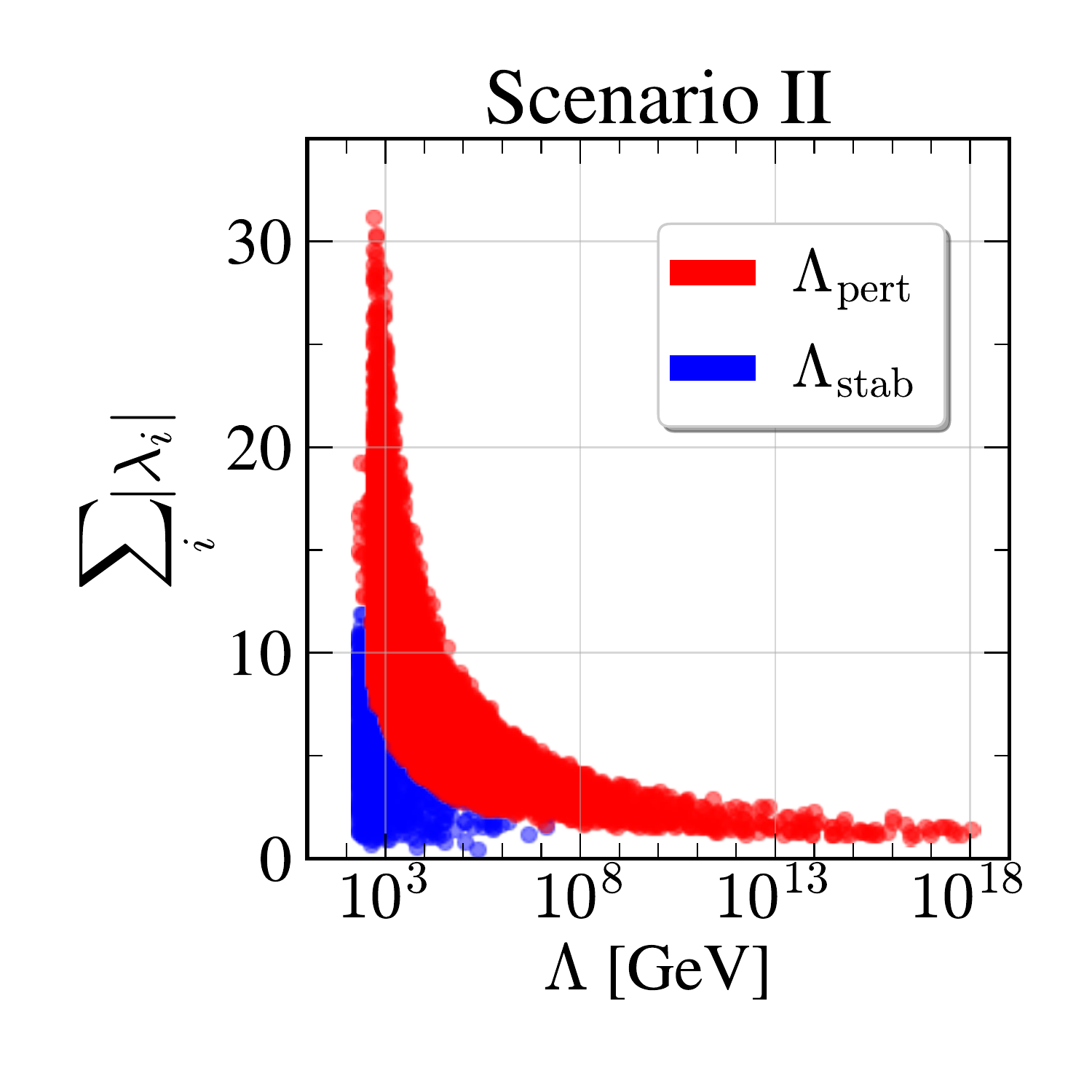}
\end{tabular}
\caption{(Left \& Middle): The breakdown energy scales of unitarity and stability as functions of the perturbativity violation scale in scenario I and II respectively. (Right): The summed magnitude of quartic couplings at the starting scale vs.\ the perturbativity and stability breakdown-energy scales.}
\label{fig:breakdowns}
\end{center}
\end{figure}

Before we investigate each respective scenario in more detail, we study at which scales the individual constraints are violated and which one gives rise to the breakdown energy $\Lambda$.
First of all, the statistics of how many parameter points that pass the constraints coming from \code{HiggsBounds} and \code{HiggsSignals} is shown in \Tab{HBHSstats} for scenario I to III.
For the points that are within these limits, we show the breakdown energies of unitarity, perturbativity and stability in \fig{breakdowns}.
As can be seen in the figure, the constraining requirement of the 2HDM during the RG running is most often unitarity; however, perturbativity violation usually follows soon after\footnote{Note that we stop the evolution when perturbativity breaks down.}.
While unitarity is broken in the evolution of more than $99\%$ of the parameter points, stability is only broken before perturbativity in $\sim 0.02\%$ ($2.5\%$) of the cases in scenario I (scenario II).
In order to illustrate the importance of the magnitude of the quartic couplings at the starting scale, \fig{breakdowns} also shows the summed quartic couplings, $\sum_i|\lambda_i|$ vs.\ the perturbativity and stability breakdown scales in scenario II.
As a quantitative example of the $\sum_i|\lambda_i|$ dependence, it is clear that to have a viable 2HDM above $10^8$ GeV one needs $\sum_i|\lambda_i|\lesssim 5$ and $10^{18}$ GeV requires $\sum_i|\lambda_i|\lesssim 2$.

\begin{figure}[h!]
\begin{center}
\begin{tabular}{cc}
\includegraphics[trim=0cm 1cm 0.5cm 0cm,clip,height=0.35\textwidth]{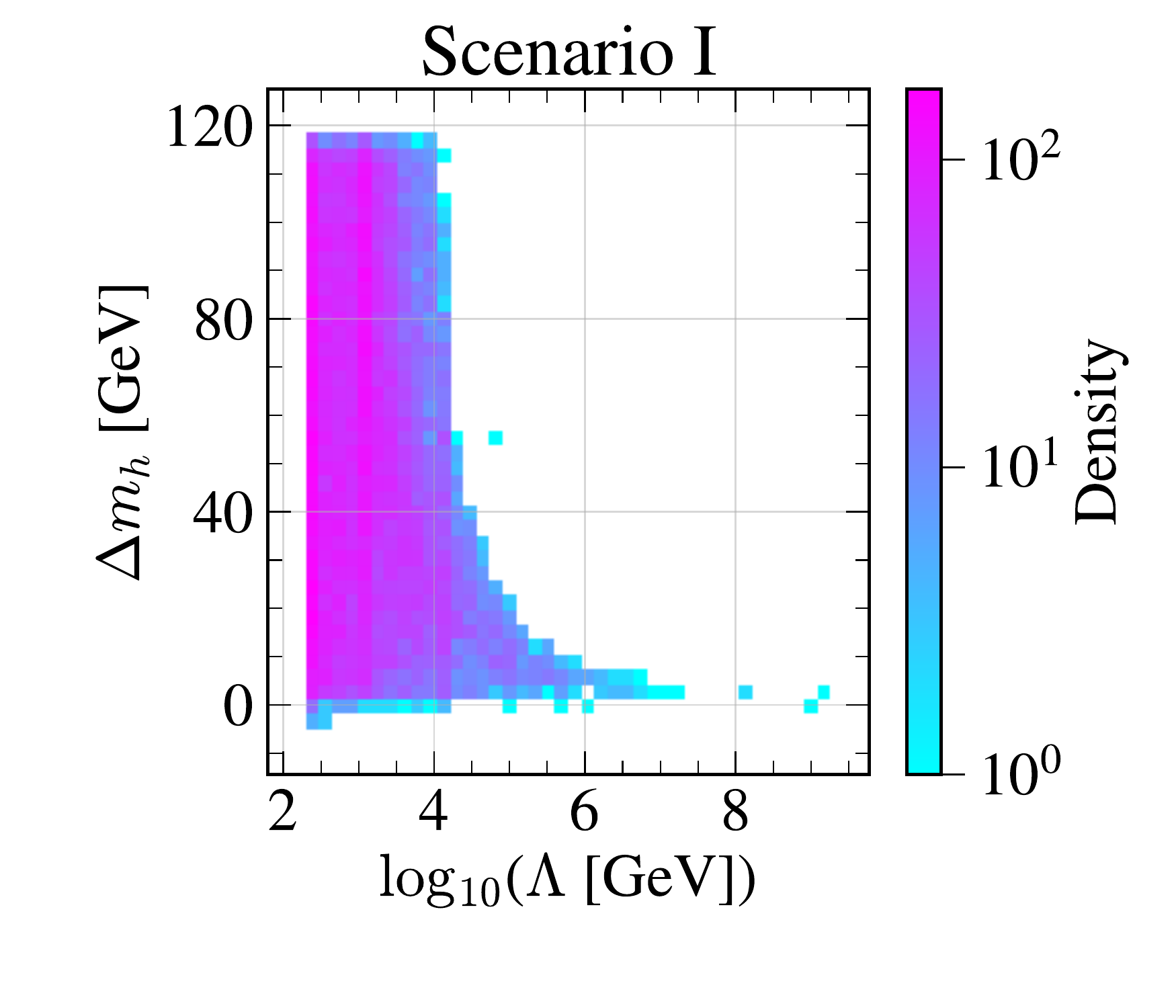} &
\includegraphics[trim=0.5cm 1cm 0.5cm 0cm,clip,height=0.35\textwidth]{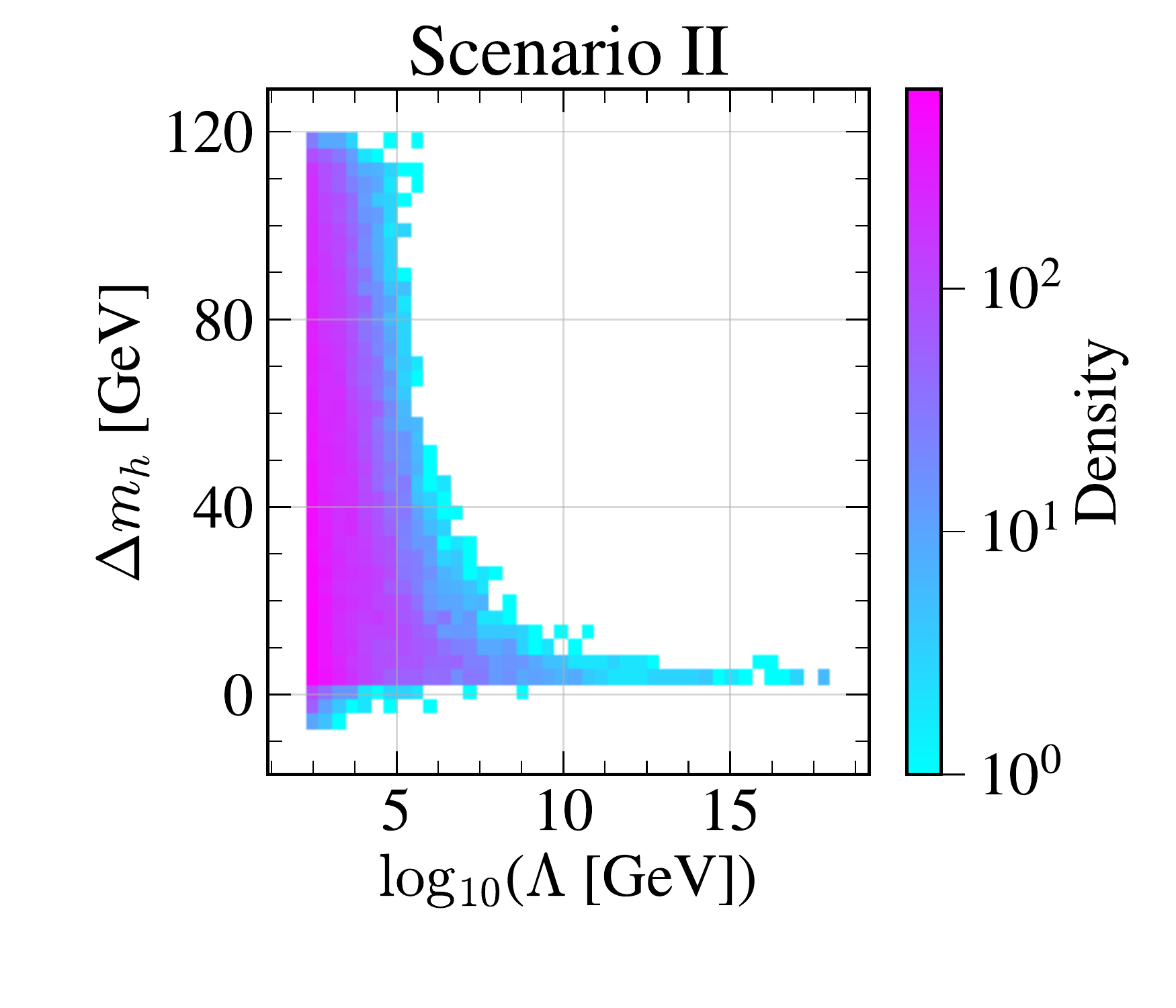}
\end{tabular}
\caption{The density of parameter points per bin as a function of $\Delta m_h$ and the RG breakdown energy $\Lambda$; where $\Delta m_h$ is defined as the difference between the 1-loop corrected and the tree-level mass, $\Delta m_h \equiv m_h - m_h^{\text{tree}}$.}
\label{fig:loopDiff}
\end{center}
\end{figure}

An additional remark, valid for all scenarios, is that some parts of the parameter space of 2HDMs exhibit very large loop corrections to the scalar masses, see \fig{loopDiff}.
This is also pointed out in \mycite{Braathen:2017jvs,Krauss:2018thf}.
It should be mentioned though that the difference between loop corrected and tree-level masses decreases, when considering models which are valid to increasing energy scales.
For example, the 1-loop contributions are significantly smaller when working with models that are valid up to energies above $\sim 10^{10}$ GeV; thereby it is sufficient to only consider 1-loop contributions in our analysis.

\subsection{Scenario I: Exact \Zsym symmetry}\label{exactScan}

The parameter space with an exact \Zsym symmetry is severely constrained and the evolution of most parameter points breaks down already at scales a few orders of magnitude higher than the EW scale.
Low Higgs boson masses are heavily favored and although we put an upper limit of a tree-level $m_H\leq$ 1 TeV, essentially no heavy scalars, $\{H,A,H^\pm\}$ above $\sim 600$ GeV, were found to be valid parameter points already at the EW scale due to perturbativity constraints.

\begin{figure}[h!]
\begin{center}
\textbf{Scenario I}\\
\begin{tabular}{cc}
\includegraphics[trim=1cm 0.5cm 4.5cm 1cm,clip,height=0.35\textwidth]{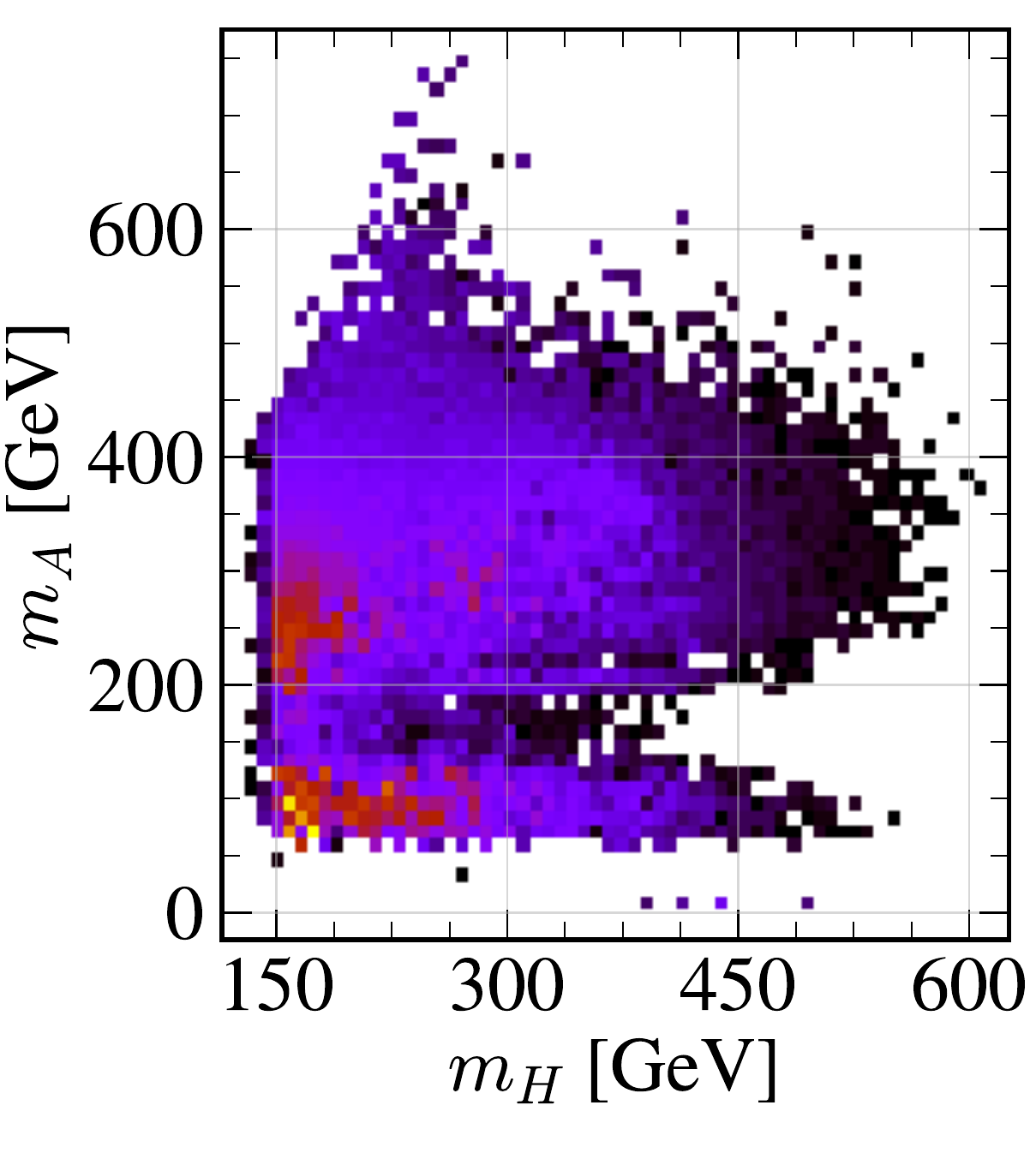}
\includegraphics[trim=1cm 0.5cm 4.5cm 1cm,clip,height=0.35\textwidth]{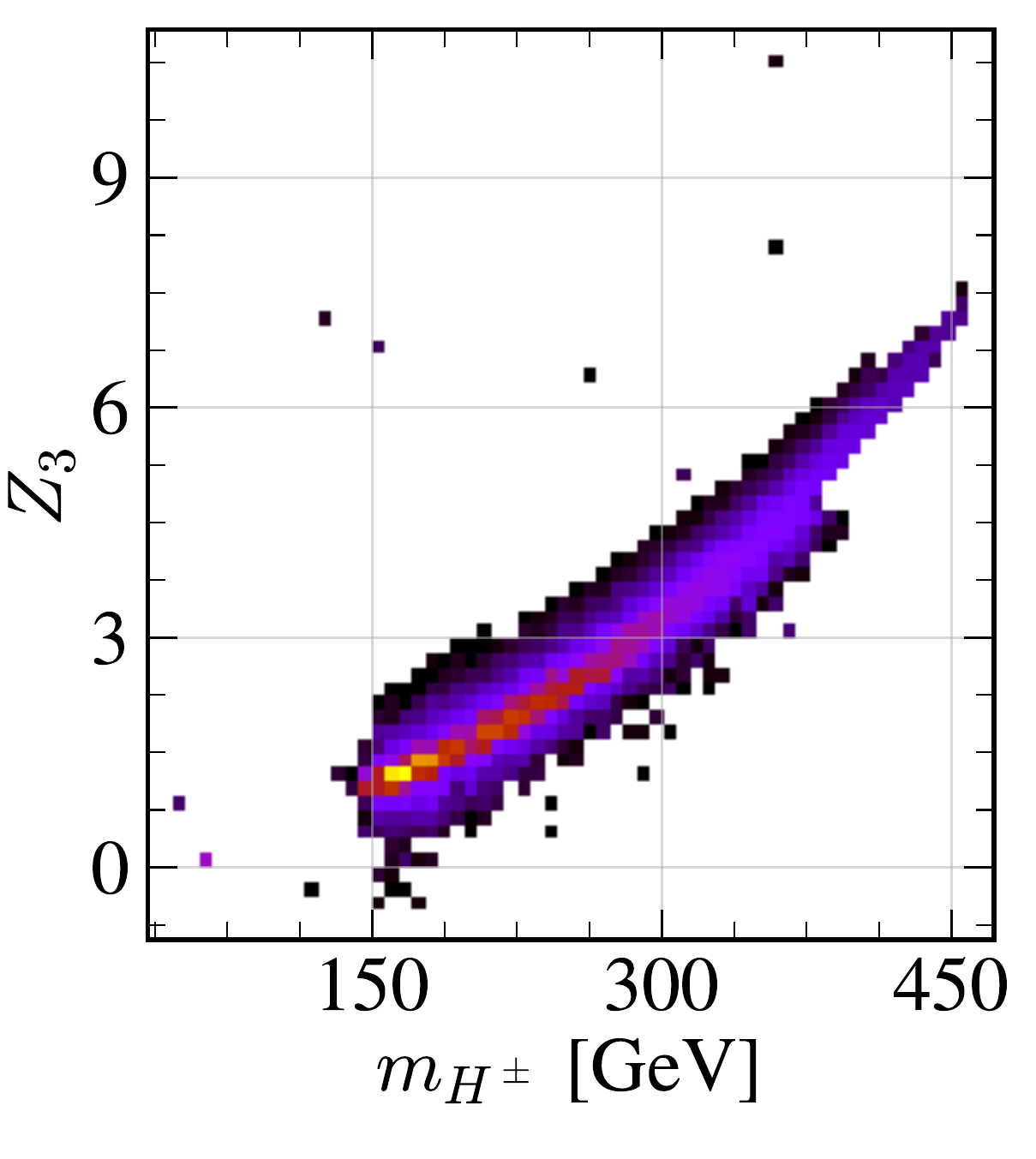}
\includegraphics[trim=1cm 0.5cm 0.5cm 1cm,clip,height=0.35\textwidth]{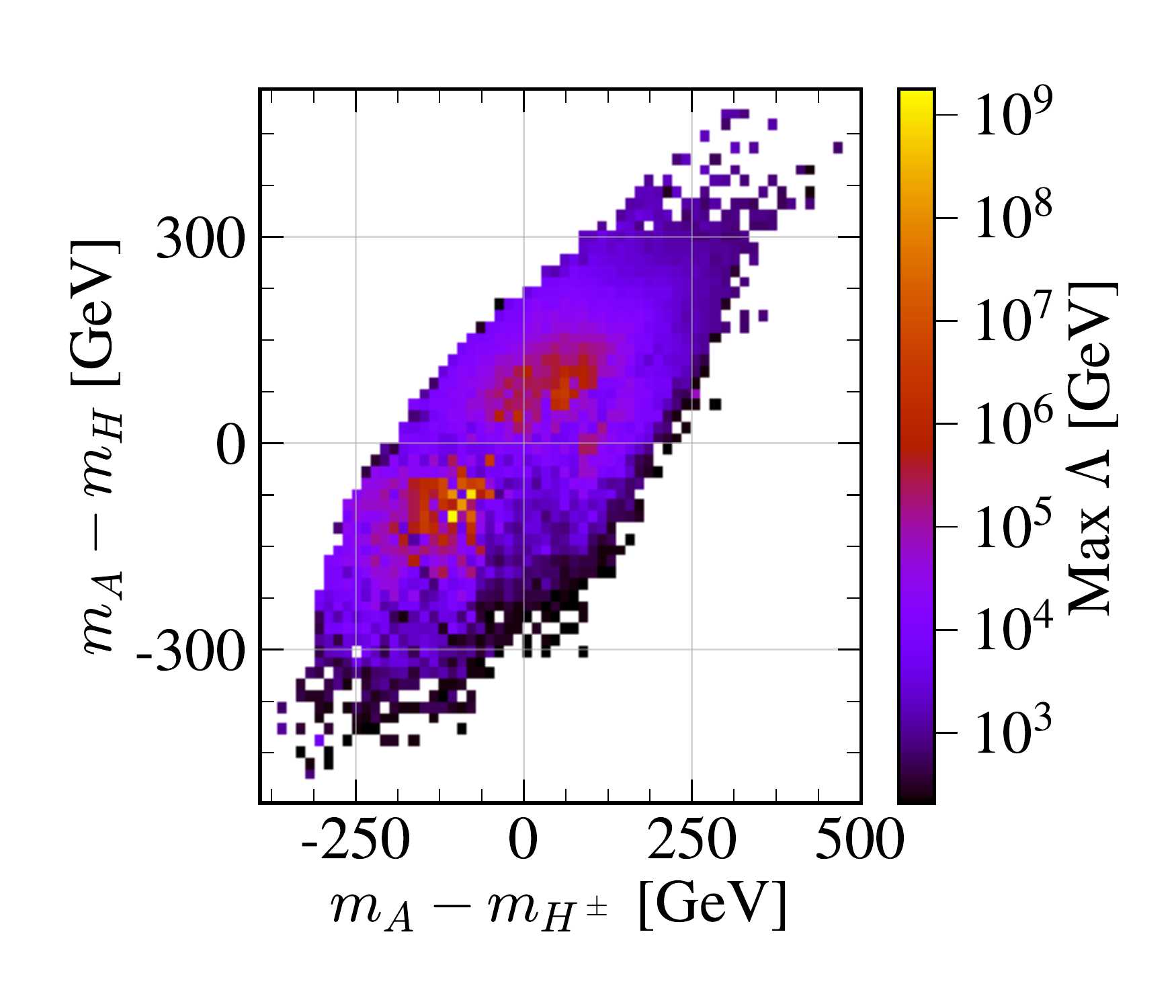}
\end{tabular}
\caption{The maximum breakdown energy scale as function of 1-loop pole masses of scalar masses in scenario I (exact \Zsym symmetry), described in \sec{exactScan}. (Left): Higgs masses $m_H$ and $m_A$. (Middle): $Z_3$ vs.\ charged Higgs mass $m_{H^\pm}$. (Right): Differences of masses.}
\label{fig:sc1_masses}
\end{center}
\end{figure}

In \fig{sc1_masses} we show the allowed masses and the correlations between them in more detail. 
From the figure we see that regions with large ($\gtrsim 500$) BSM Higgs masses break down already at TeV scales and that smaller ($\lesssim 250$ GeV) are favored because of the corresponding small quartic couplings. 
From the leftmost figure we also see that regions with $m_A \sim 150$ GeV are disfavored, which is due to experimental constraints, and the same applies to $m_A \approx m_H$ and $m_A \approx m_{H^\pm}$ shown in the rightmost panel. 
To illustrate that larger masses are directly correlated to larger quartic couplings, we show in the middle panel the correlation between $Z_3$ and $m_{H^\pm}$, where the $Z_3$ coupling is fixed and essentially determined by all other quartic couplings, as can be seen in \eq{Z3sym}. 
Choosing any other BSM mass on the $x$-axis would produce a similar plot. 
Furthermore, in \fig{sc1_masses} one also sees that all masses lie at roughly the same scale, since large mass differences are also connected to large quartic couplings.

In short, we conclude that imposing an exact \Zsym symmetry on the 2HDM requires large quartic couplings to be able to obtain large scalar masses.
This results in rapid RGE running, which makes it essentially impossible to find models that are valid all the way up to the Planck scale without some sort of fine-tuning; the highest energy reached in the scan was $\sim 10^{13}$ GeV.
Similar findings have been found in \mycite{Das:2015mwa}.

\subsection{Scenario II: Softly broken \Zsym parameter scan}\label{softScan}

\begin{figure}[h!]
\begin{center}
\textbf{Scenario II}\\
\begin{tabular}{cc}
\includegraphics[trim=1cm 0.5cm 4.6cm 0.8cm,clip,height=0.35\textwidth]{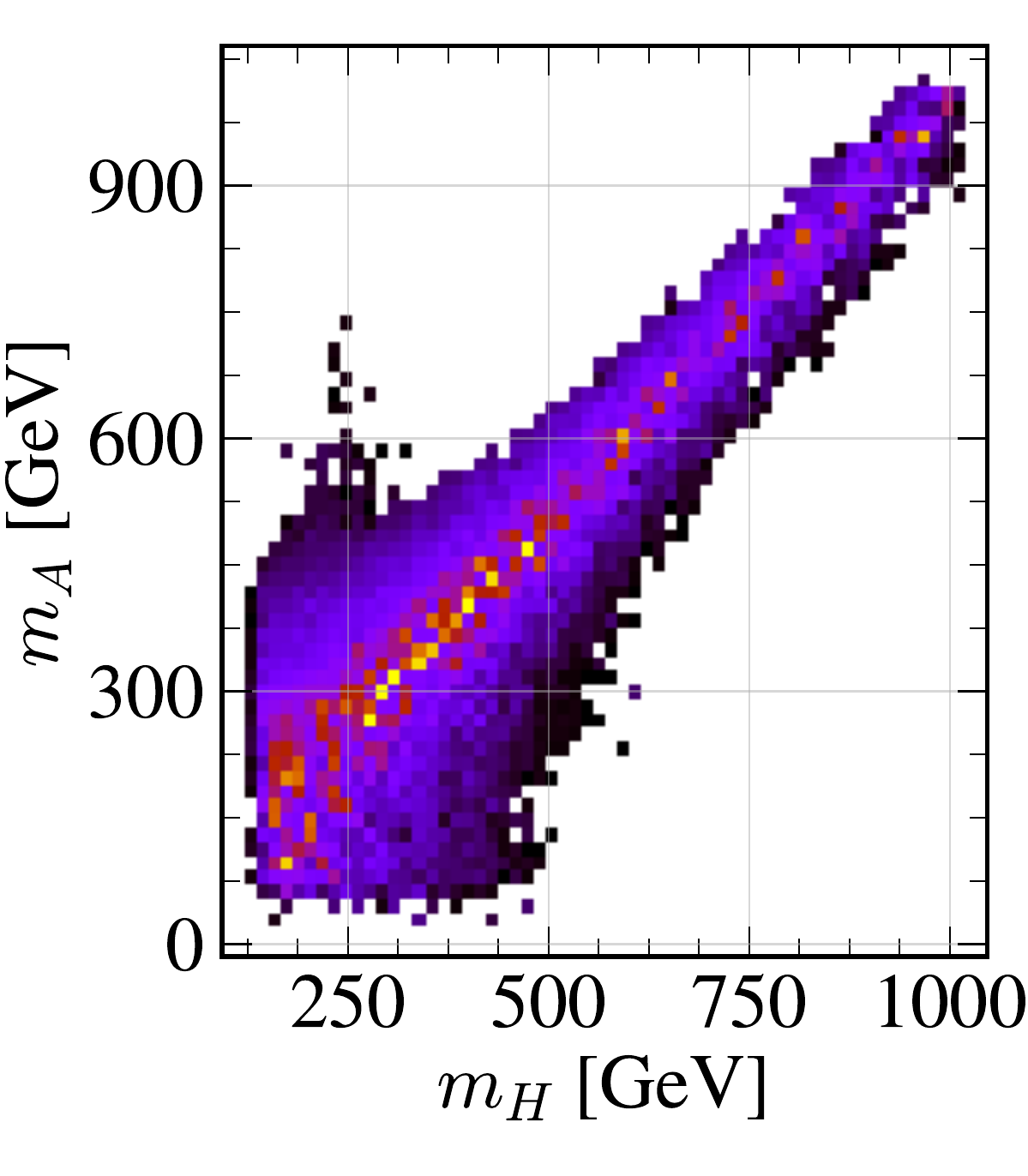}
\includegraphics[trim=1cm 0.5cm 4.8cm 0.8cm,clip,height=0.35\textwidth]{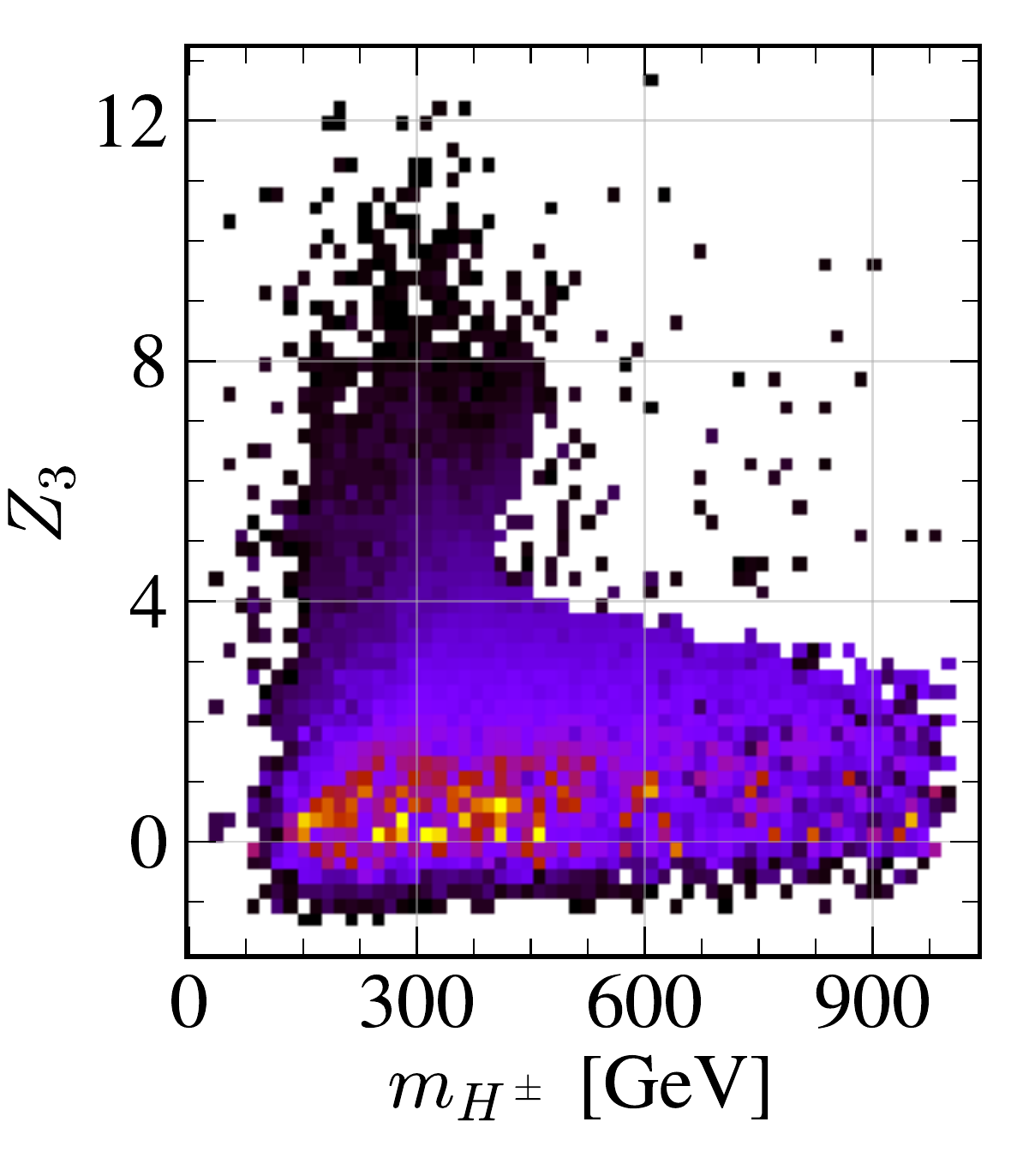}
\includegraphics[trim=1cm 0.5cm 0.5cm 0.8cm,clip,height=0.35\textwidth]{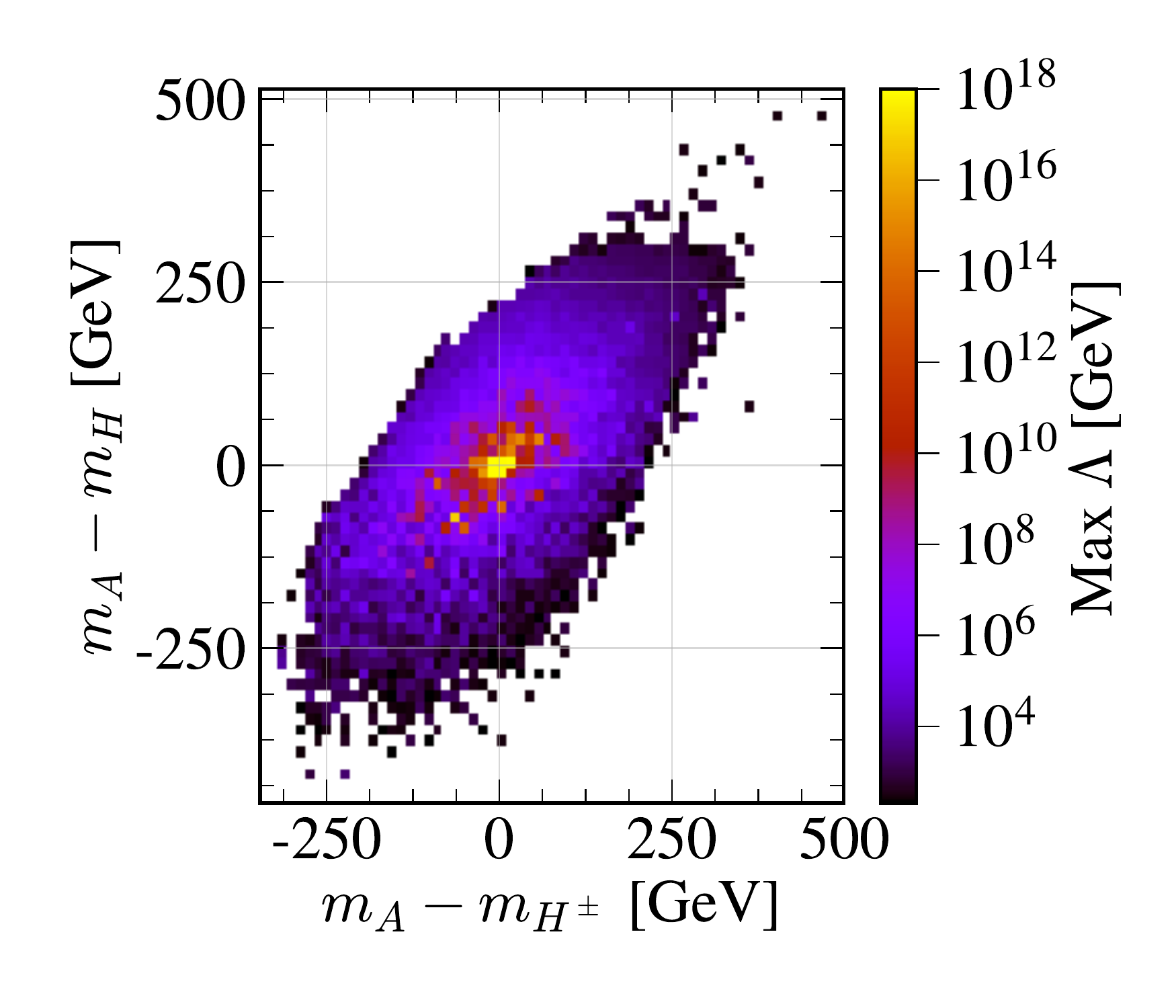}
\end{tabular}
\caption{Maximum breakdown energy scale as function of scalar masses in scenario II. (Left): Masses of scalar $H$ and pseudo-scalar $A$ Higgs bosons. (Middle): $Z_3$ vs.\ charged Higgs mass $m_{H^\pm}$. (Right): Mass differences.}
\label{fig:softMasses}
\end{center}
\end{figure}

\begin{figure}[h!]
\hspace{-0.08\textwidth}
\begin{tabular}{cc}
\includegraphics[trim=0.5cm 1cm 0cm 0cm,clip,height=0.35\textwidth]{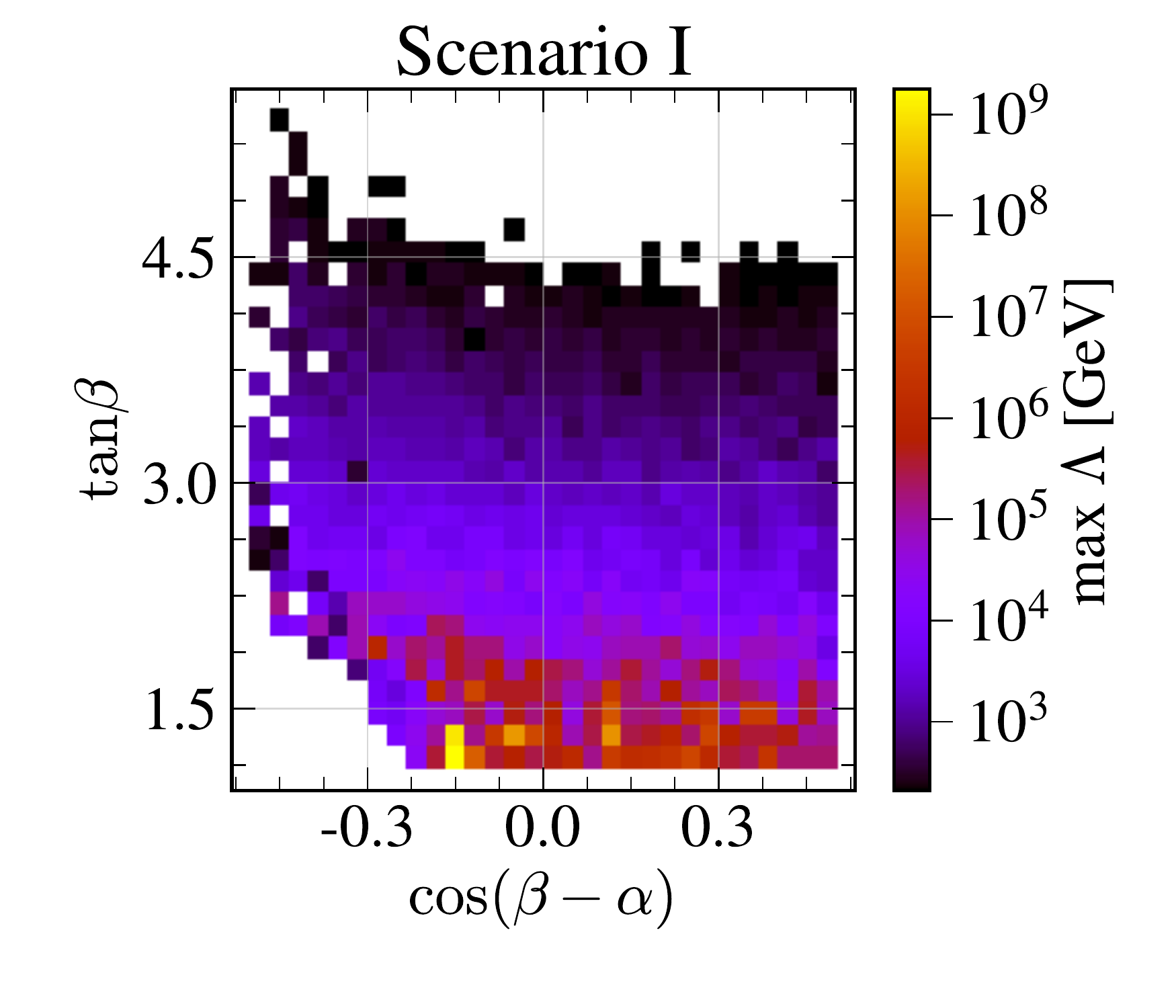}
\includegraphics[trim=0.5cm 1cm 5cm 0cm,clip,height=0.35\textwidth]{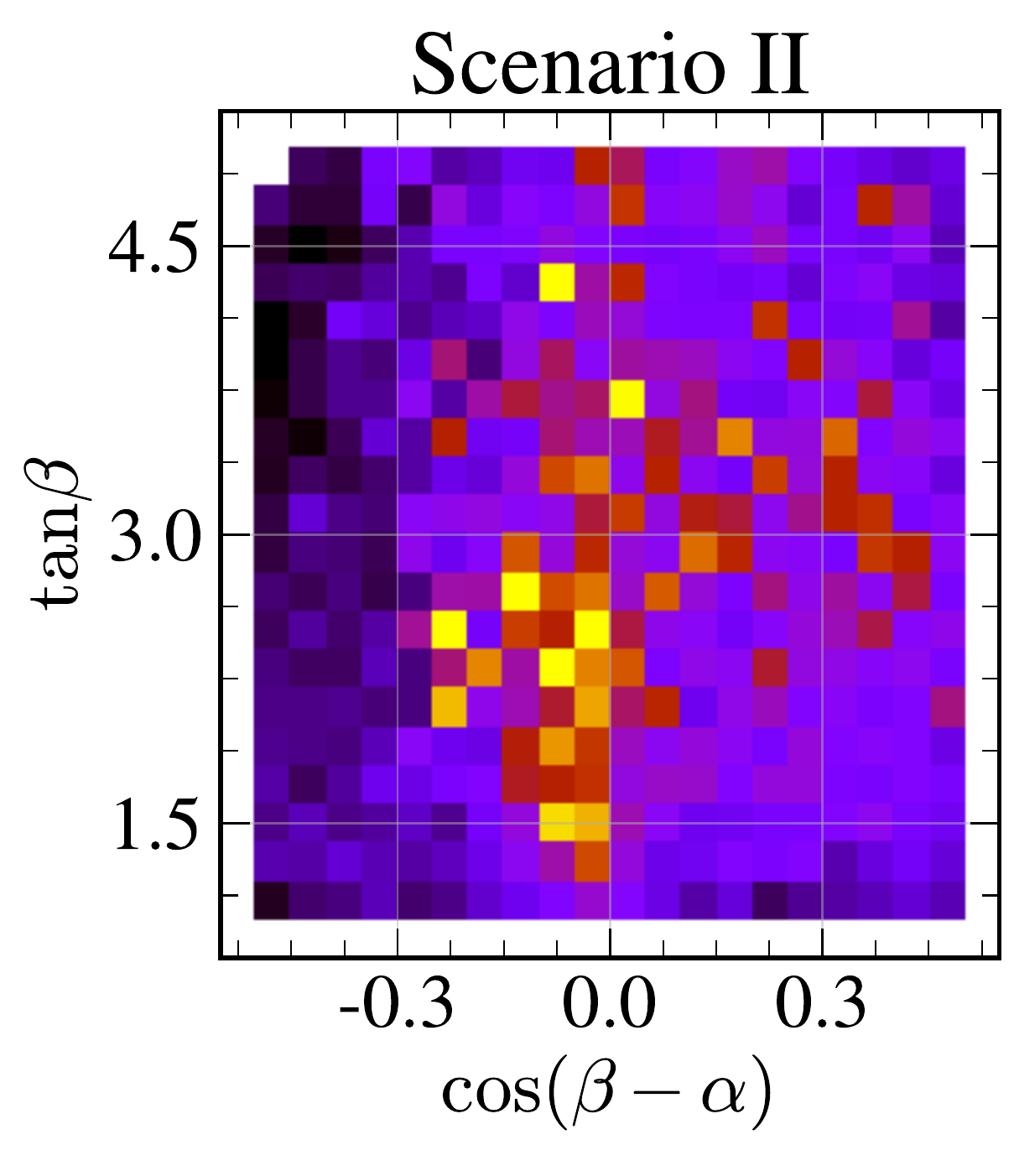}
\includegraphics[trim=1cm 1cm 0cm 0cm,clip,height=0.35\textwidth]{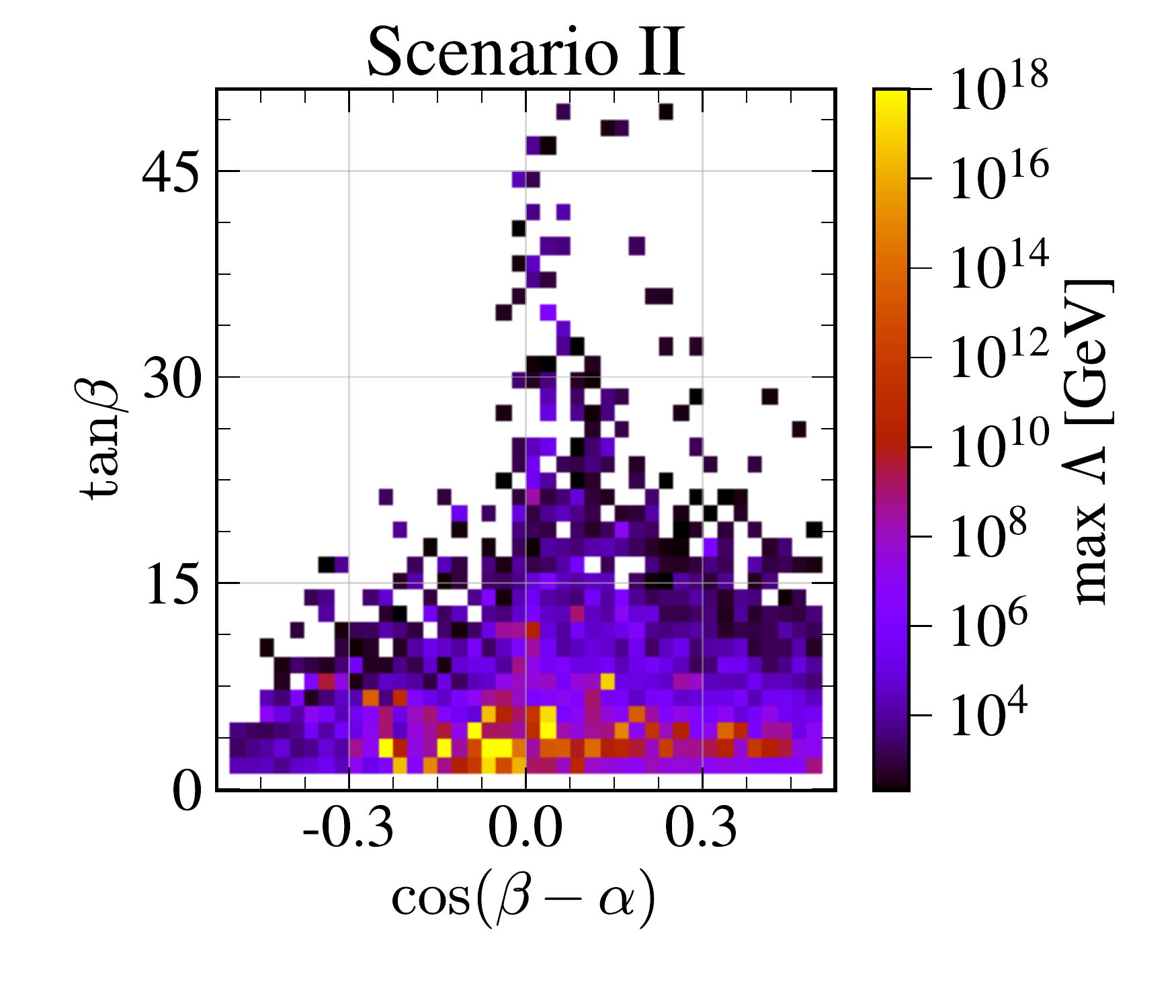}
\end{tabular}
\caption{Maximum breakdown energy as function of $\tan\beta$ and $\cos(\beta-\alpha)$. (Left): Scenario I (Middle \& Right): Scenario II. The middle figure is a zoomed in region of the right one. Note the difference in scale of the breakdown energy in the two scenarios.}
\label{fig:softAngles}
\end{figure}

As seen in the previous section, imposing an exact \Zsym symmetry on the general 2HDM makes it hard, if not impossible, to find parameter points that correspond to a valid model all the way up to the Planck scale; furthermore, demanding large BSM Higgs masses makes it even more difficult.
The main effect from breaking the \Zsym softly is a positive one; it opens up regions in parameter space that are much more stable during RG evolution.
The introduction of $m_{12}^2$ helps to give larger BSM masses without increasing the magnitude of the quartic couplings; thus, the strong correlation of large Higgs boson masses and large quartic couplings is reduced when comparing to the case of an exact \Zsym symmetry.
This can be seen by comparing \fig{sc1_masses} to \fig{softMasses}, which shows the larger Higgs masses that are viable in scenario II.
Also, experimental constraints are not as constraining in scenario II.
Similarly to scenario I, there is still a strong correlation between the BSM masses for parameter space points that are valid all the way up to the Planck scale.
In order to illustrate the dependence on couplings to other particles we show plots of the maximum breakdown energy as a function of $\cos(\beta-\alpha)$ and $\tan\beta$ in \fig{softAngles}.
From the figure, we note that the region of valid parameter points in $\tan\beta$ is larger in scenario II compared to scenario I.
Similar plots for all the quartic couplings in the generic basis, as well as the Higgs basis, are collected in \app{higgsBasisPlots}.
Although it has been checked that the specific Yukawa \Zsym symmetry type does not matter for our conclusions about \Zsym breaking effects in this case, the plots in \fig{softAngles} differs quite a bit if one were to consider a type-II symmetry instead of a type-I. 
In addition, one should then also include constraints from $B$-physics such as $b\rightarrow s \gamma$.

Note that there are irregular patterns in the regions where some parameter points are valid all the way to the Planck scale in figures \ref{fig:softMasses}, \ref{fig:softAngles} and also \ref{fig:hard1} below; the scale of nearby bins differ several orders of magnitude.
This effect is purely statistical in nature and a larger data sample and bin choice would smooth out the plots.

\begin{figure}[h!]
\hspace{-0.03\textwidth}
\begin{tabular}{cc}
\includegraphics[trim=1cm 1cm 0cm 0cm,clip,height=0.35\textwidth]{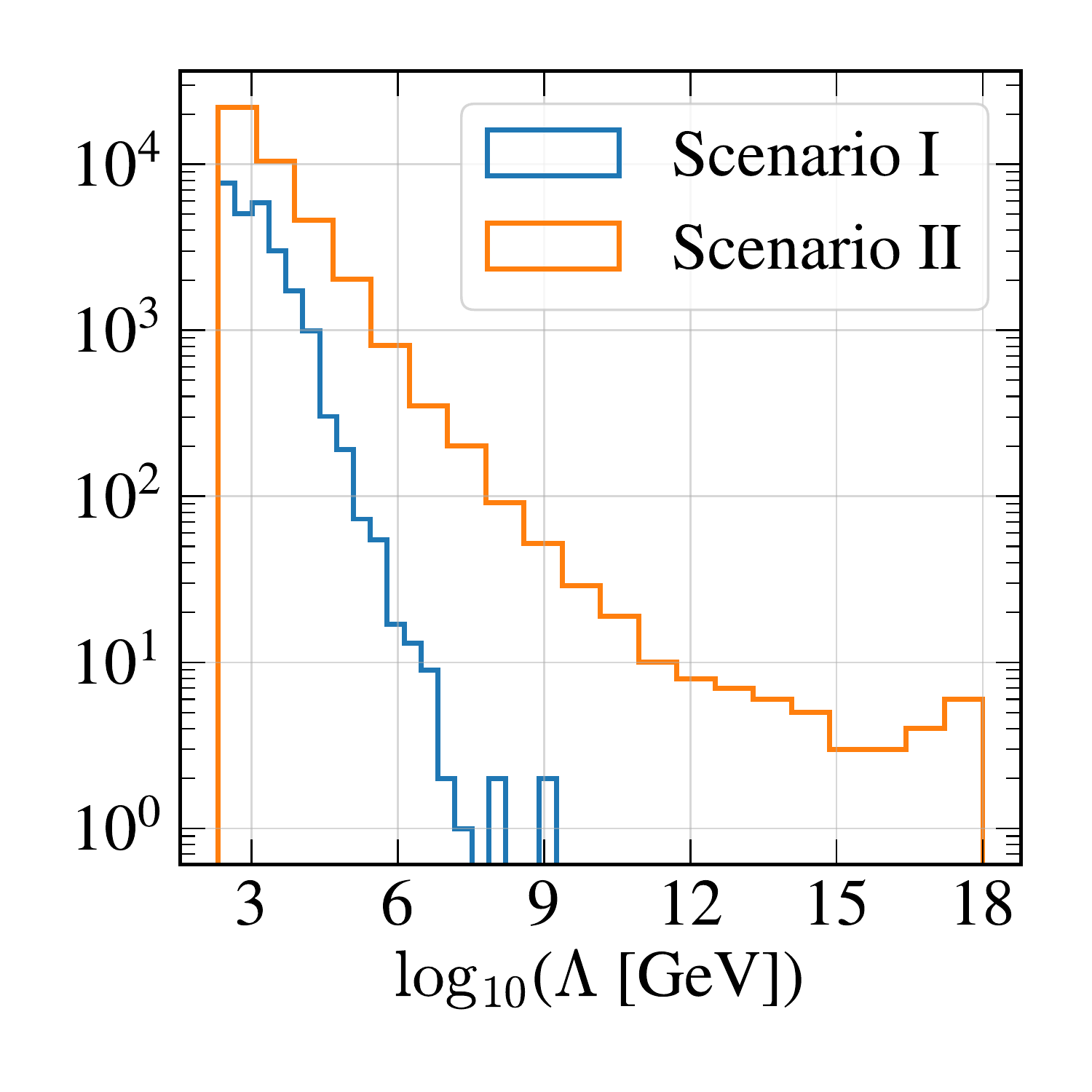}
\includegraphics[trim=1cm 1cm 0cm 0cm,clip,height=0.35\textwidth]{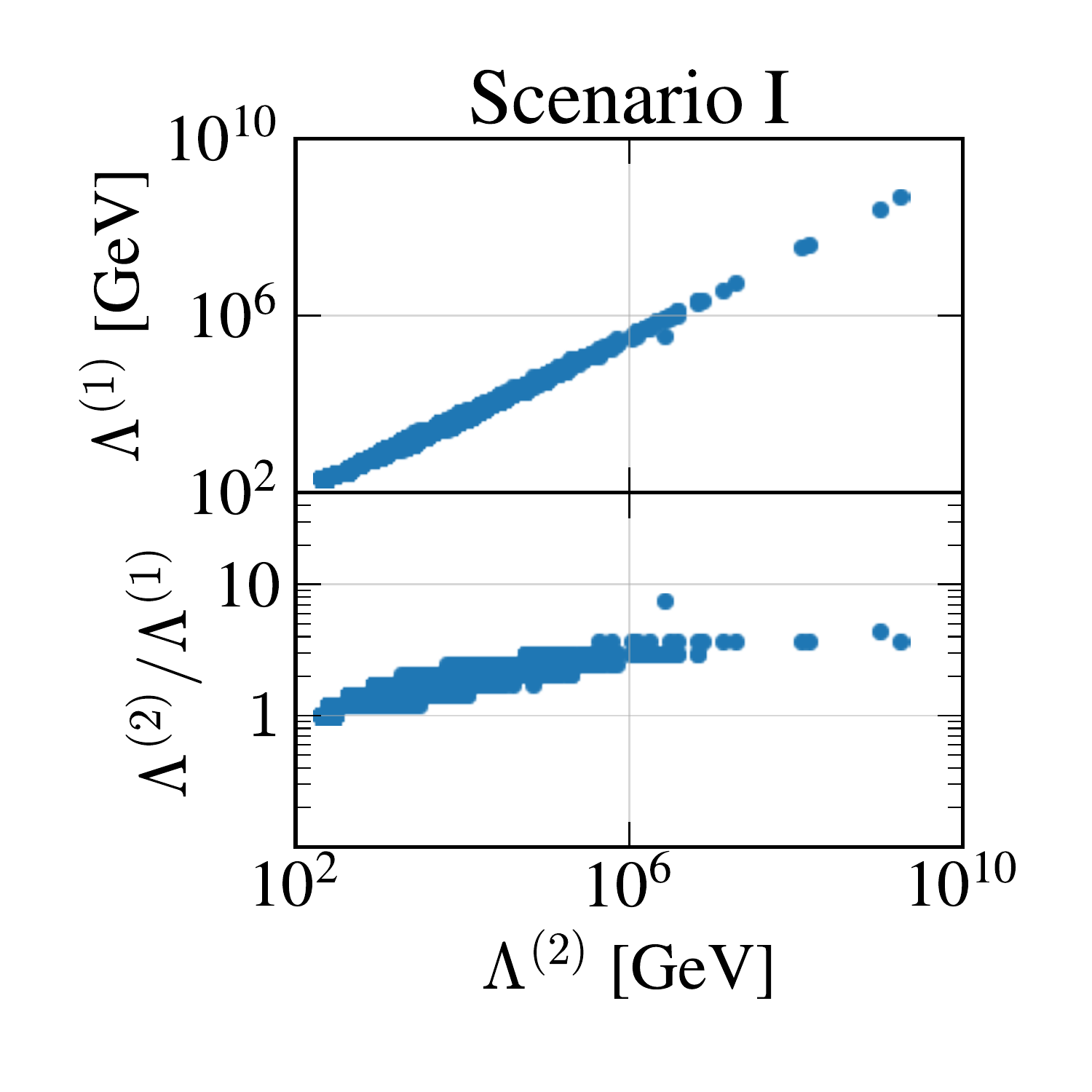}
\includegraphics[trim=1cm 1cm 0cm 0cm,clip,height=0.35\textwidth]{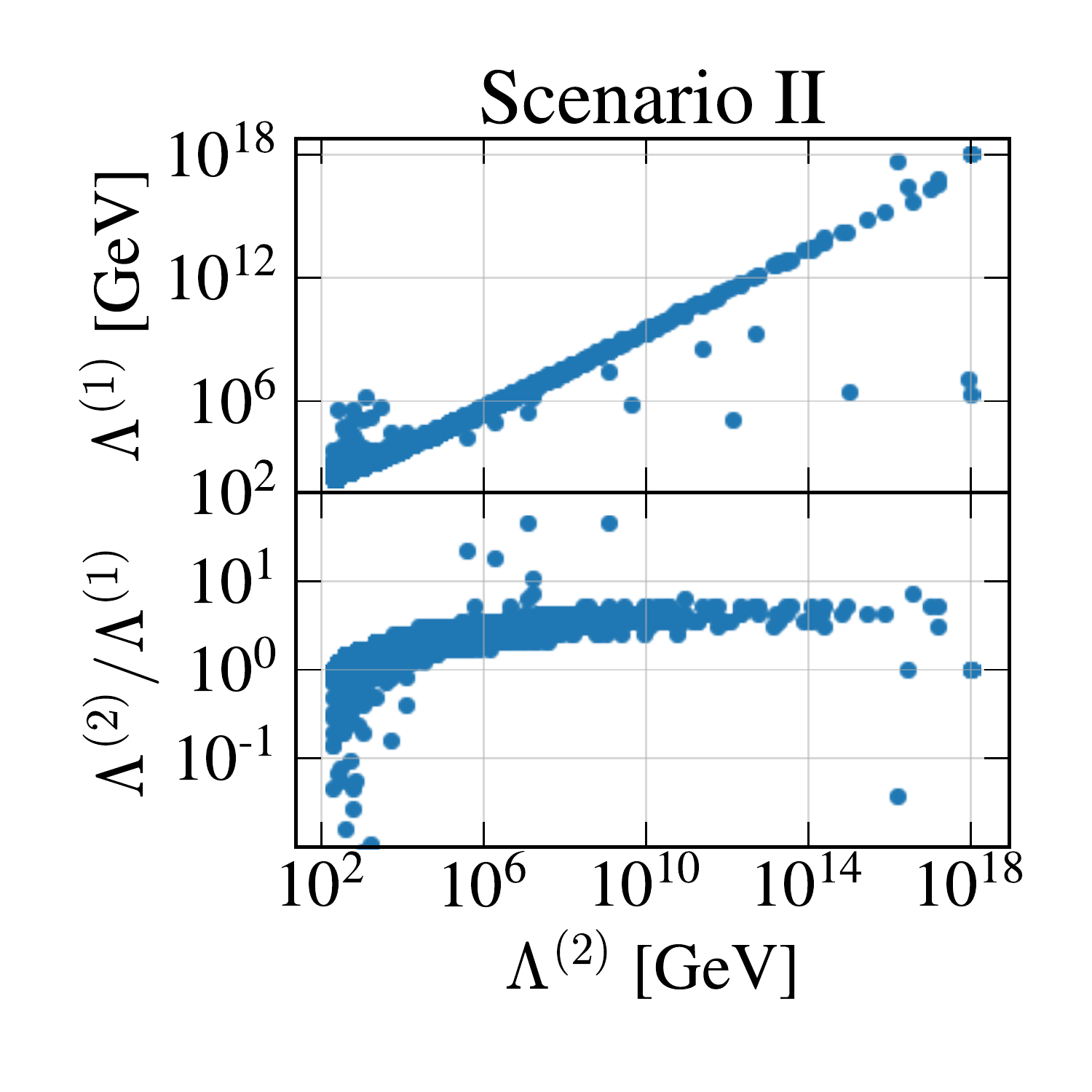}
\end{tabular}
\caption{(Left): Histograms of breakdown energy, $\Lambda$, in the parameter scans of scenario I \& II. Breaking the \Zsym softly opens up a parameter region where models are valid all the way up to the Planck scale. (Middle \& Right): Comparison of breakdown energy scales using 1-loop and 2-loop RGEs in scenario I \& II respectively. We define $\Lambda^{(n)}$ as the energy scale where either perturbativity, unitarity or stability is violated using the $n$-loop RGEs.}
\label{fig:efHist}
\end{figure}

To illustrate the difference induced by allowing soft \Zsym breaking, a histogram of parameter points sorted by breakdown energy is shown in \fig{efHist} for both scenario I and II. 
If we compare the different scenarios, we see that breaking the \Zsym softly opens up an entire new region of the 2HDM, where the model is theoretically viable all the way to the Planck scale.

A comparison of the 2-loop and 1-loop breakdown scales is also shown for both scenario I and II in \fig{efHist}.
In general, using the 2-loop RGEs increases the breakdown energy scale, $\Lambda$, with a factor of $\mathcal{O}(1)$ to $\mathcal{O}(10)$; although there are a few parameter points with a larger difference, usually related to the point being close to stability boundaries.
This is a consequence from the fact that RG running of the parameters slows down when going to 2-loop RGEs, which can also be seen in the example scenario in \fig{gen2Demo}.

\subsection{Scenario III: Hard scalar \Zsym breaking}\label{hardScan}

In the last section, we studied the scenario of a soft broken \Zsym symmetry.
A softly broken \Zsym does not spread in the RG evolution in that no further \Zsym breaking parameters are generated; since the mass parameters do not enter the RGEs for the dimensionless parameters.
Now, we break the symmetry hard by having small non-zero $\lambda_6, \lambda_7$ that can have major implications for the RG evolution.
In general, if there is no kind of fine-tuning present, non-zero $\lambda_6, \lambda_7$ speed up the running of the other quartic couplings.
At 2-loop order these \Zsym breaking parameters induce \Zsym breaking in the Yukawa sector as well, giving rise to FCNCs at tree-level.

\begin{figure}[h!]
\begin{center}
\textbf{Scenario III}
\begin{tabular}{cc}
\includegraphics[trim=0.5cm 1cm 0cm 0cm,clip,height=0.35\textwidth]{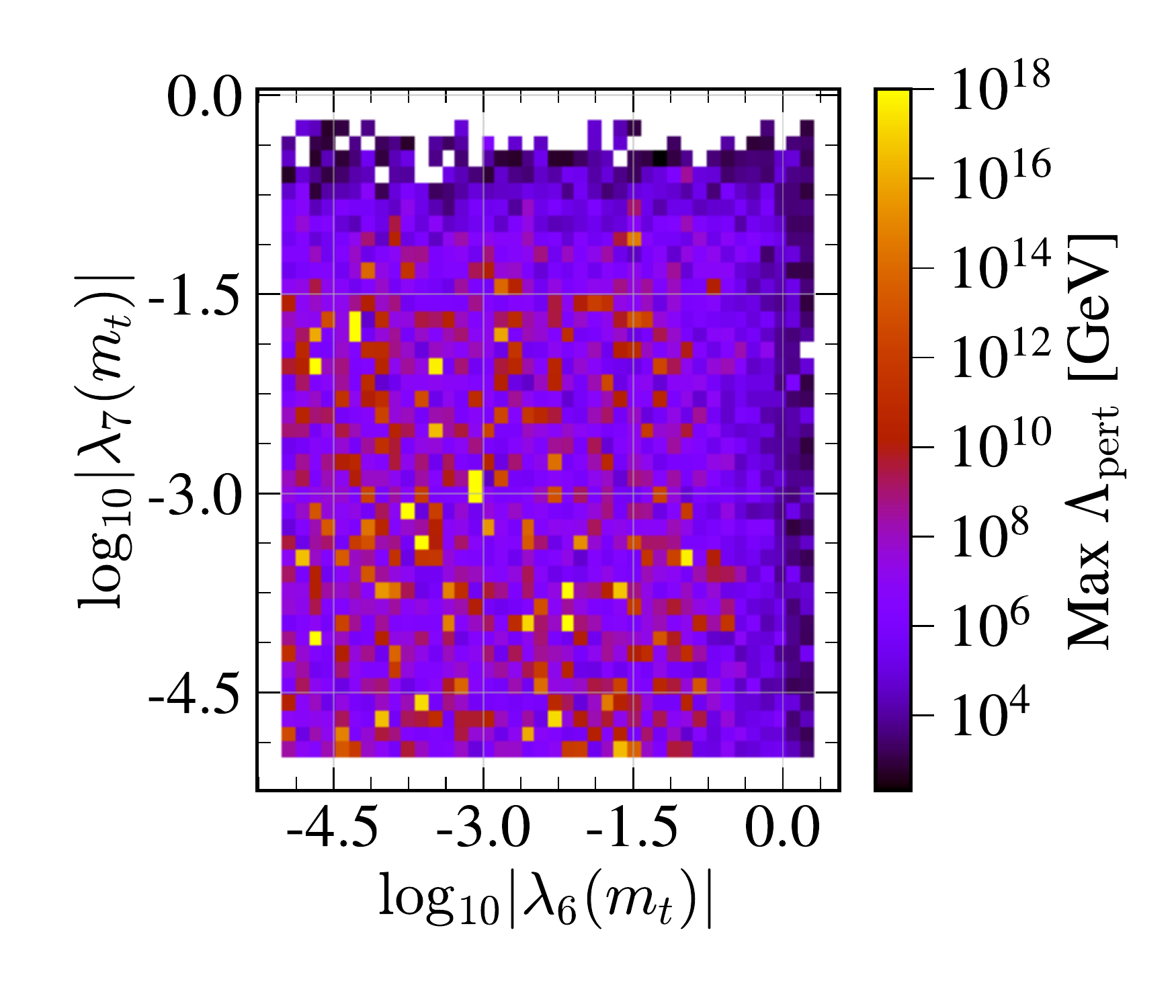}
\includegraphics[trim=0.5cm 1cm 0cm 0cm,clip,height=0.35\textwidth]{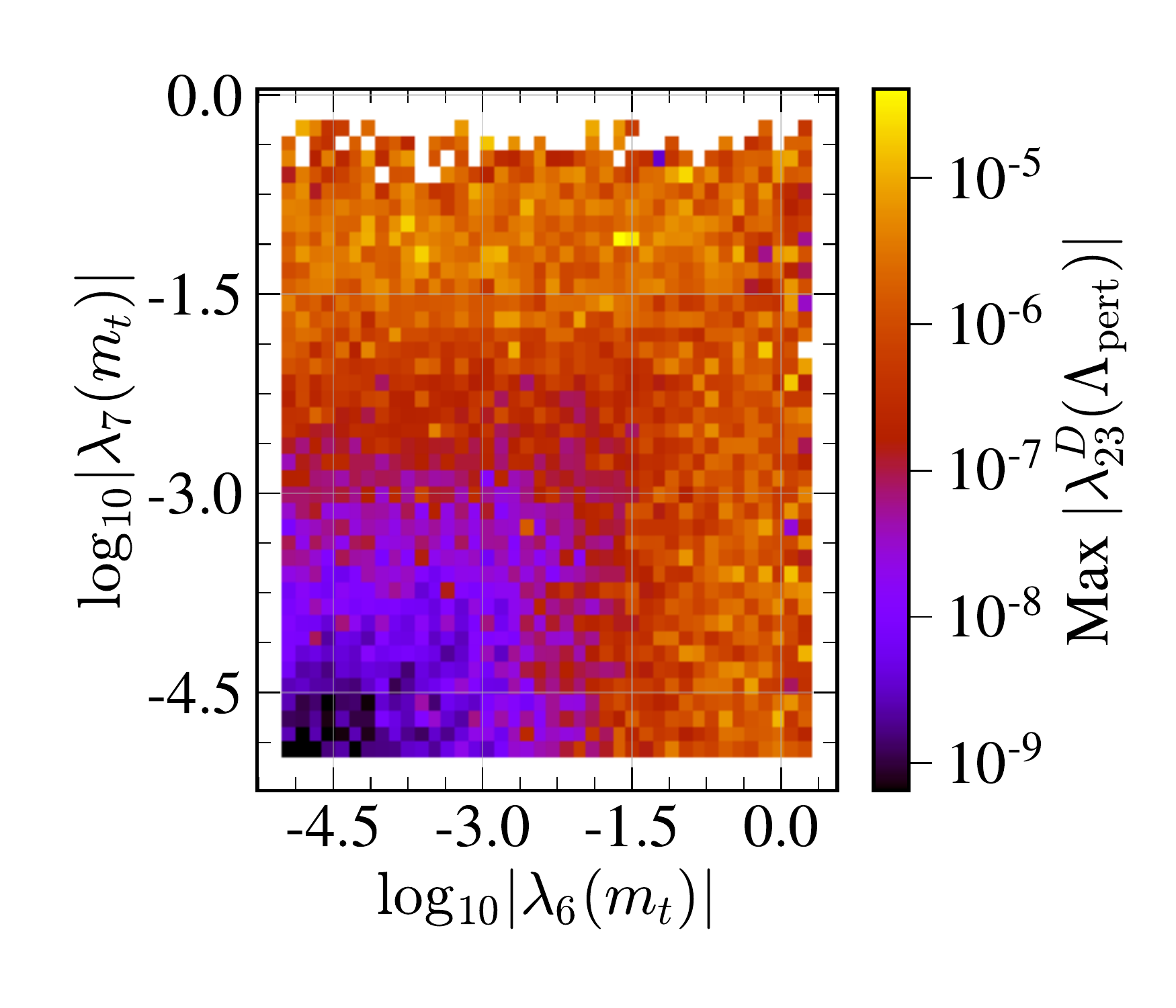}
\end{tabular}
\caption{(Left): Perturbativity breakdown energy scale as function of \Zsym breaking parameters at top mass scale. (Right): The generated Yukawa element $\lambda_{23}^D = \rho^D_{23} \sqrt{v^2/(2m_s m_b)}$.}
\label{fig:hard1}
\end{center}
\end{figure}

Since we expect the induced FCNCs to be very suppressed, we here ignore if the parameter points break stability to be able to generate as large FCNCs as possible.
For simplicity, we also ignore the unitarity constraint; it is usually violated very close to the perturbativity breakdown energy and thus has minor implications for the analysis.
However, even with these relaxed conditions, we find that no sizable FCNCs are generated, as we will now show.

To see the dependency on the magnitude of $\lambda_{6,7}$, the breakdown scale of perturbativity as functions of $\lambda_6$ and $\lambda_7$ is shown in \fig{hard1}.
From that plot, we deduce that to have a model that is good all the way up to $10^{18}$ GeV one needs $\lambda_{6,7}\lesssim 0.1$ and models that have $\lambda_{6,7}\gtrsim 1$ break down already at scales $\sim 10^5$ GeV.
 
To investigate the FCNCs that are induced, we plot the $\lambda_{23}^D$ element at the perturbativity breakdown energy, which is the largest element in more than $99\%$ of the cases, as a function of $\lambda_{6,7}(m_t)$ in \fig{hard1}.
Even though, as already mentioned, there are no strict limits on the FCNCs at energies which have not been probed experimentally, we assume that the values of the Yukawa couplings should not be widely different at disparate scales.
Therefore, we will use a generic limit on the non-diagonal Yukawa elements as a measure of the induced FCNCs.
A discussion of neutral meson mixings can be found in \cite{Bijnens:2011gd} and we will use the resulting limit of sizable FCNCs to be $\lambda_{i\neq j}^F \lesssim 0.1$.
As will be presented below, we see that the generated Yukawa element is not problematic in this scenario; the model breaks down before any significant FCNCs are generated.
In a way this is natural since the hard \Zsym breaking in the potential only affects the Yukawa sector at 2-loop level, whereas the other potential parameters are affected already at 1-loop level.

\begin{figure}[h!]
\begin{center}
\textbf{Example from scenario III}
\begin{tabular}{ccc}
\hspace{-1.3cm}
\includegraphics[trim=0cm 1cm 0.5cm 0.5cm,clip,height=0.32\textwidth]{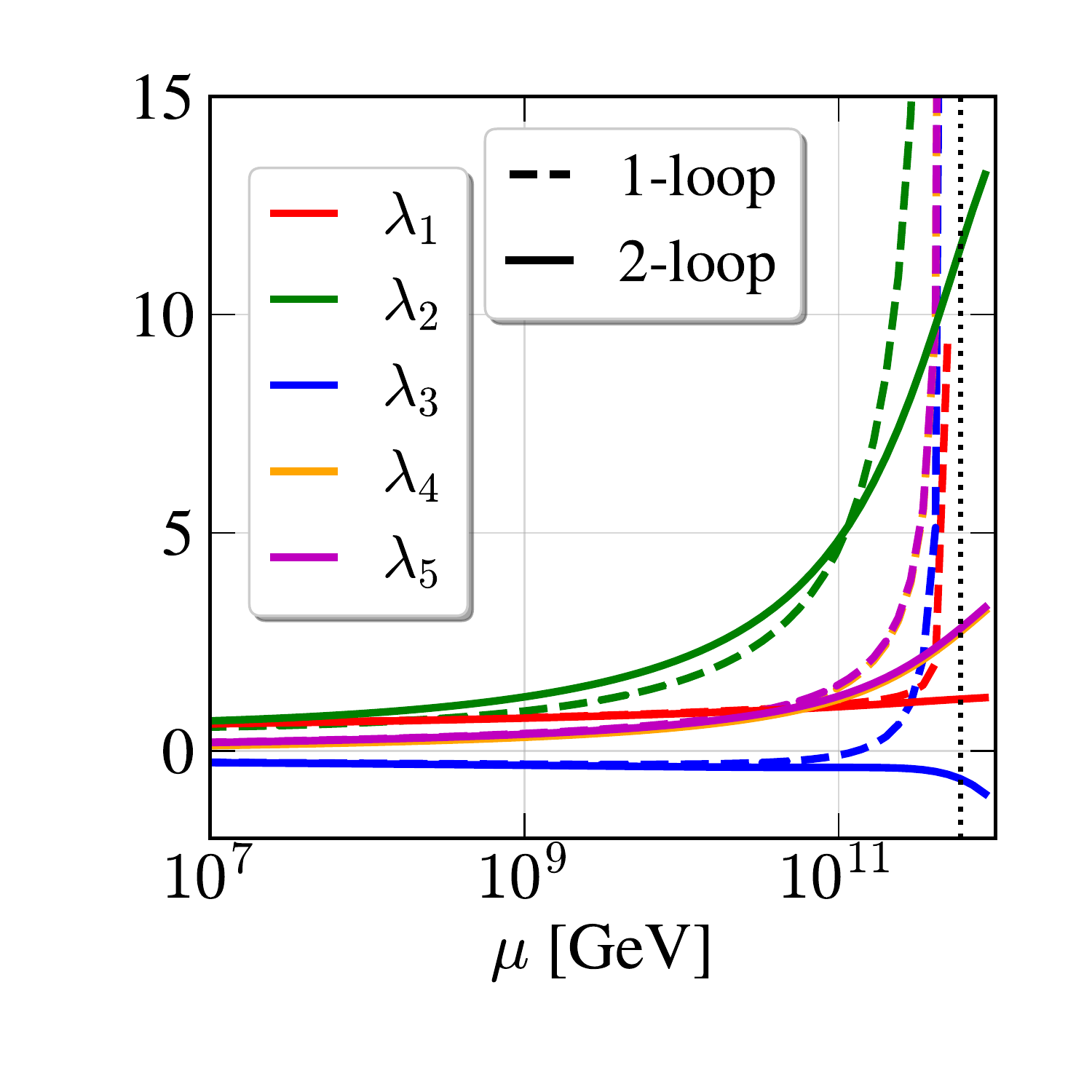}&
\includegraphics[trim=0.5cm 1cm 0.5cm 0.5cm,clip,height=0.32\textwidth]{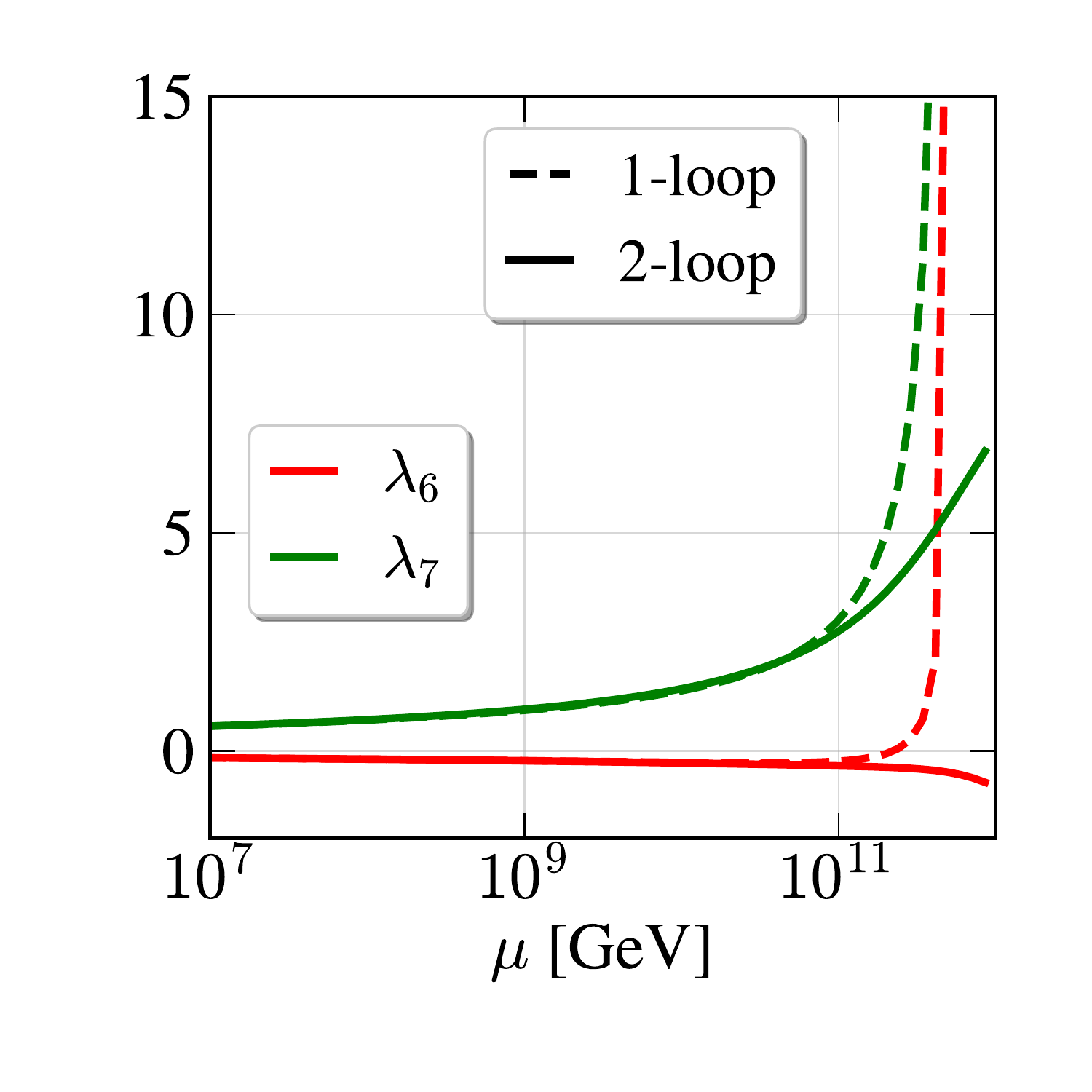}&
\includegraphics[trim=0.5cm 1cm 0cm 0.5cm,clip,height=0.32\textwidth]{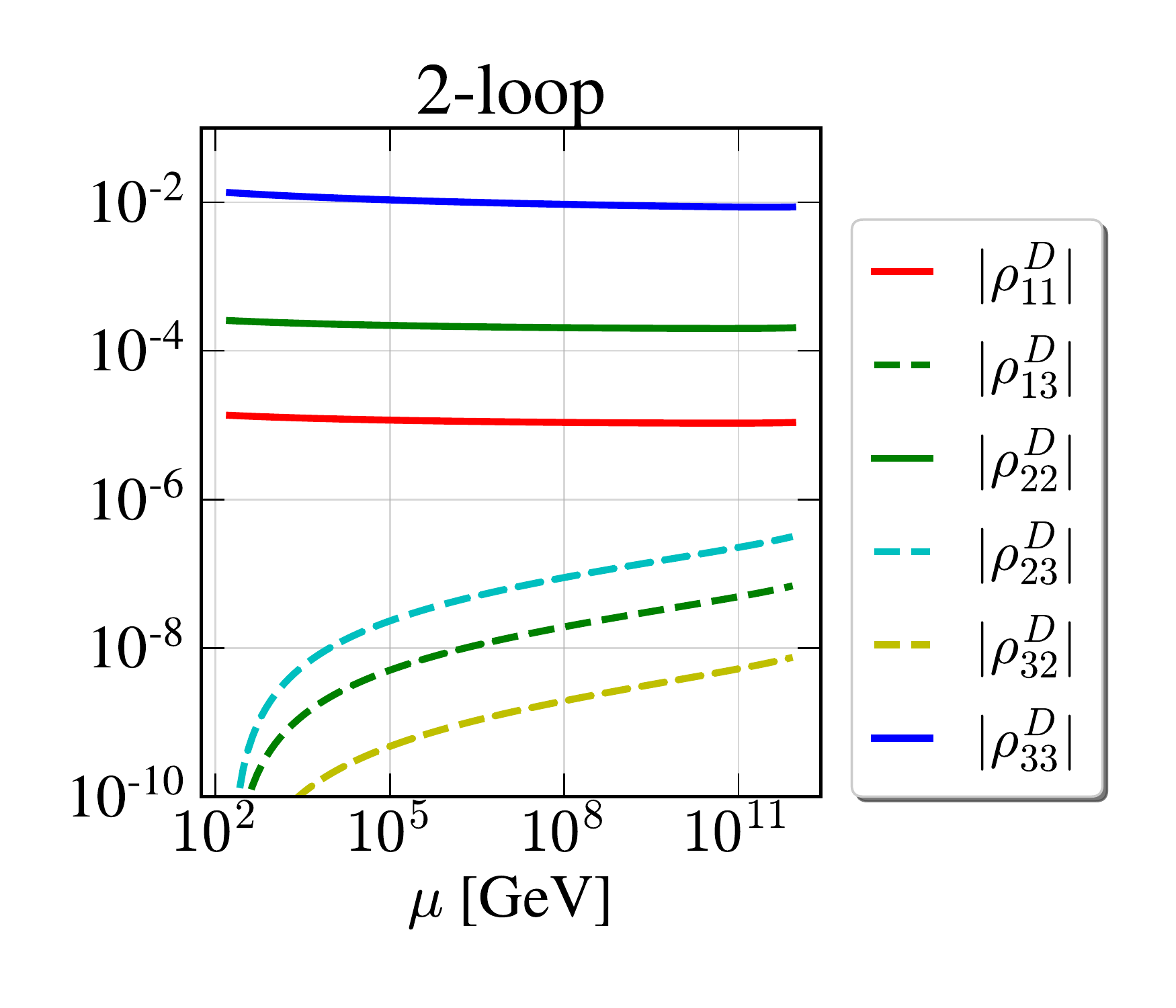}
\end{tabular}
\caption{Evolution of the parameter point in \eq{hard1}; which is an example point from scenario III (hard broken \Zsym in the potential).}
\label{fig:hard11D}
\end{center}
\end{figure}

As an example of this scenario, we plot the evolution of the parameter point 
\begin{align}\label{eq:hard1}
  \tan\beta =~&  1.17069, &&&  M_{12}^2 =~&   11435.4\text{ GeV}^2, &&& \lambda_1 =~&  0.468466,\nn
    \lambda_2 =~&  0.754759, &&& \lambda_3 =~&  -0.150675, &&& \lambda_4 =~& 0.0035644,\nn
  \lambda_5 =~&  0.0692585, &&& \lambda_6 =~& -0.0831947, &&& \lambda_7 =~& 0.206476,
\end{align}
in \fig{hard11D}.
This parameter point is chosen since it induces the largest FCNC.
At the scale of perturbativity breakdown, $\sim 10^{12}$ GeV, the non-diagonal Yukawa element is $\lambda_{23}^D\sim 10^{-4}$.
For completeness, it should be noted that stability and unitarity are violated at $\approx 5$ TeV  and  $\approx 10^{11}$ GeV respectively for this point, but as argued above we ignore this since we are looking for maximal ${\cal Z}_2$-breaking effects.

\subsection{Scenario IV: Yukawa alignment}\label{yukScan}

The ansatz of flavor alignment in the Yukawa sector as in \eq{alignmentAnsatz} is not stable under RG running, except for the \Zsym symmetric values of the $a^F$ parameters in \Tab{Z2symmetries}.
In this section we investigate how large the generated FCNCs can become by varying these $a^F$ parameters.
Since the flavor alignment is an example of a \Zsym breaking, it will also spread to the scalar sector and induce non-zero $\lambda_{6,7}$, which could be a problem in the evolution of the potential.

\begin{figure}[h!]
\begin{center}
\textbf{Scenario IV.a}
\hspace{-1cm}
\begin{tabular}{cc}
\includegraphics[trim=0cm 0.5cm 0cm 0cm,clip,height=0.35\textwidth]{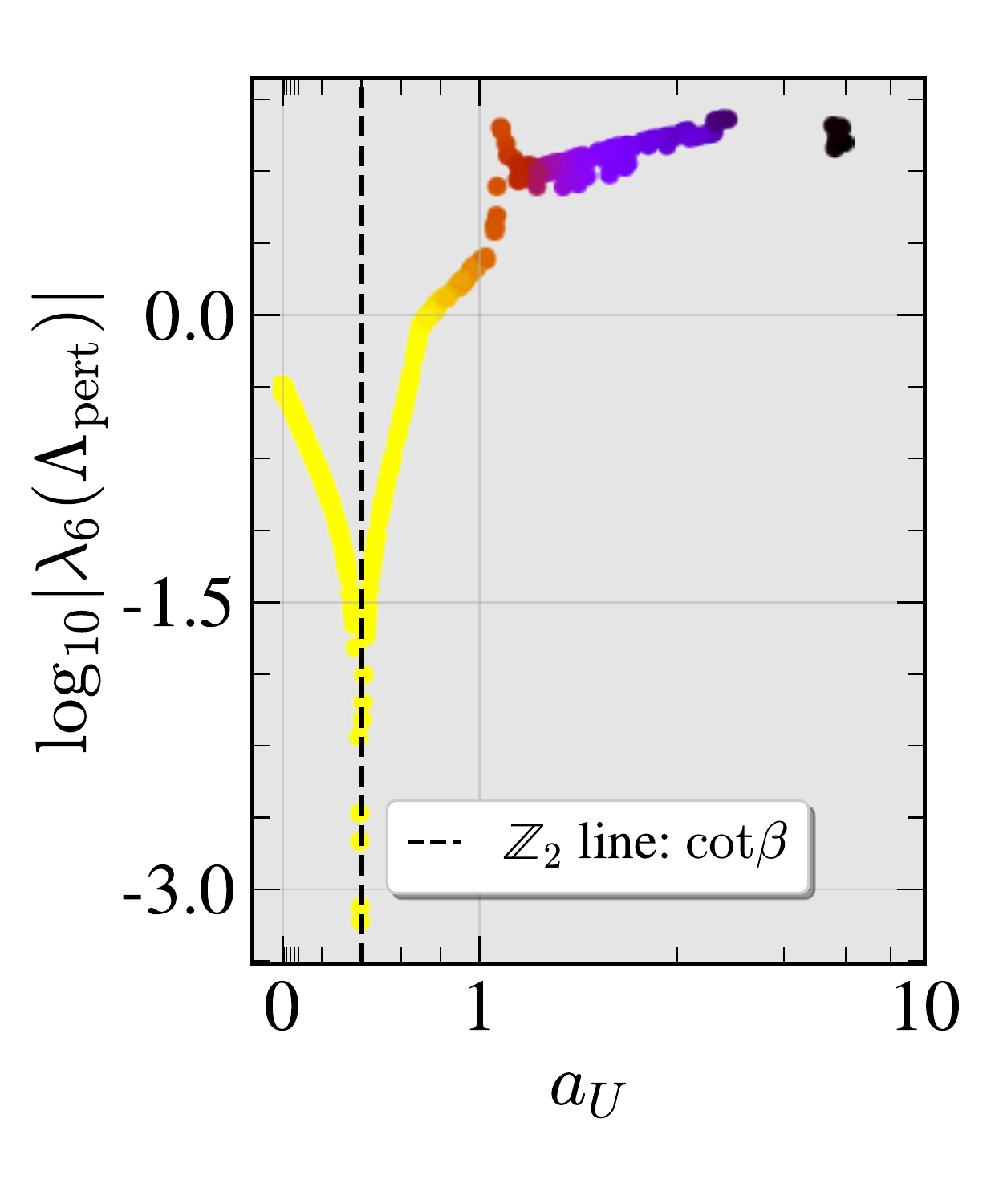}
\includegraphics[trim=0cm 0.5cm 0cm 0cm,clip,height=0.35\textwidth]{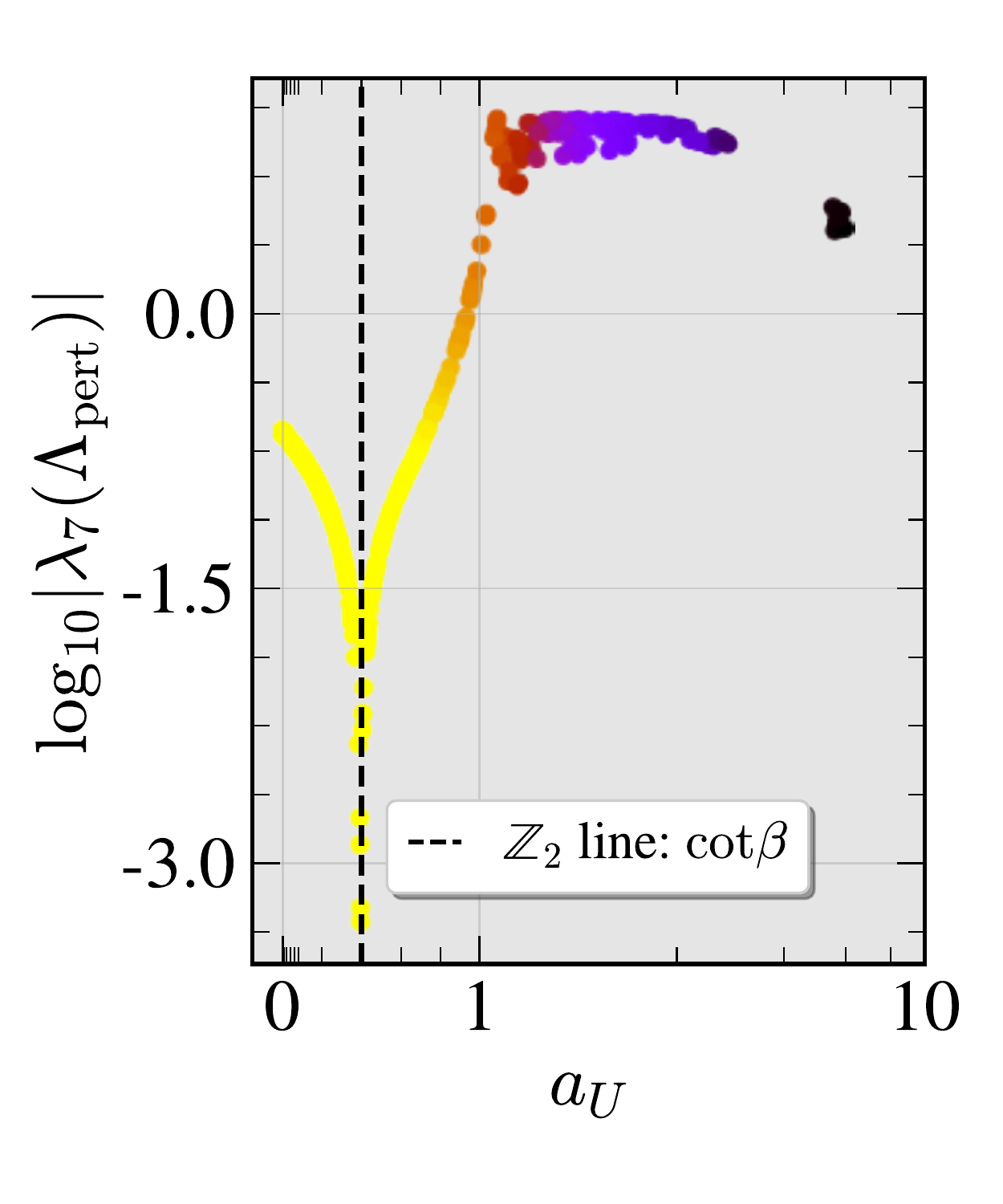}
\includegraphics[trim=0cm 0.5cm 0.5cm 0cm,clip,height=0.35\textwidth]{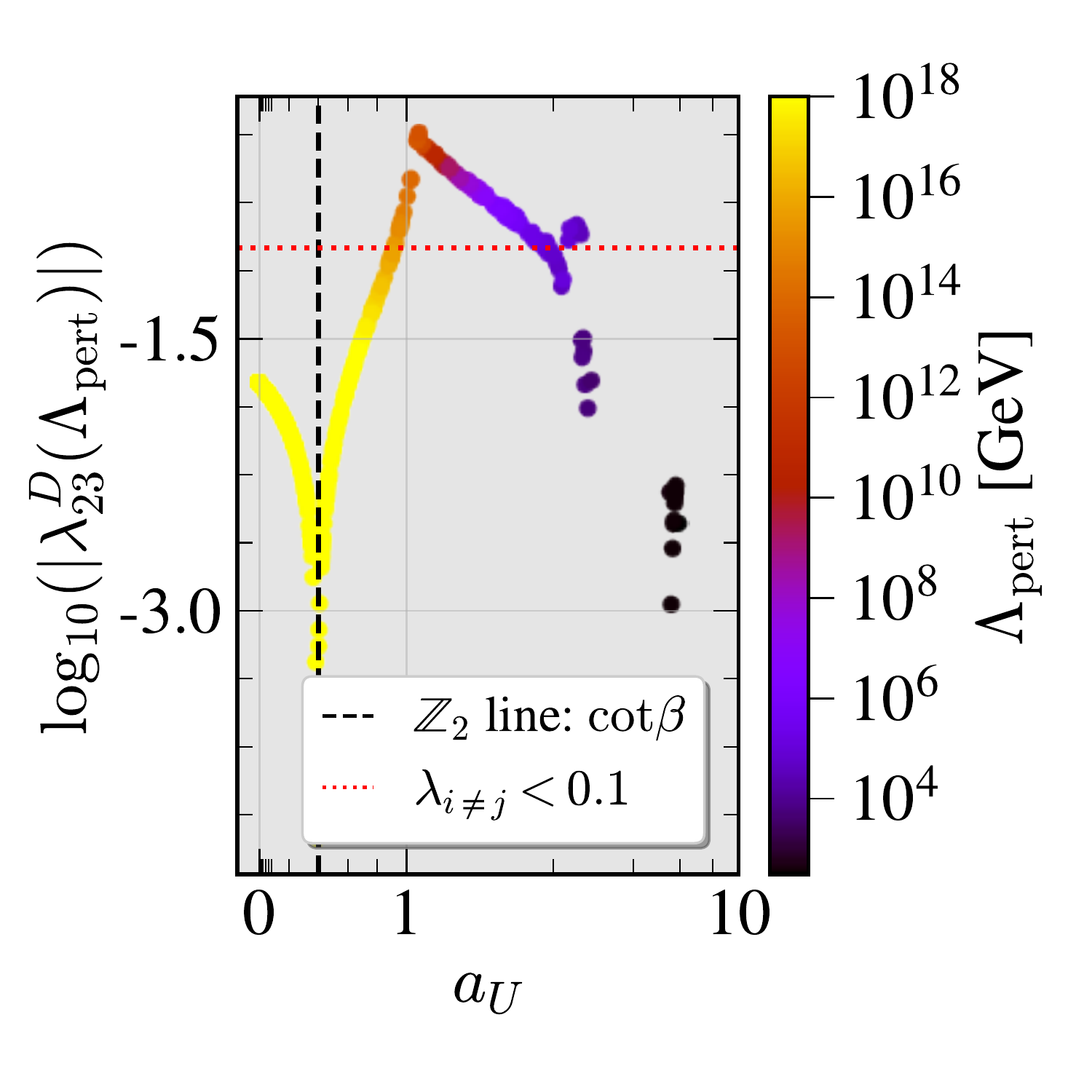}\\
\includegraphics[trim=0cm 0.5cm 0cm 0cm,clip,height=0.35\textwidth]{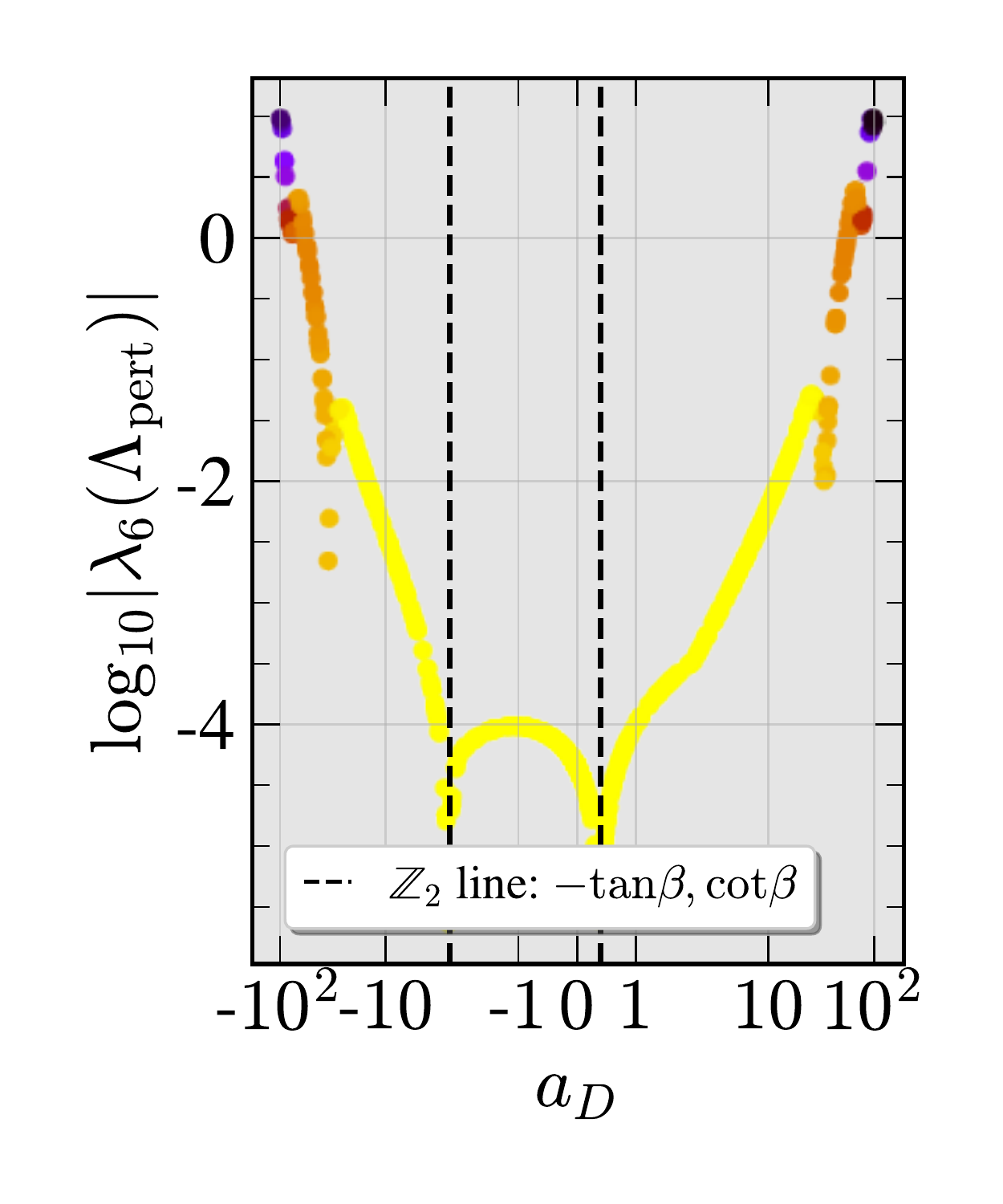}
\includegraphics[trim=0cm 0.5cm 0cm 0cm,clip,height=0.35\textwidth]{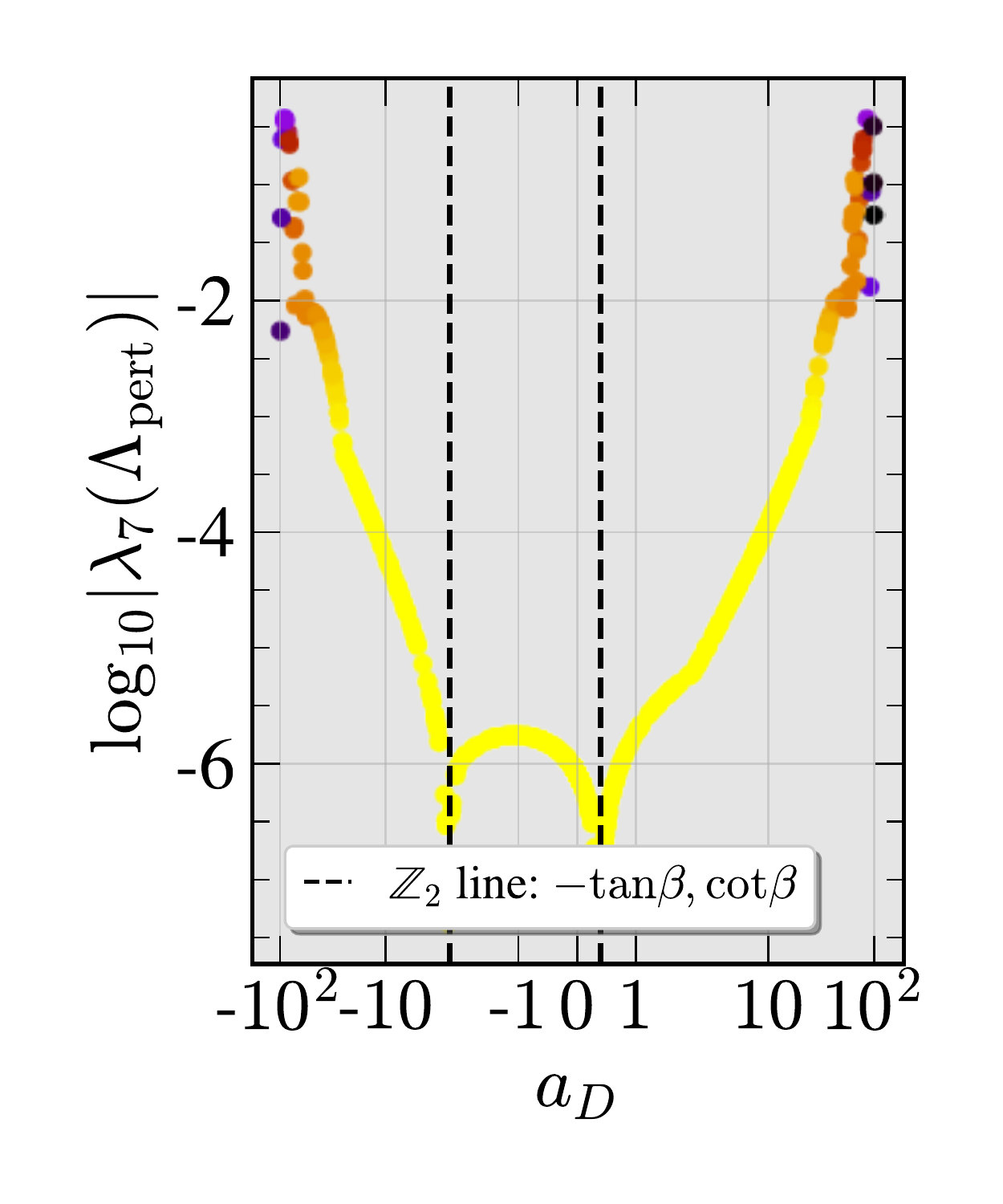}
\includegraphics[trim=0cm 0.5cm 0.5cm 0cm,clip,height=0.35\textwidth]{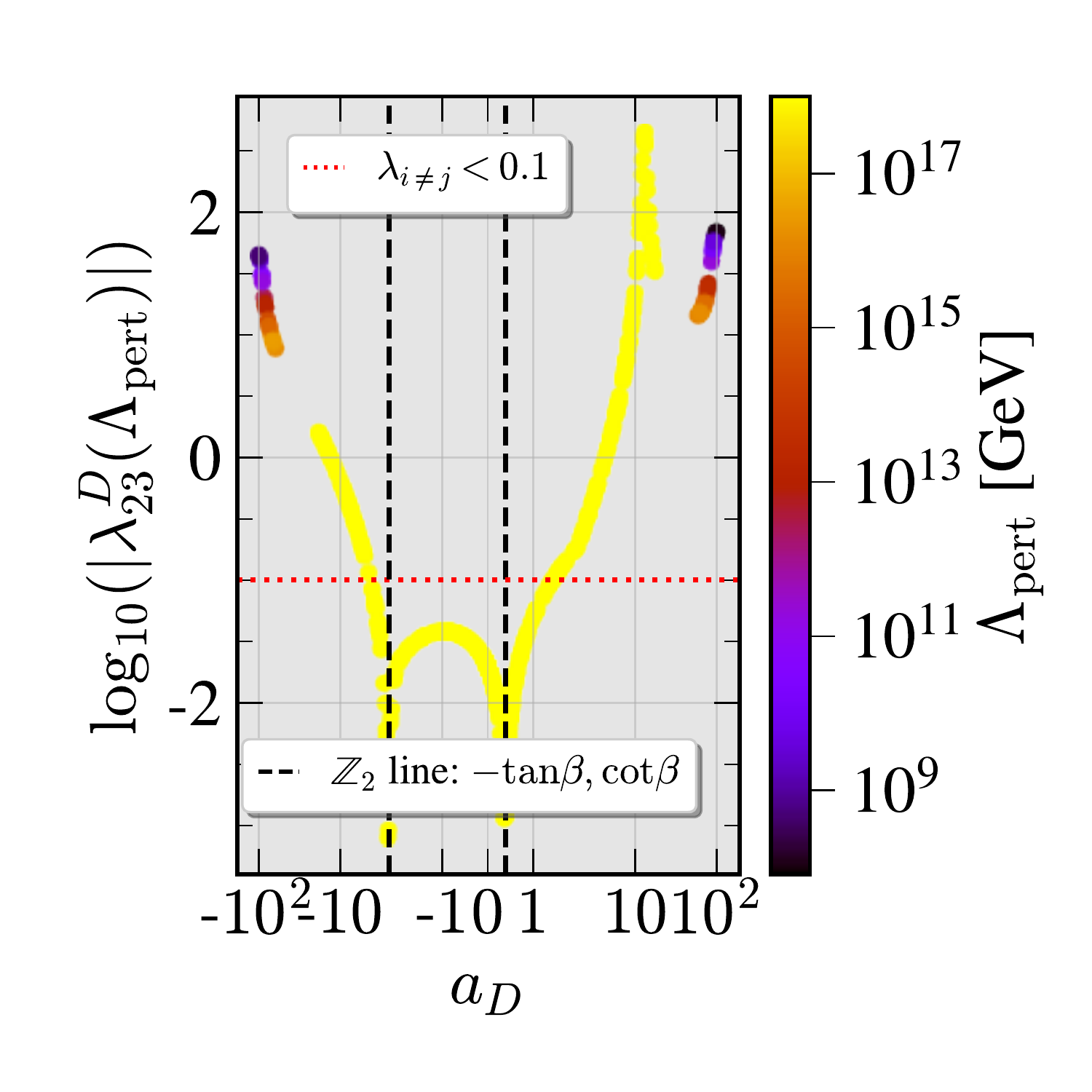}
\end{tabular}
\caption{Separate variation of the alignment parameters $a_U$ and $a_D$ with the scalar potential in \eq{gen2}. $a_D(a_U)=\cot\beta\approx 0.4$ in upper (lower) row.}
\label{fig:yukA}
\end{center}
\end{figure}

In scenario IV.a we separately vary each $a^F$ one-by-one, while keeping the other fixed, with the scalar potential in \eq{gen2}, see \fig{yukA}.
The \Zsym symmetric values, $a^U=\cot\beta$ and $a^{D}=\cot\beta$ or $-\tan\beta$, are clearly visible and are displayed as vertical lines.
As expected, the up-sector is much more sensitive because of the large top Yukawa coupling.
sizable non-diagonal FCNCs can be generated by deviating from the \Zsym symmetric values in both the up and down sector.
We especially note that, perturbativity in the scalar potential easily breaks down for large deviations of $a^U \gtrsim 1$ in the up sector.

When varying $a^D$, the maximum generated FCNC occurs for a value $a^D \approx 5/ a^U$, where $a^U=\cot\beta$.
In this region, $\lambda_{23}^D$ can become very large, $\ordo{10^2}$, without perturbativity breaking down.
The phenomenon that the maximum occurs when $a^D \propto 1/a^U$ is general and is more clearly visible in \fig{yuklamD23} below.

\begin{figure}[h!]
\begin{center}
\textbf{Example from scenario IV.a}
\begin{tabular}{cc}
\includegraphics[trim=0cm 1cm 0.5cm 0.5cm,clip,height=0.3\textwidth]{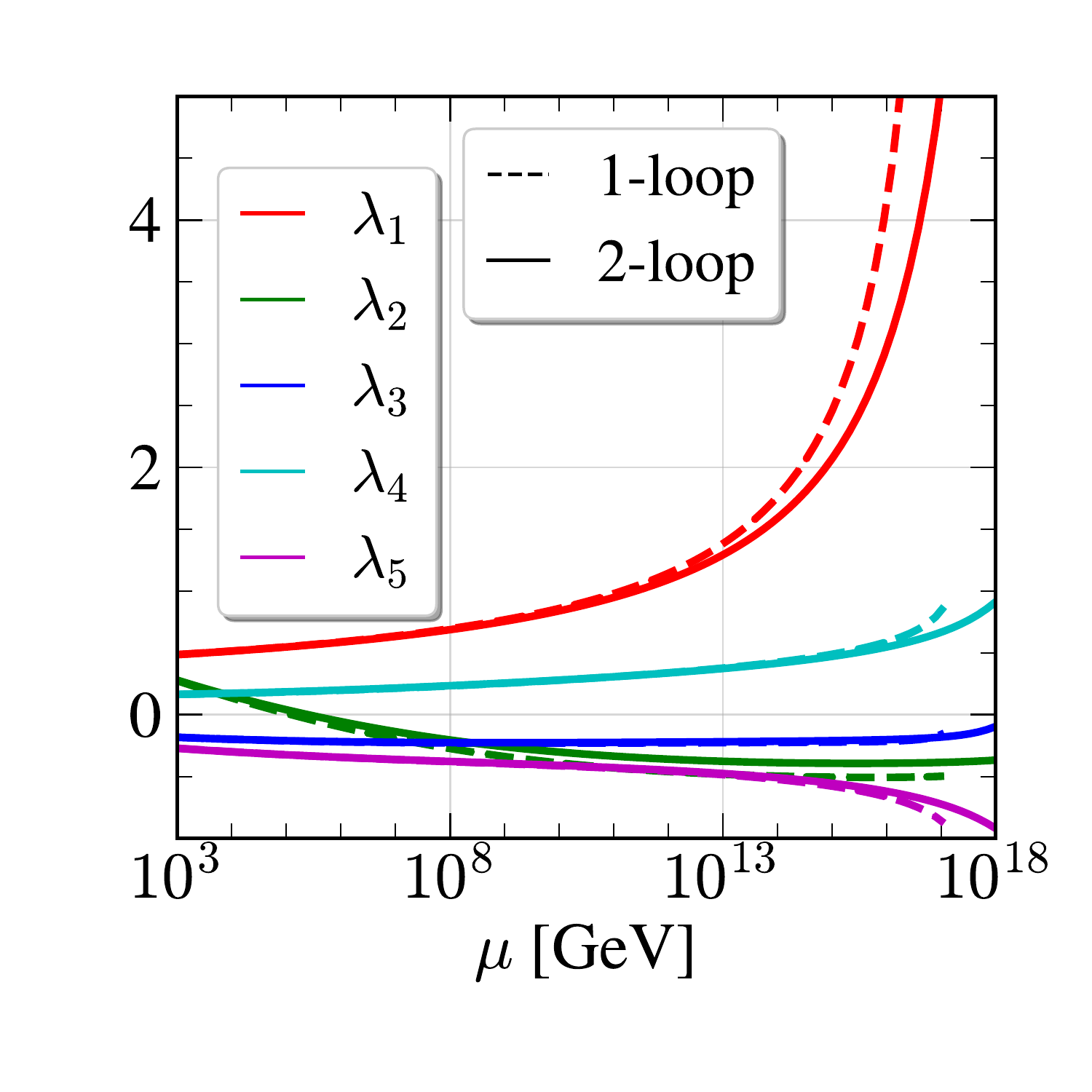}
\includegraphics[trim=0.5cm 1cm 0.5cm 0.5cm,clip,height=0.3\textwidth]{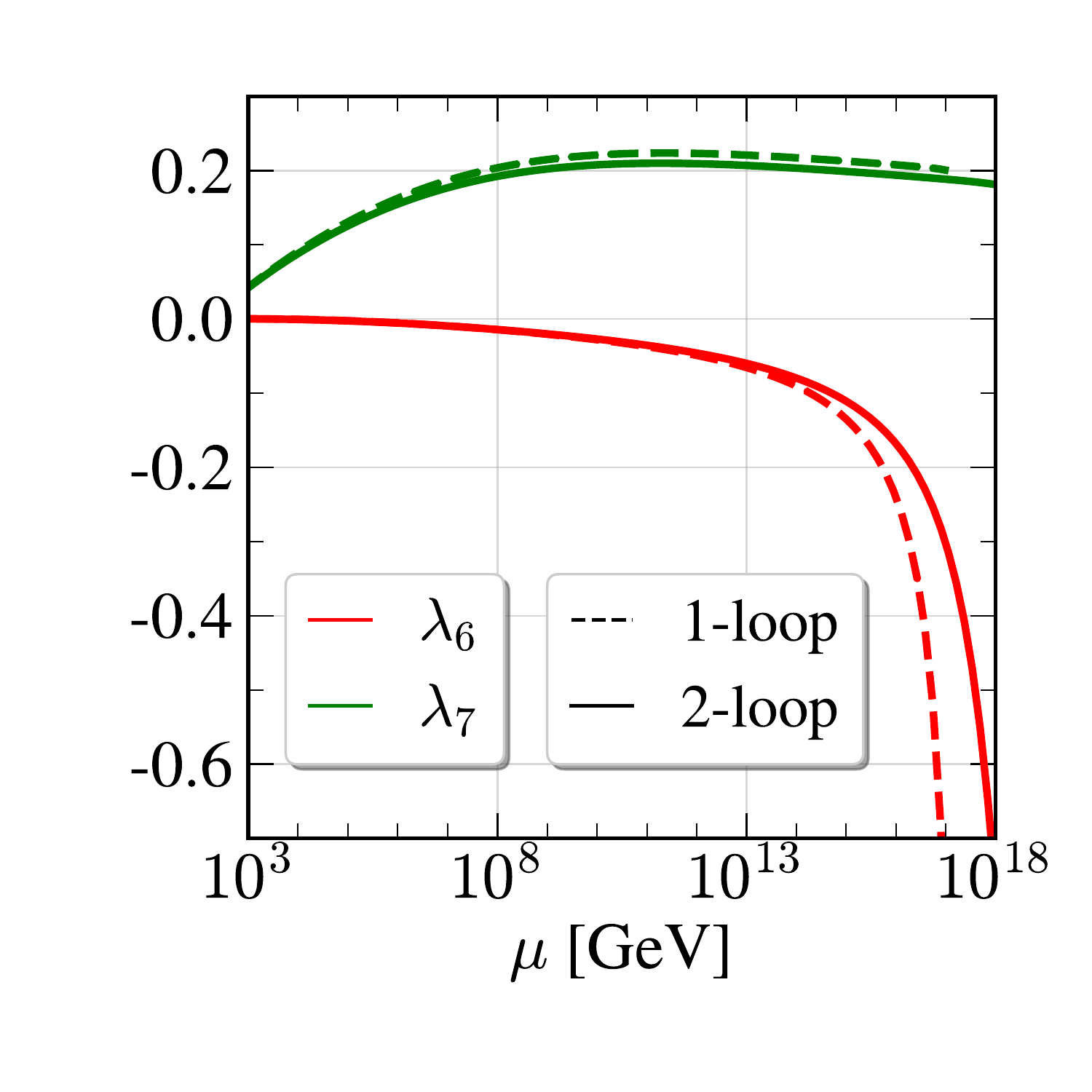}
\includegraphics[trim=0.5cm 1cm 0cm 0.5cm,clip,height=0.3\textwidth]{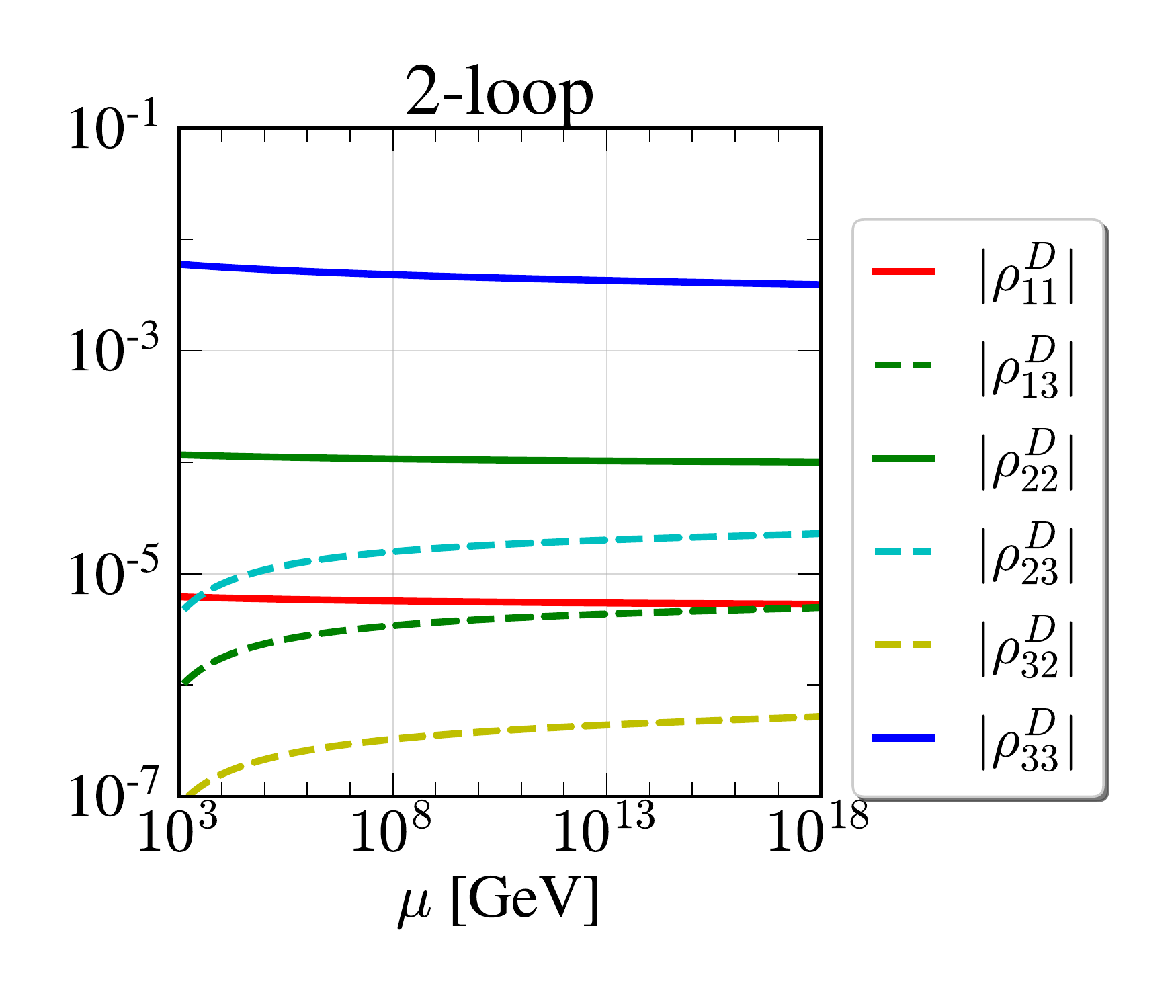}
\end{tabular}
\caption{Evolution of the parameter point in \eq{gen2} with an aligned Yukawa sector, $a^U=0.8$, $a^D=a^L=\cot\beta\approx 0.4$.}
\label{fig:aligned11D}
\end{center}
\end{figure}

To illustrate the RG evolution, an example parameter point from scenario IV.a with the alignment coefficients $a^U=0.8$, $a^D=a^L=\cot\beta\approx 0.4$ is shown in \fig{aligned11D}.
Stability breaks down for this parameter space point already at 1 TeV, but as before we ignore that.

\begin{figure}[h!]
\begin{center}
\textbf{Scenario IV.b}
\begin{tabular}{cc}
\includegraphics[trim=0cm 1cm 0.5cm 0.5cm,clip,height=0.35\textwidth]{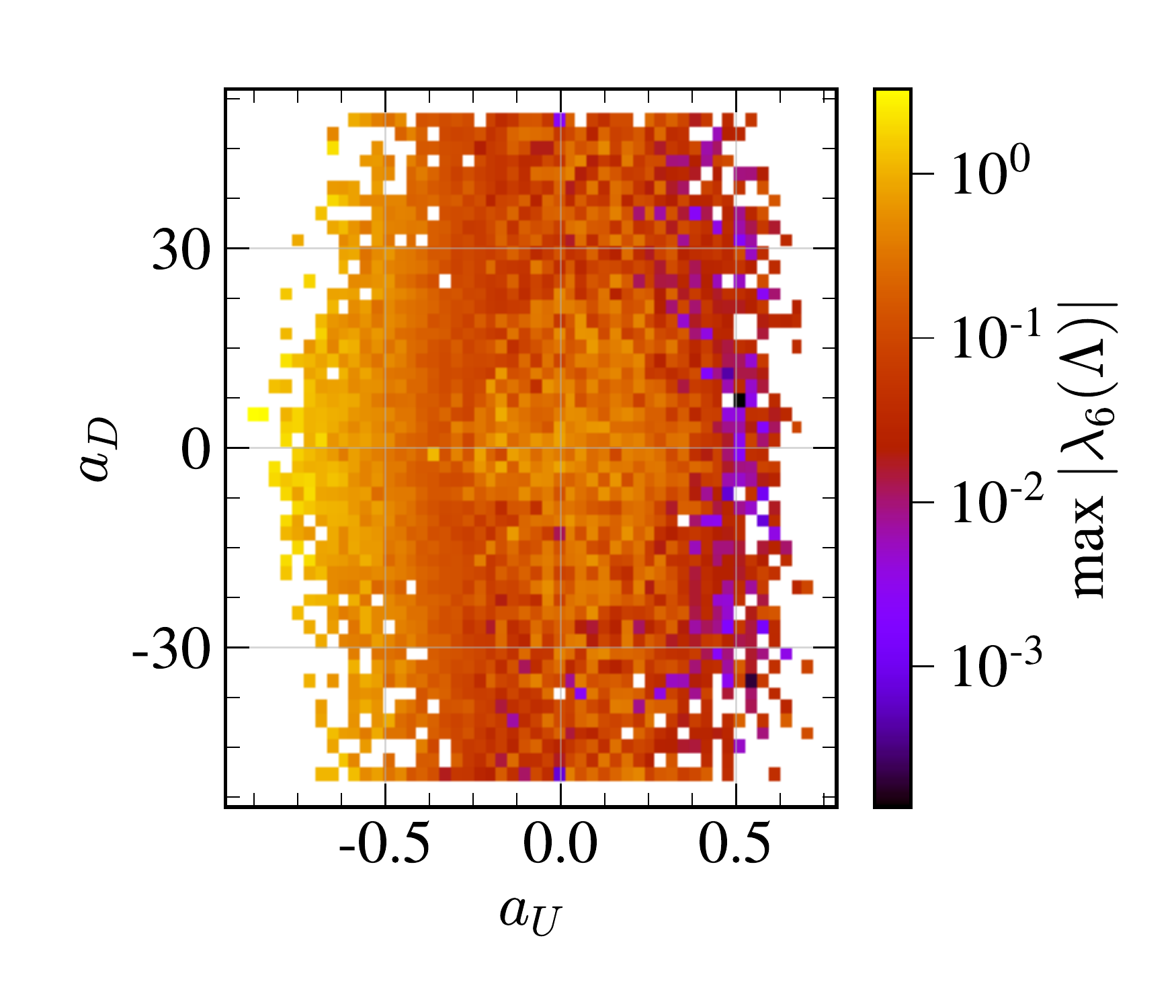}
\includegraphics[trim=0.5cm 1cm 0.5cm 0.5cm,clip,height=0.35\textwidth]{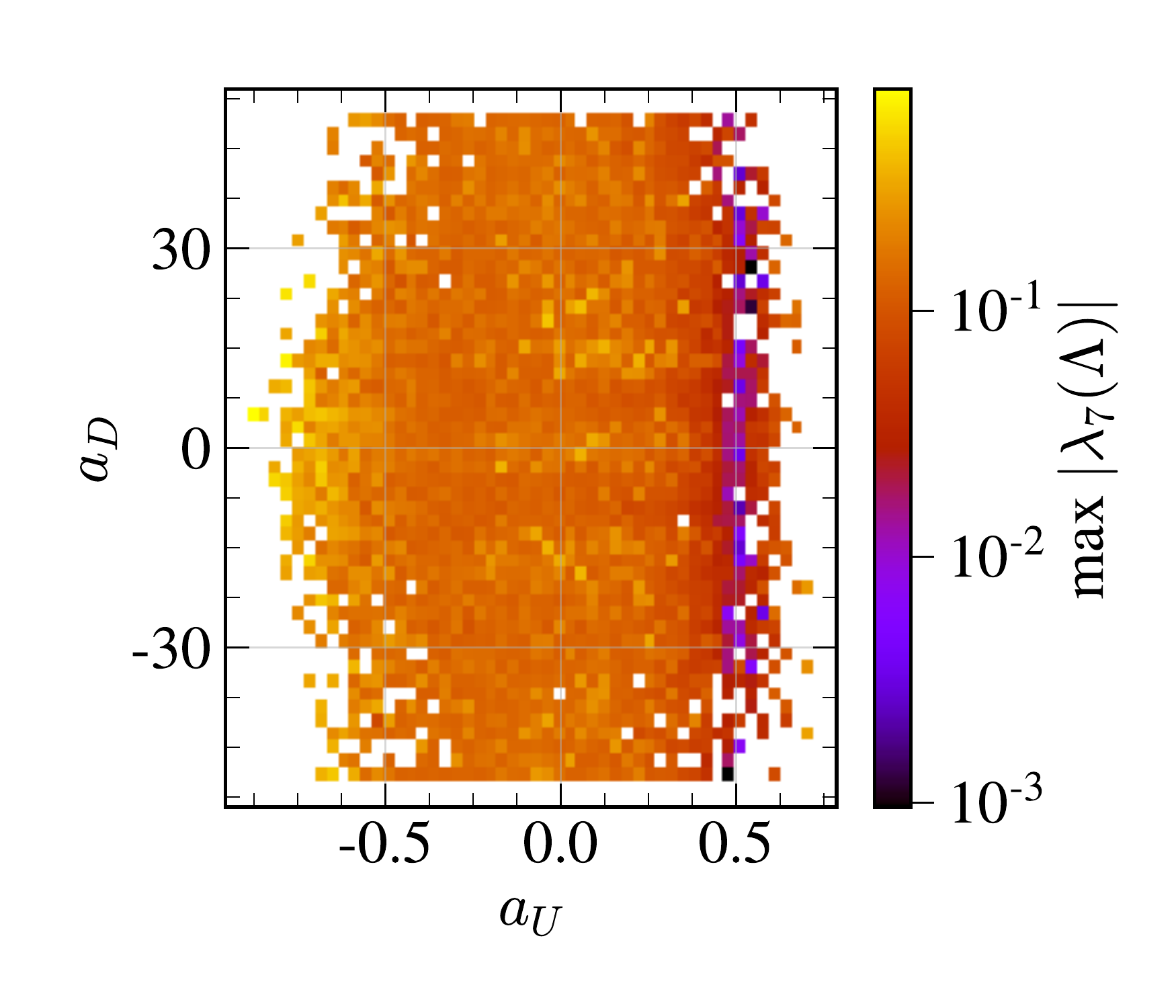}
\end{tabular}
\caption{Generated $\lambda_6$ (Left) and $\lambda_7$ (Right) at the scale $\Lambda=10^{10}$ GeV as a function of the alignment parameters $a_U$ and $a_D$.}
\label{fig:yukLambda67}
\end{center}
\end{figure}

In scenario IV.b, we perform a random parameter scan of a 2HDM with a softly broken \Zsym potential, just as in scenario II, but fix $\tan\beta = 2$.
Furthermore, here we also vary the alignment parameters in the quark sector, $a^{U,D}$.
This way we get a quantitative estimate of the induced \Zsym breaking parameters $\lambda_{6,7}$ and $\lambda_{i\neq j}^F$.
As discussed earlier, we only keep parameter points that satisfies all theoretical constraints during the entire RG running up to $10^{10}$ GeV.

The induced $\lambda_{6,7}$ are shown for a scan of $10^4$ surviving parameter points in \fig{yukLambda67}.
It is clear that the generated \Zsym breaking parameters are more sensitive to variations in $a^U$, because of the large top Yukawa coupling.
Note the vertical area around $a^U=\cot\beta=1/2$ which gives the smallest $\lambda_{6,7}$.

As previously mentioned, it is the $\lambda_{23}^D = \rho^D_{23} \sqrt{v^2/(2m_s m_b)}$ which is the largest generated non-diagonal element in more than $99\%$ of all cases; therefore we use it as a measure of the induced FCNCs in \fig{yuklamD23}.
The dark regions, where the smallest FCNCs are generated, arise since the relations $a^U=a^D$ and $a^D = -1/a^U$ correspond to an aligned Yukawa sector that is 1-loop stable under RG evolution \cite{Ferreira:2010xe}.
Even though these relations imply that there is a \Zsym symmetric Yukawa sector in one particular basis, at 2-loop order the scalar and Yukawa sectors mix and induce FCNCs if the $a^F$ do not take on the correct \Zsym symmetric values set by the scalar potential.
We also note the region $a^D\approx 10/a^U$, where the maximum FCNCs are generated; similar to what was seen when varying $a^D$ separately in \fig{yukA}. 

\begin{figure}[h!]
\begin{center}
\begin{tabular}{cc}
\includegraphics[trim=0cm 0.5cm 0.5cm 0cm,clip,height=0.35\textwidth]{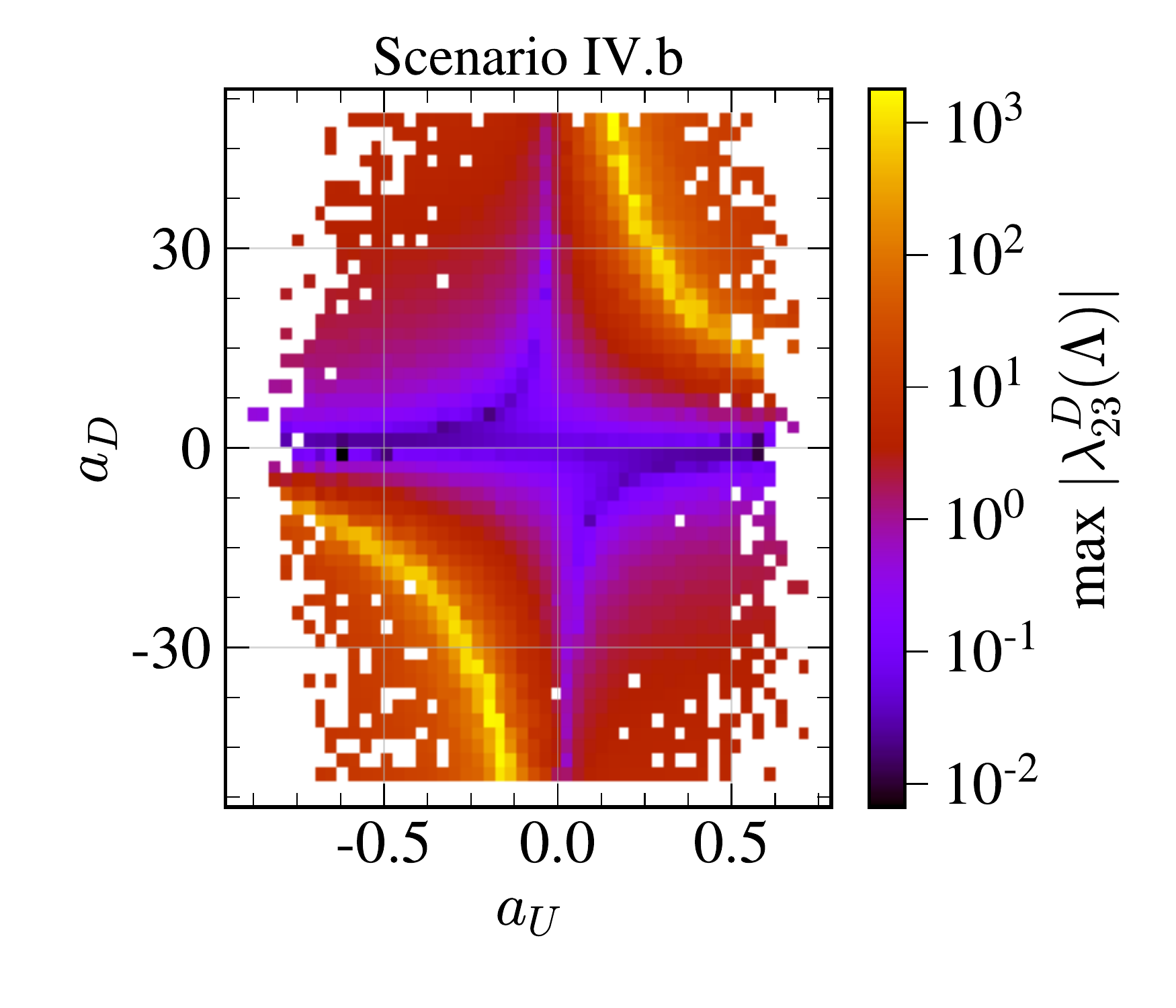}
\end{tabular}
\caption{Generated $\lambda_{23}^D = \rho^D_{23} \sqrt{v^2/(2m_s m_b)}$ at the scale $\Lambda=10^{10}$ GeV as a function of the alignment parameters $a_U$ and $a_D$.}
\label{fig:yuklamD23}
\end{center}
\end{figure}

\section{Conclusions}\label{conclusion}

We have derived the complete set of 2-loop RGEs for a general, potentially complex, 2HDM and implemented them in the \code{C++} code \code{2HDME} \cite{Oredsson:2018vio}.
Using this software, we performed a RG analysis of the \CP-conserving 2HDM and investigated the parameter space with different levels of \Zsym symmetry (exact, soft or hard breaking).
The case of a softly broken \Zsym symmetry is the most unconstrained type of 2HDM, since it allows for large scalar masses, larger values of $\tan\beta$ and we found regions in the parameter space with models that are valid all the way up to the Planck scale.

Breaking the \Zsym symmetry hard at the EW scale induces dimensionless \Zsym breaking parameters during the RG running which can be potentially dangerous and severely limit the validity of the model.
We have looked at two scenarios, where we have either broken the \Zsym symmetry hard in the scalar potential, by having non-zero $\lambda_{6,7}$, or in the Yukawa sector, by making a flavor Yukawa ansatz at the EW scale.

We have found that for $\lambda_{6,7} \lesssim 0.1$ at the EW scale, it is possible to find models that are valid all the way up to the Planck scale at $10^{18}$ GeV; while models break down already at $\sim 10^5$ GeV if $\lambda_{6,7} \gtrsim 1$.
The generated FCNCs, that spread from the scalar sector first at 2-loop order, are however heavily  suppressed.

A flavor ansatz can be a viable solution to the FCNC problem, even though it is not protected during RG evolution.
Deviations of $a^F$ from the \Zsym symmetric limit, $a^D=a^U=\tan\beta$, $a^D=a^U=\cot\beta$ or $a^D=-1/a^U$, should however not be too large.
The up sector is especially sensitive when it comes to induced $\lambda_{6,7}$; which contribute to a rapid growth of the quartic couplings.
In the down sector, the induced FCNCs pose the primary limitation.
Finally we observe that there is a region in the flavor alignment parameter space, $a^D\propto 1/a^U$, which gives the maximum FCNCs.
The origin of this effect is unclear and requires further study.

\begin{acknowledgments}

The authors would like to thank Hugo Serodio and Johan Bijnens for useful discussions.

This work is supported in part by the Swedish Research Council grants
contract numbers 621-2013-4287 and 2016-05996 and by the European Research Council (ERC) under the European Union's Horizon 2020 research and innovation programme (grant agreement No 668679).

\end{acknowledgments}

\newpage

\appendix

\section{Quartic couplings in scenario II}\label{higgsBasisPlots}

\begin{figure}[h!]
  \begin{center}
  \textbf{Generic basis}
  \begin{tabular}{cc}
  \hspace{-1cm}
  \includegraphics[trim=0.5cm 1cm 5cm 0.5cm,clip,height=0.35\textwidth]{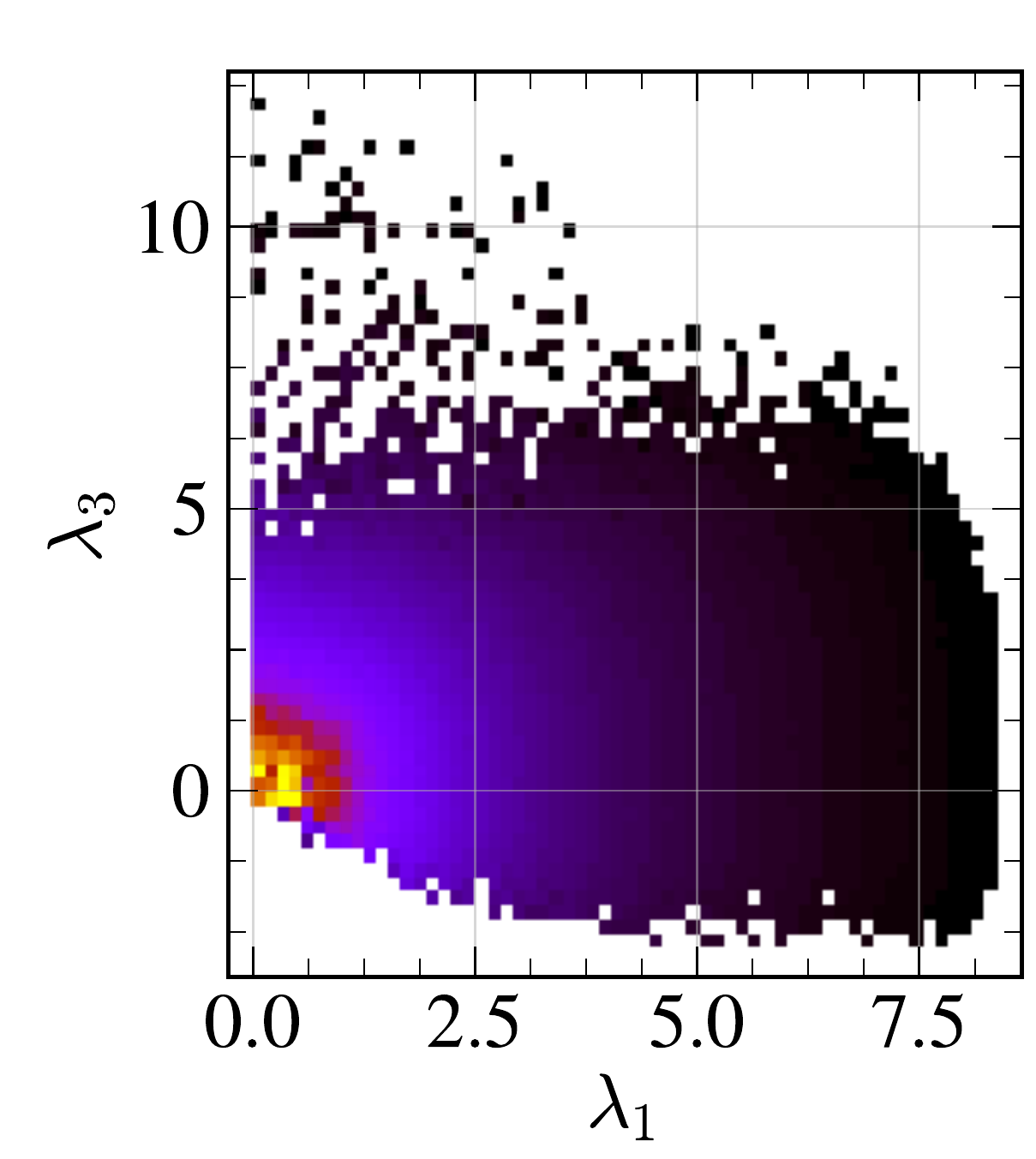}
  \includegraphics[trim=0.5cm 1cm 5cm 0.5cm,clip,height=0.35\textwidth]{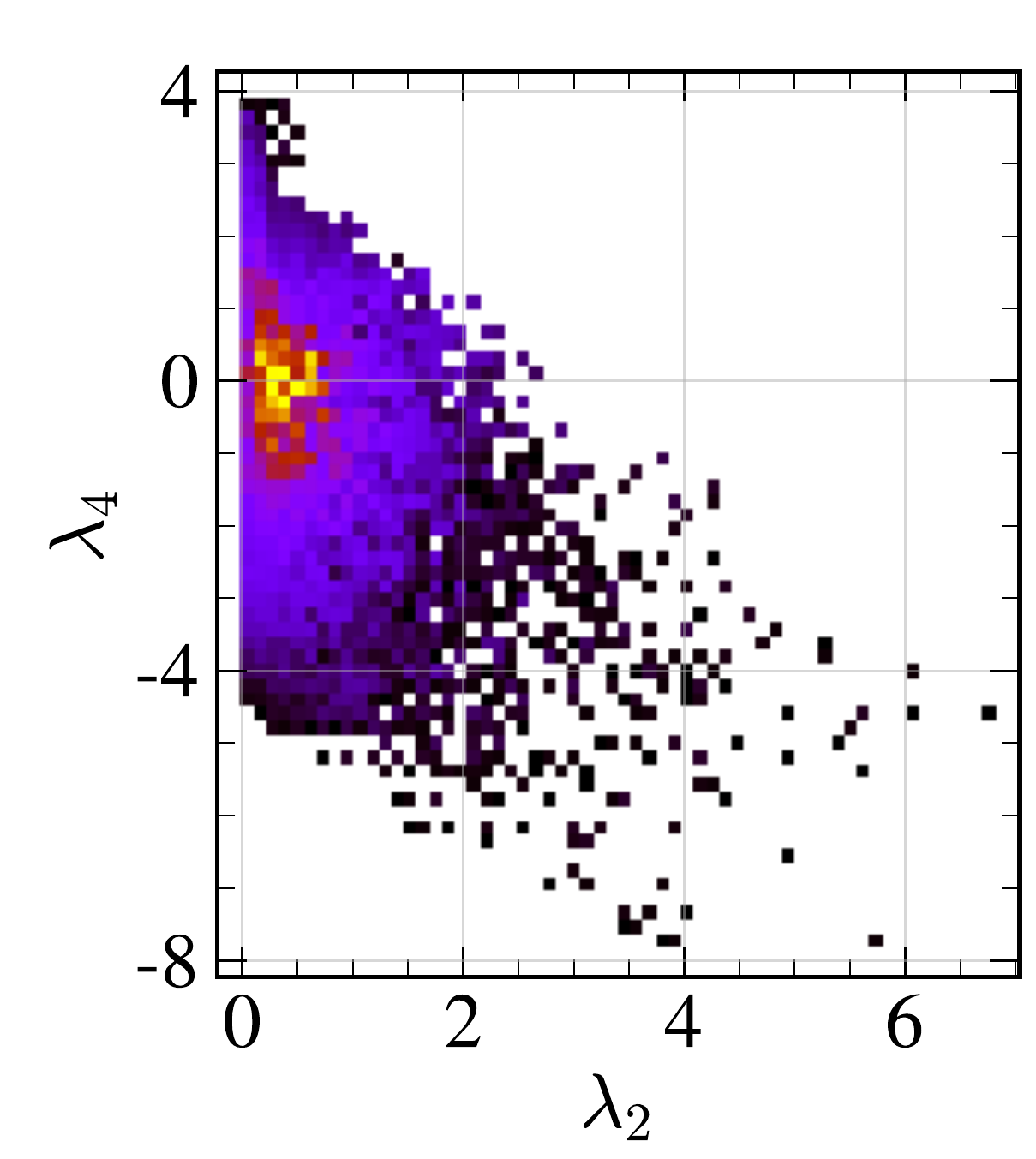}
  \includegraphics[trim=0.5cm 1cm 0cm 0.5cm,clip,height=0.35\textwidth]{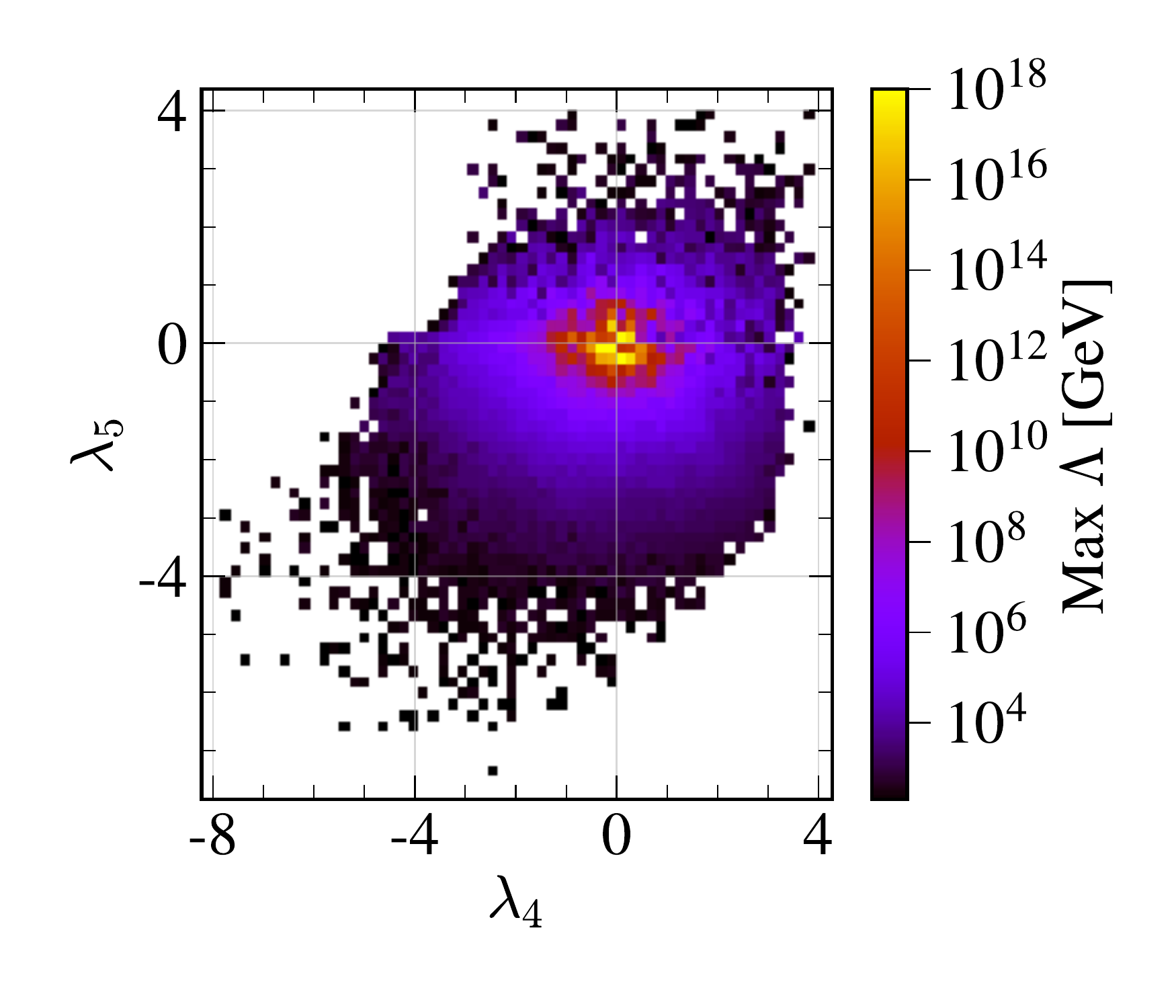}.
  \end{tabular}
  \caption{The maximum breakdown energy scale as a function of quartic couplings in the generic basis in the parameter scan of scenario II.}
  \label{fig:HBHS_softGeneric}
  \end{center}
  \end{figure}

\begin{figure}[h!]
  \begin{center}
  \textbf{Higgs basis, Scenario II}\\
  \begin{tabular}{cc}
  \includegraphics[trim=0.5cm 1cm 5cm 0.5cm,clip,height=0.35\textwidth]{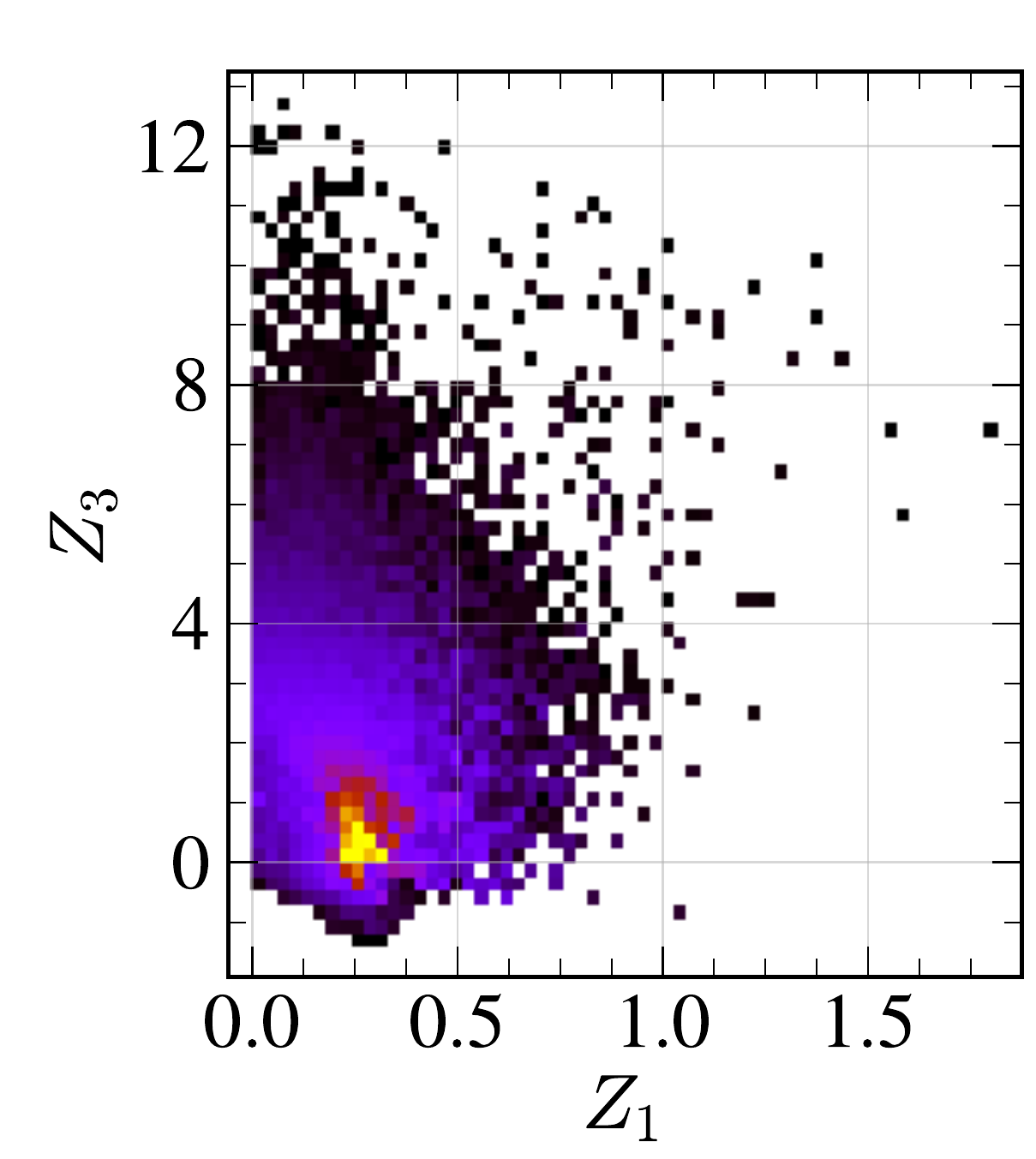}
  \includegraphics[trim=0.5cm 1cm 0.5cm 0.5cm,clip,height=0.35\textwidth]{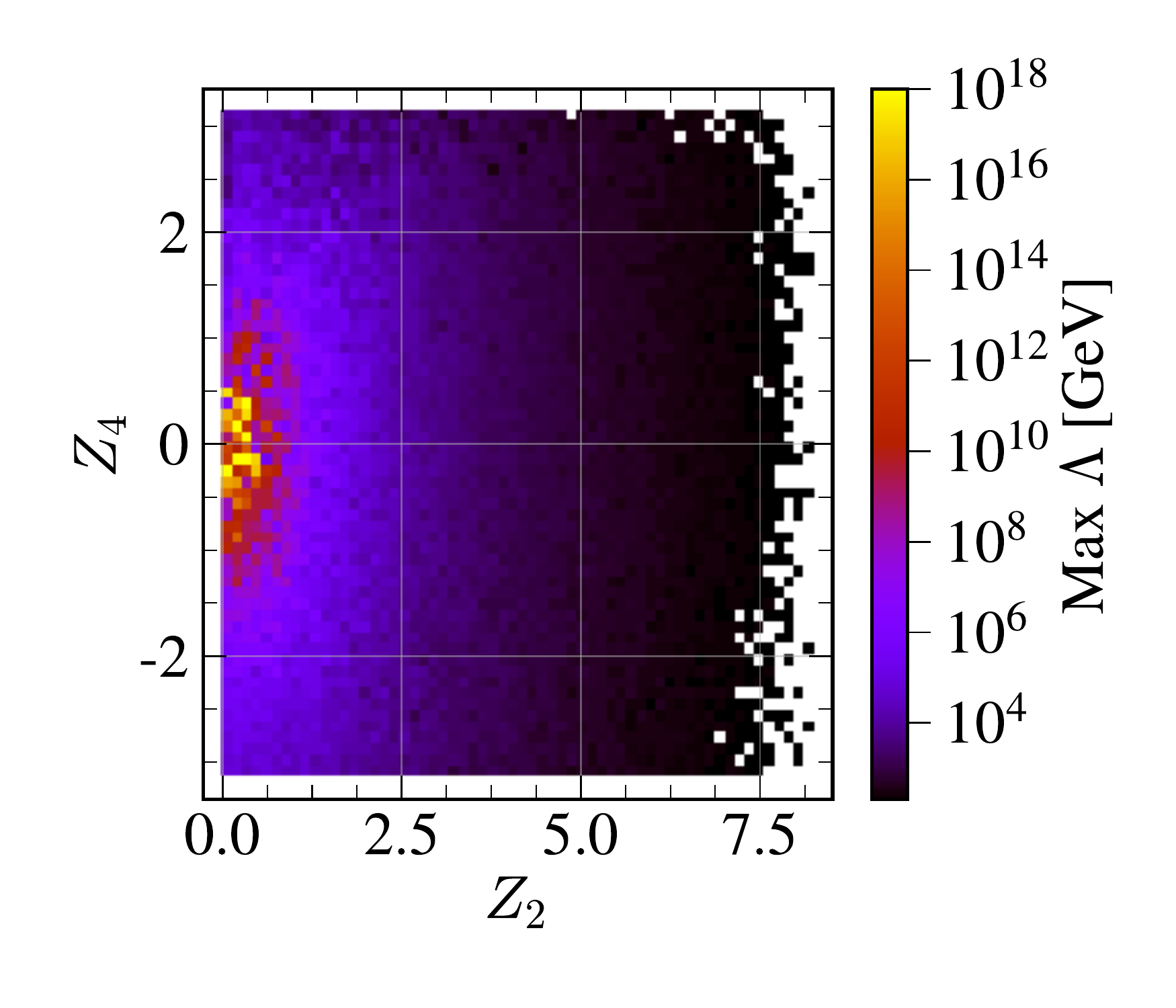}\\
  \includegraphics[trim=0.5cm 1cm 5cm 0.5cm,clip,height=0.35\textwidth]{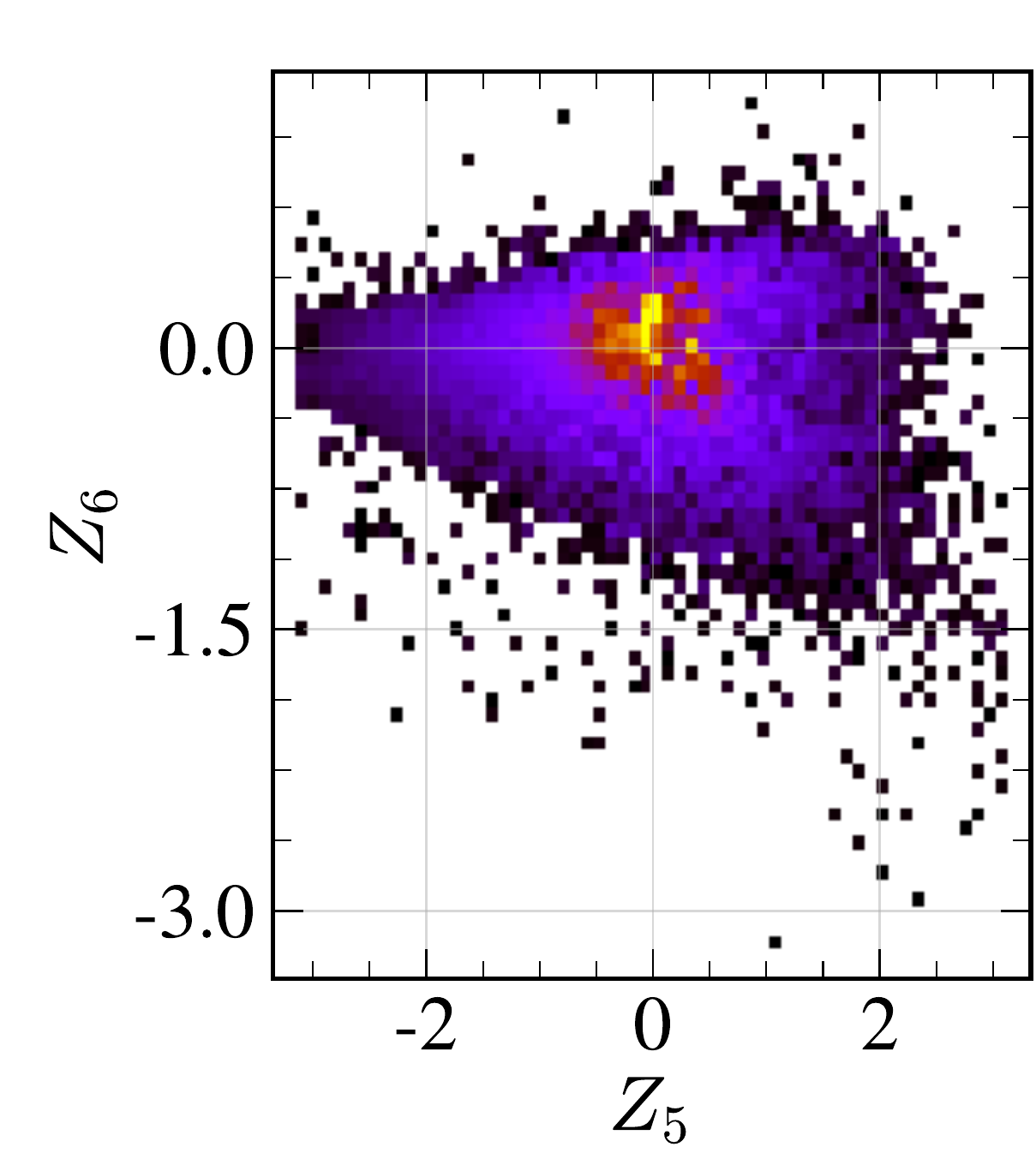}
  \includegraphics[trim=0.5cm 1cm 0.5cm 0.5cm,clip,height=0.35\textwidth]{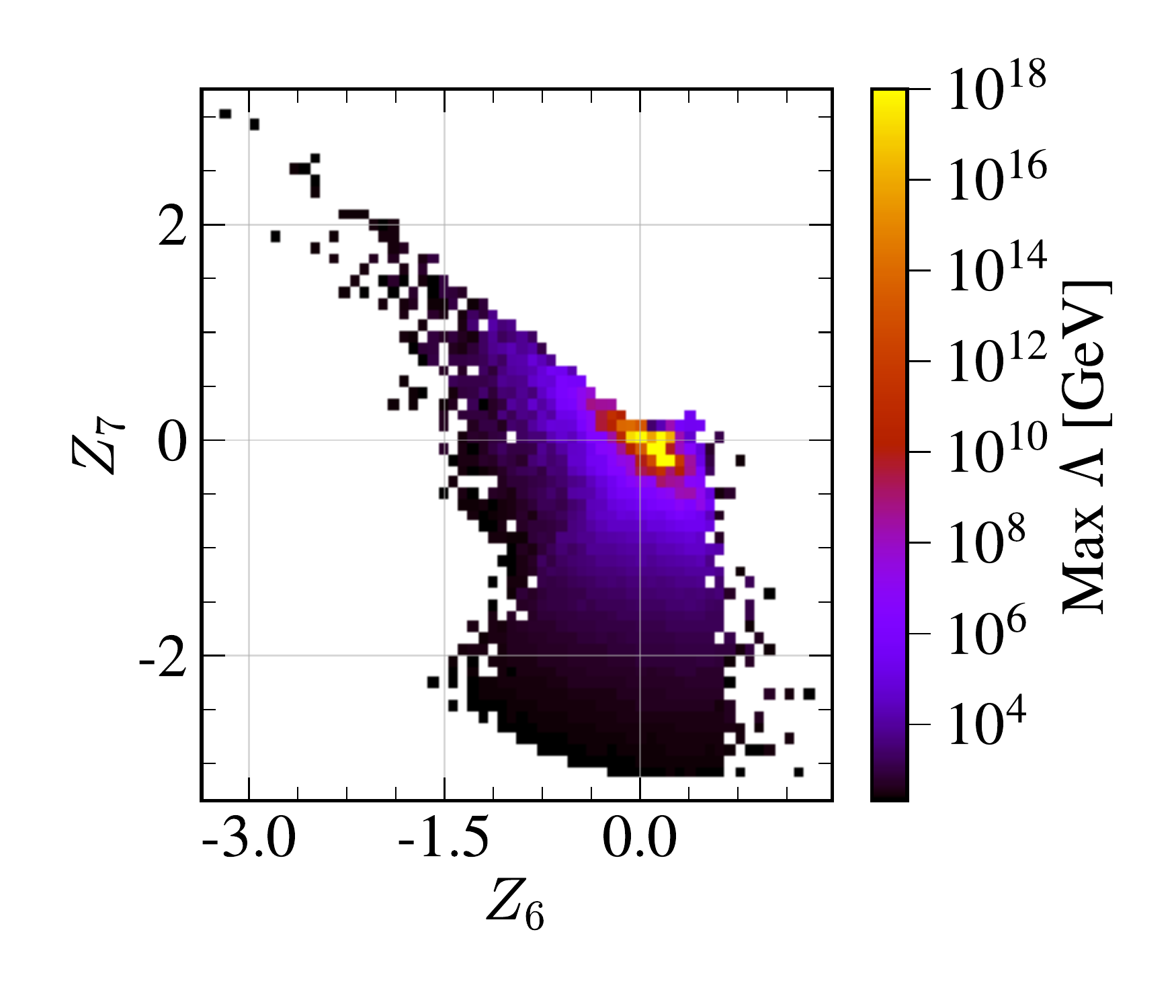}
  \end{tabular}
  \caption{The maximum breakdown energy scale as a function of quartic couplings in the Higgs basis in the parameter scan of scenario II.}
  \label{fig:HBHS_softHiggs}
  \end{center}
  \end{figure}

The quartic couplings in the random parameter scan of scenario II with a softly broken \Zsym symmetry, described in \sec{softScan}, as functions of breakdown energy in RG evolution are shown in the generic basis in \fig{HBHS_softGeneric} and in the Higgs basis in \fig{HBHS_softHiggs}.
The color coding is the maximum breakdown-energy from perturbativity, unitarity or stability in each bin.

\section{Tree-level unitarity conditions} \label{unitarity}

The tree-level unitarity conditions for a general 2HDM have been worked out in \mycite{Ginzburg:2005dt}.
There, they work out the following scattering matrices:
\begin{align}
	\Lambda_{21} \equiv~ & \left( \begin{array}{ccc} \lambda_1 & \lambda_5 & \sqrt{2}\lambda_6 \\
										\lambda_5^* & \lambda_2 & \sqrt{2}\lambda_7^* \\
										\sqrt{2}\lambda_6^* & \sqrt{2}\lambda_7 & \lambda_3 + \lambda_4 \end{array}\right),\\
	\Lambda_{20} \equiv~ & \lambda_3 - \lambda_4,\\
	\Lambda_{01} \equiv~ & \left( \begin{array}{cccc} \lambda_1 & \lambda_4 & \lambda_6 & \lambda_6^* \\
										\lambda_4 & \lambda_2 & \lambda_7 & \lambda_7^* \\
										\lambda_6^* & \lambda_7^* & \lambda_3 & \lambda_5^* \\
										\lambda_6 & \lambda_7 & \lambda_5 & \lambda_3 \end{array}\right),\\	
	\Lambda_{00} \equiv~ & \left( \begin{array}{cccc} 3\lambda_1 & 2\lambda_3 + \lambda_4 & 3\lambda_6 & 3\lambda_6^* \\
										2\lambda_3 + \lambda_4 & 3\lambda_2 & 3\lambda_7 & 3\lambda_7^* \\
										3\lambda_6^* & 3\lambda_7^* & \lambda_3 + 2\lambda_4 & 3\lambda_5^* \\
										3\lambda_6 & 3\lambda_7 & 3\lambda_5 & \lambda_3 + 2\lambda_4 \end{array}\right).	
\end{align}
In the end, the unitarity constraint puts upper limits on the absolute value of the eigenvalues, $\Lambda_i$, of these matrices,
\begin{align}
	\abs{\Lambda_i} < 8\pi.
\end{align}

\section{Tree-level stability} \label{stability}

Here, we give the conditions for the tree-level scalar potential to be bounded from below, as worked out in \mycite{Ivanov:2006yq, Ivanov:2007de}.

When working out these conditions, \mycite{Ivanov:2006yq, Ivanov:2007de} constructed a Minkowskian formalism of the 2HDM that uses gauge-invariant field bilinears,
\begin{align}
	r^0 \equiv~& \Phi_1^\dagger \Phi_1 + \Phi_2^\dagger \Phi_2,\\
	r^1 \equiv~& 2\real{\Phi_1^\dagger \Phi_2},\\
	r^2 \equiv~& 2\imag{\Phi_1^\dagger \Phi_2},\\
	r^3 \equiv~& \Phi_1^\dagger \Phi_1 - \Phi_2^\dagger \Phi_2.
\end{align}
These can be used to create a four-vector $r^\mu = (r^0, \vec{r})$; where one can raise and lower the indices as usual with the flat Minkowski metric $\eta^{\mu\nu} = \text{diag}(1,-1,-1,-1)$.
In this formalism, the scalar potential is conveniently written as
\begin{align}
	V = -M_\mu r^\mu + \frac{1}{2} r^\mu \Lambda_\mu^\nu r_\nu,
\end{align}
where 
\begin{align}
	M_\mu = \left( -\frac{1}{2}(Y_1+Y_2), \real{Y_3}, -\imag{Y_3}, -\frac{1}{2}(Y_1-Y_2)\right)
\end{align}
and
\begin{align}\small
	\Lambda_\mu^\nu = \frac{1}{2} \left( \begin{array}{cccc} \frac{1}{2}(Z_1 + Z_2) + Z_3 & -\real{Z_6 + Z_7} & \imag{Z_6 + Z_7} & -\frac{1}{2}(Z_1-Z_2) \\
	\real{Z_6 + Z_7} & -Z_4-\real{Z_5} & \imag{Z_5} & -\real{Z_6-Z_7}\\
	-\imag{Z_6+Z_7} & \imag{Z_5} & -Z_4 +\real{Z_5} & \imag{Z_6-Z_7}\\
	\frac{1}{2}(Z_1-Z_2) & -\real{Z_6-Z_7} & \imag{Z_6 - Z_7} & -\frac{1}{2}(Z_1+Z_2)+Z_3\end{array}\right).
\end{align}
The scalar potential is bounded from below \textit{if and only if} all of the below requirements are fulfilled:
\begin{itemize}
	\item All the eigenvalues of $\Lambda_\mu^\nu$ are real.
	\item There exists a largest eigenvalue that is positive, $\Lambda_0>\{\Lambda_1, \Lambda_2, \Lambda_3\}$ and $\Lambda_0>0$.
	\item There exist four linearly independent eigenvectors; one $V^{(a)}$ for each eigenvalue $\Lambda_a$.
	\item The eigenvector corresponding to the largest eigenvalue is timelike, while the others are spacelike,
	\begin{align}
		V^{(0)}\cdot V^{(0)} =~& \left(V^{(0)}_0\right)^2 - \sum_{i=1}^3\left(\vec{V}^{(0)}_i\right)^2>0,\\
		V^{(i)}\cdot V^{(i)} =~& \left(V^{(i)}_0\right)^2 - \sum_{j=1}^3\left(\vec{V}^{(i)}_j\right)^2<0.
	\end{align}
\end{itemize}

\bibliographystyle{jhep}
\bibliography{2HDMbib}

\end{document}